\documentclass[11pt]{article}
\usepackage{amsfonts}
\usepackage{amssymb}
\usepackage{amsmath}
\usepackage{amsxtra}
\usepackage{graphicx}
\usepackage{psfrag}
\usepackage{mathrsfs}
\usepackage{natbib}
\usepackage{stmaryrd}
\usepackage{subfig}
\usepackage{tikz}
\usepackage{empheq}
\usepackage{url}
\usepackage[normalem]{ulem}
\usepackage{color}

\def\bone{\mbox{\boldmath$1$}}
\def\bzero{\mbox{\boldmath$0$}}

\def\bF{\mbox{\boldmath$ F$}}

\def\bN{\mbox{\boldmath$ N$}}

\def\bP{\mbox{\boldmath$ P$}}

\def\bk{\mbox{\boldmath$ k$}}

\def\bn{\mbox{\boldmath$ n$}}

\def\bv{\mbox{\boldmath$ v$}}

\def\bx{\mbox{\boldmath$ x$}}

\begin{document}
\title{Perspectives on the mathematics of biological patterning and morphogenesis}
\author{Krishna Garikipati\thanks{Mechanical Engineering, \& Mathematics, University of Michigan, {\tt krishna@umich.edu}}}
\maketitle
\begin{abstract} 
A central question in developmental biology is how size and position are determined. The genetic code carries instructions on how to control these properties in order to regulate the pattern and morphology of structures in the developing organism. Transcription and protein translation mechanisms implement these instructions. However, this cannot happen without some manner of sampling of epigenetic information on the current patterns and morphological forms of structures in the organism. Any rigorous description of space- and time-varying patterns and morphological forms reduces to one among various classes of spatio-temporal partial differential equations. Reaction-transport equations represent one such class. Starting from simple Fickian diffusion, the incorporation of reaction, phase segregation and advection terms can represent many of the patterns seen in the animal and plant kingdoms. Morphological form, requiring the development of three-dimensional structure, also can be represented by these equations of mass transport, albeit to a limited degree. The recognition that physical forces play controlling roles in shaping tissues leads to the conclusion that (nonlinear) elasticity governs the development of morphological form. In this setting, inhomogeneous growth drives the elasticity problem. The combination of reaction-transport equations with those of elasto-growth makes accessible a potentially unlimited spectrum of patterning and morphogenetic phenomena in developmental biology. This perspective communication is a survey of the partial differential equations of mathematical physics that have been proposed to govern patterning and morphogenesis in developmental biology. Several numerical examples are included to illustrate these equations and the corresponding physics, with the intention of providing physical insight wherever possible.
\end{abstract}
\section*{Keywords} Reaction; diffusion; phase segregation; nonlinear elasticity; buckling

\section{Introduction and background}
\label{sec:Intro}
Developmental biology is concerned with the development of patterns and morphological form (morphogenesis) in organisms. It is useful to make these terms precise at the outset in order to enable a mathematical physics-centered discussion: The term \emph{pattern} will be applied here to a scalar field in one-, two- or three-dimensional manifolds. \emph{Morphological form} will be taken to refer to the vector placement of material points of the developing organism, in the spirit of D'Arcy Thompson \citep{Thompson1917}. The central quest for mathematical physics in this context is to seek a quantitative description that governs patterning and morphogenesis. While the plan for development of an organ, system or entire organism is executed by gene expression, this cannot happen without sampling of ``positional'' information. This may come as the geometry, which is the same as the morphological form, itself evolving, upon which patterns develop. We note that the literature in developmental biology or even in some quarters of biophysics, may be restricted to temporal evolution only, for which purpose the arguments can be presented in a homogeneous setting, without considering local gradients. However, since patterns and morphology do develop heterogeneously over finite-sized regions of the organ, system or organism, spatial variation cannot be ignored, and insofar as the geometry is continuous, the only rigorous description is that of the partial differential equation. This perspective communication is concerned with the partial differential equations of mathematical physics that have been proposed to govern patterning and morphogenesis in developing organisms. It emphasizes broad observations on the nature of each type of partial differential equation and on how it can translate to both robust and precise patterns and morphological form.

\subsection{Patterning by reaction-transport phenomena; size and position}
\label{sec:sizepattern}

Scalar fields that define patterns are governed by reaction-transport equations. Alan Turing famously lay the groundwork for this description in his landmark paper \citep{Turing1952}. In it, he considered the dynamics of two or more chemical species, or morphogens, evolving by diffusion, and by either enhancing or suppressing the production rates of themselves and of each other. He theorized that the patterns laid down by these morphogens would drive growth by gene expression and cell differentiation. A linearized analysis reveals that, depending on the combination of coefficients, diffusion can drive an instability in the pattern, leading to growth in certain modes, and the emergence of patterns. \cite{Gierer1972} built on Turing's ideas by showing that for an activator-inhibitor species pair, short-range activation combined with long-range inhibition could give rise to primary patterns, which, if they guided cell differentiation, could lead to tissue patterning. A wide range of patterns can be obtained with nonlinear reactions, and some of them bear striking resemblance to markings on animal skins, seashells, plant leaves and petals to name but a few examples \citep{Murray2002,Murray2003}. 

\begin{figure}
\begin{minipage}[]{0.5\textwidth}
\centering
\subfloat[{\tiny Cheetah, Mark Probst, Wiki Images, Creative Commons BY-SA 2.0}]{\includegraphics[width=\textwidth]{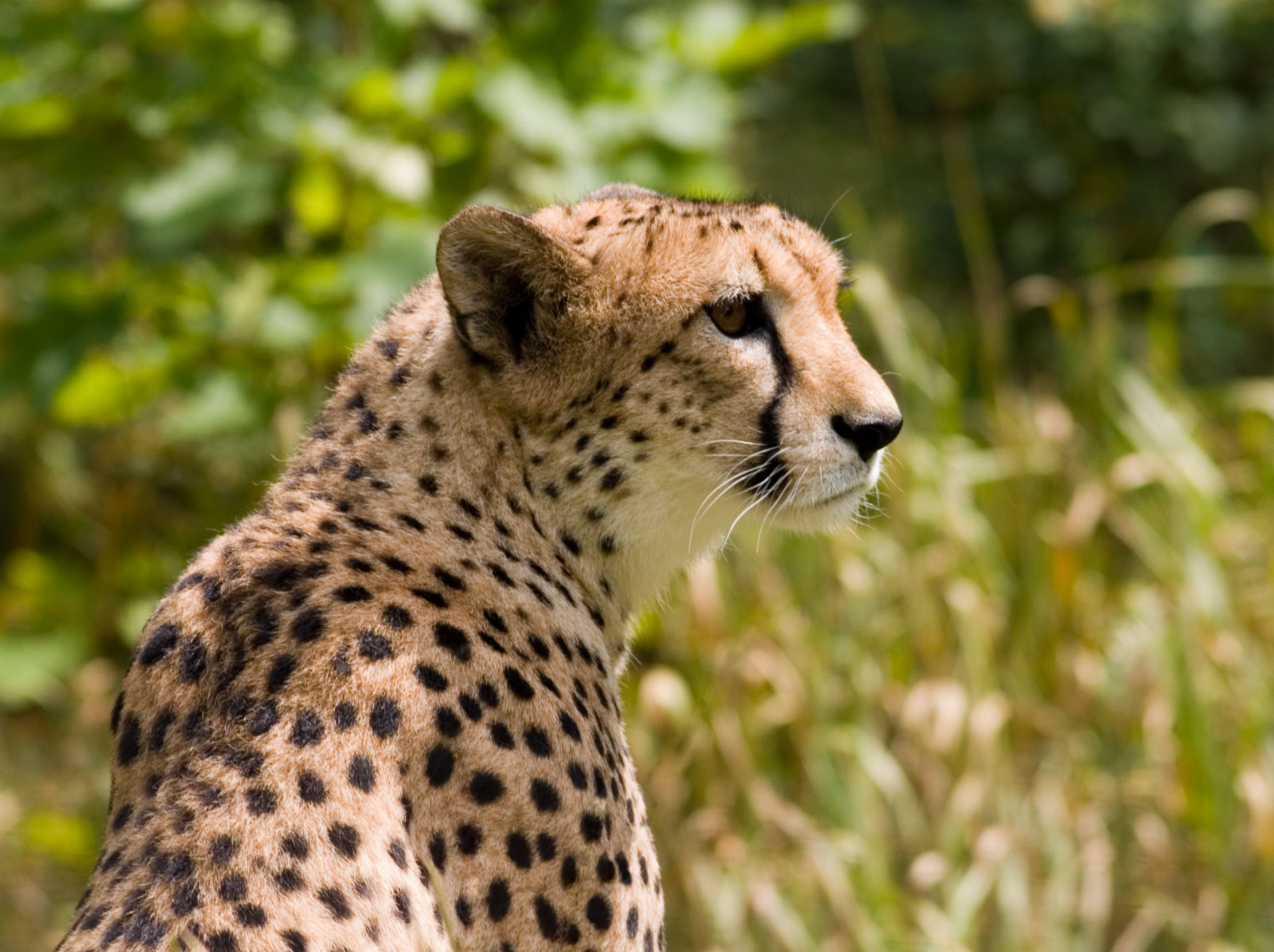}}
\end{minipage}
\begin{minipage}[]{0.5\textwidth}
\centering
\subfloat[{\tiny Adder, Steve Jurvetson, Wiki Images, Creative Commons BY 2.0}]{\includegraphics[width=\textwidth]{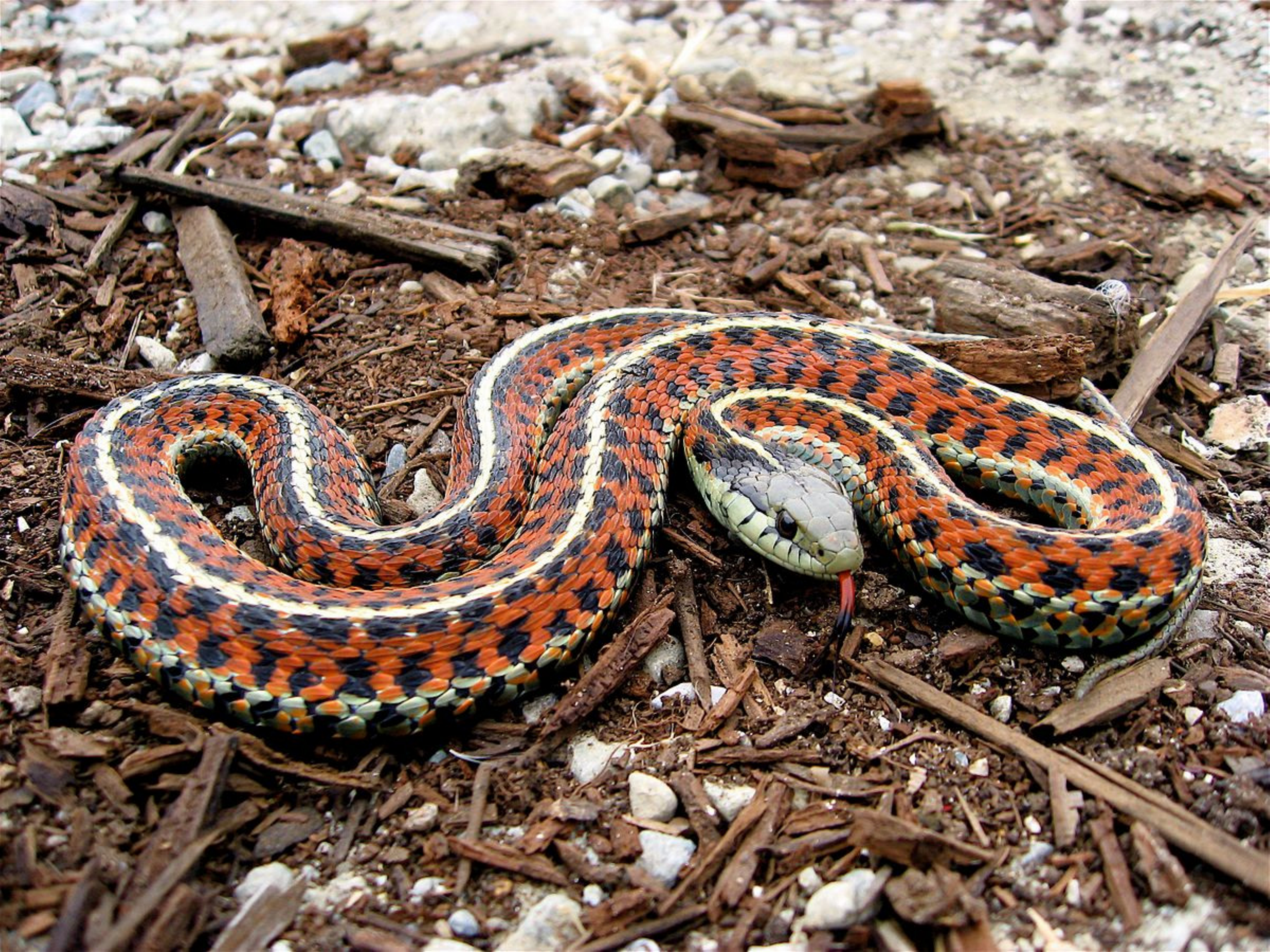}}
\end{minipage}
\begin{minipage}[]{0.5\textwidth}
\centering
\subfloat[{\tiny Tricyrtis, Andr\'{e} Karwath, Wiki Images, Creative Commons BY-SA 2.5}]{\includegraphics[width=\textwidth]{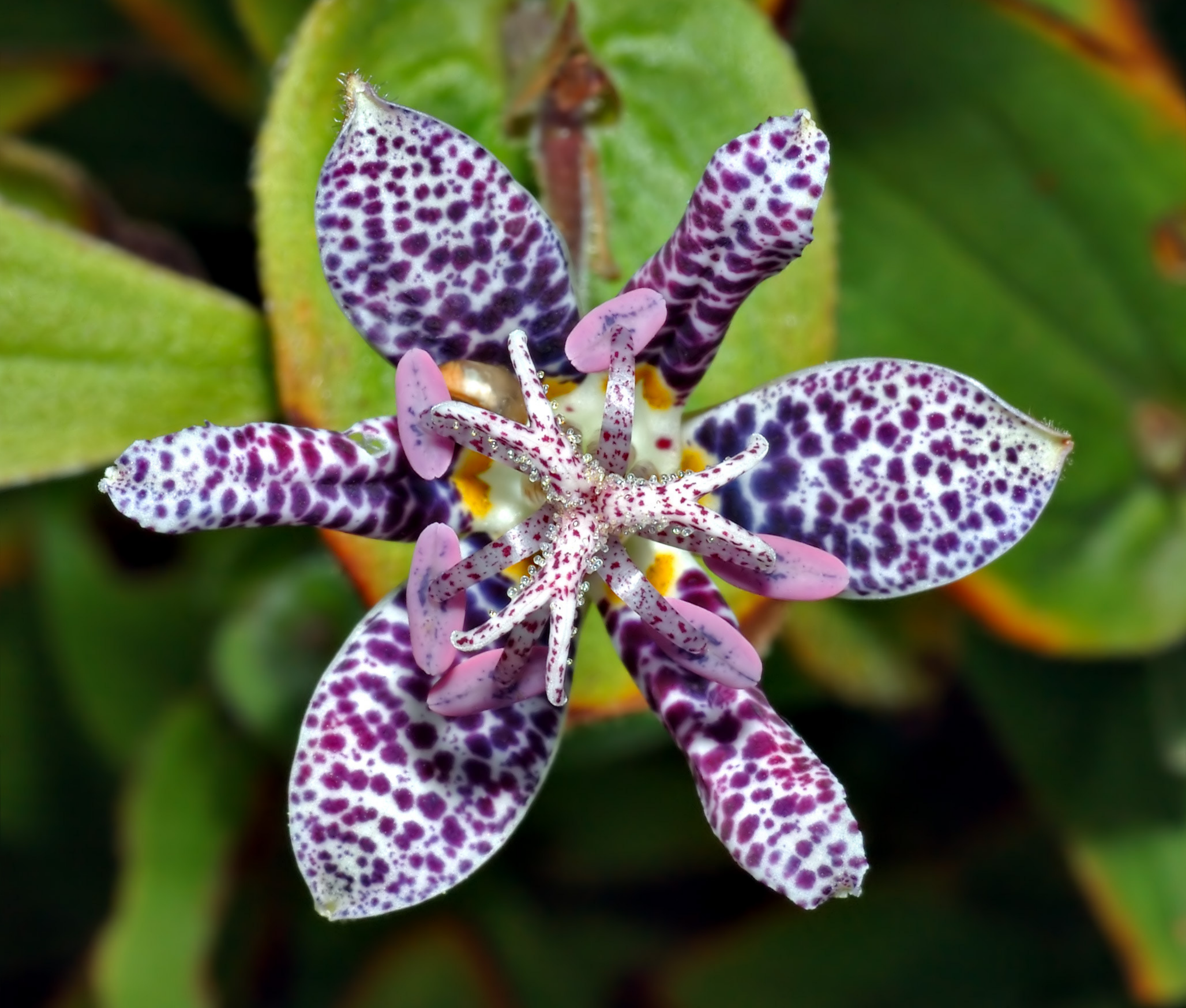}}
\end{minipage}
\begin{minipage}[]{0.5\textwidth}
\centering
\subfloat[{\tiny Villi, University of Nottingham School of Veterinary Medicine, Creative Commons}]{\includegraphics[width=\textwidth]{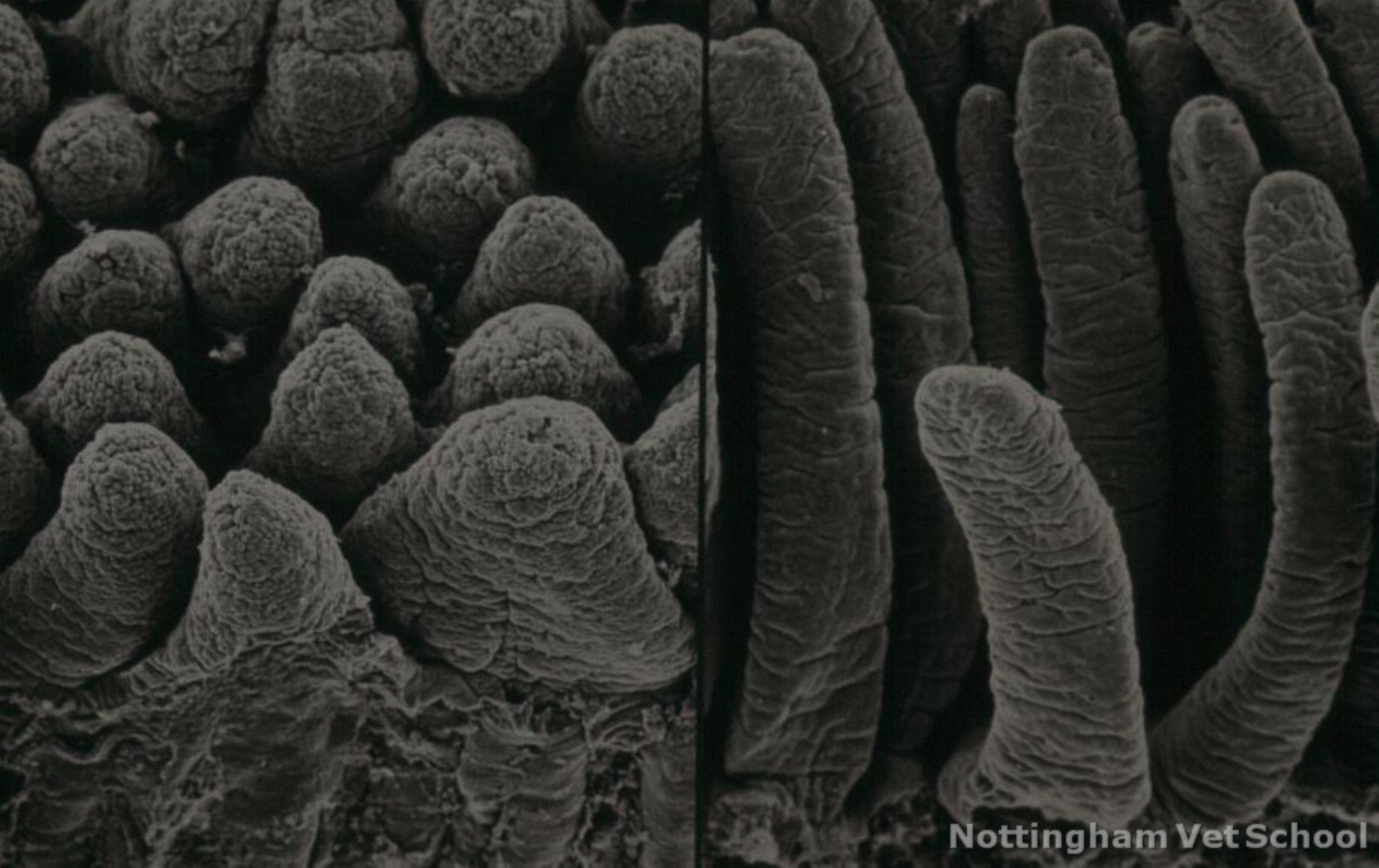}}
\end{minipage}
\caption{Animal and plant patterns of robust sizes but variable position.}
\label{fig:patternSize}
\end{figure}

\begin{figure}[hbt]
\centering
{\includegraphics[width=0.6\textwidth]{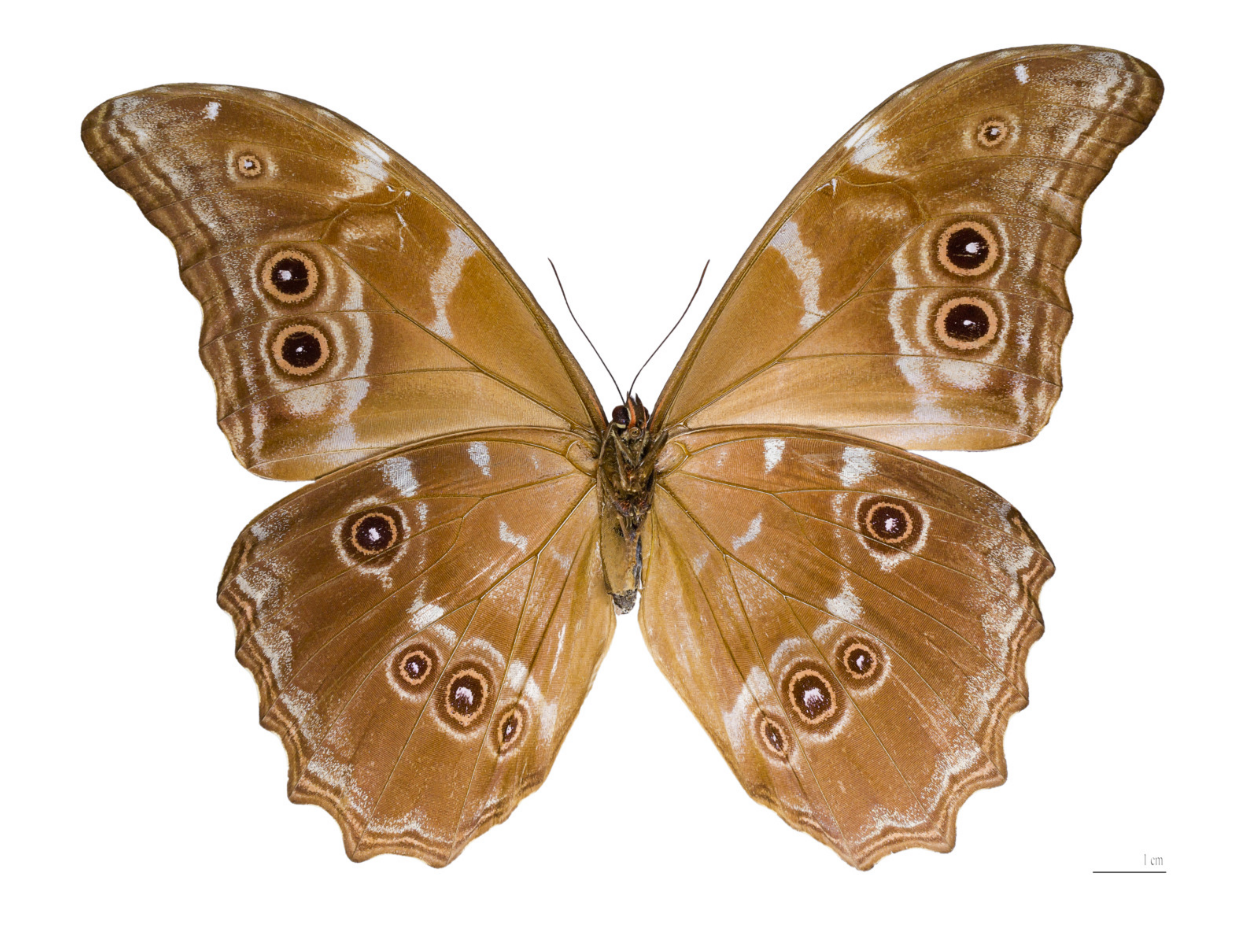}}
\caption{\emph{Morpho didius} wing pattern of robust sizes and position. (Didier Descouens, Wiki Images, Creative Commons BY-SA 4.0)}
\label{fig:patternSizePosition}
\end{figure}

\begin{figure}[hbt]
\centering
{\includegraphics[width=0.6\textwidth]{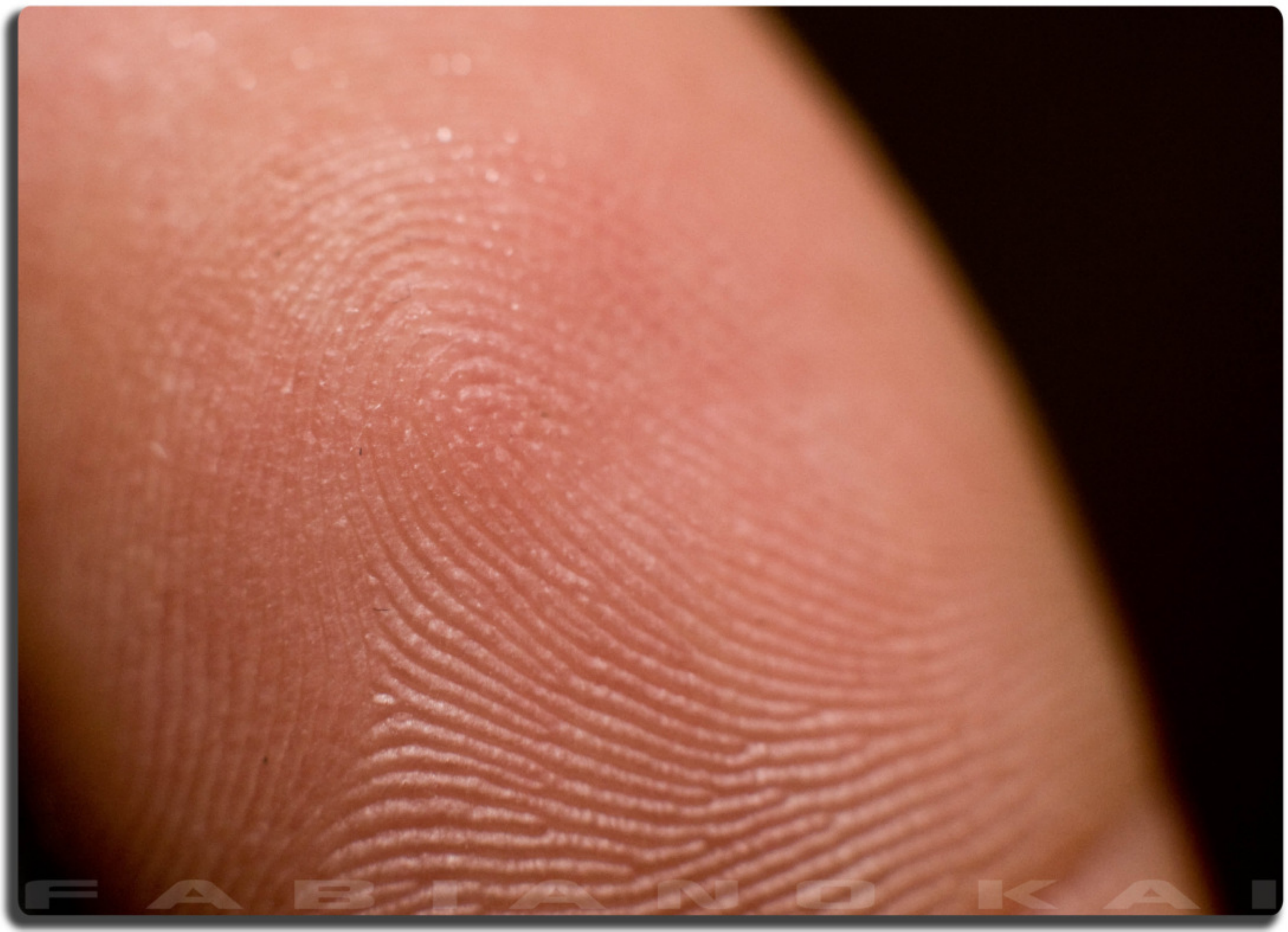}}
\caption{Fingerprints have robustness of size, but have robustness of only certain aspects of position. (Fabiano Kai, Creative Commons)}
\label{fig:patternSizepartPosition}
\end{figure}

Many biological patterns typically show robustness in feature size. This is true of the patterns appearing in Figures \ref{fig:patternSize}, \ref{fig:patternSizePosition} and \ref{fig:patternSizepartPosition}. However, given two individuals of a species with similar or (hypothetically) identical size, the field values that define the patterns on each are not the same functions of position for the examples in Figure \ref{fig:patternSize}. In the case of Figure \ref{fig:patternSizePosition}, however, the argument can be made that the field values of the patterns on (hypothetical) individuals of identical size must also be identical. 

This may not appear to be a very consequential set of observations when applied to the animal and plant patterns and morphologies  in Figure \ref{fig:patternSize}. In mathematical terms, whereas a frequency analysis of these patterns would yield the same spectrum, any Hilbert norm of the difference in field values would differ significantly from zero. However, when applied to the butterfly wing pattern in the example of  Figure \ref{fig:patternSizePosition}, it has consequences. It is reasonable to require that Hilbert norms of the difference in field values of patterns between similarly sized individuals of the species, here \emph{Morpho didius}, must not be very large. When applied to morphological form, this statement brings mathematical precision to the requirement that body features appear in the same positions and be similarly sized. An intermediate example is Figure \ref{fig:patternSizepartPosition}. Fingerprints demonstrate robust size of pattern, and of the position of the central whorl, even though the fingerprint is otherwise unique, implying that, in this sense, the pattern is not robust across individuals.

Turing's ideas have seeded a flourishing tradition of reaction-diffusion models of patterning in mathematical biology. Prominent among these are applications to butterfly and mammalian markings \citep{Murray1981}, fish patterns \citep{KondoAsai1995,Barrio1999}, seashells \citep{Meinhardt1987,Meinhardt2009} and studies of patterns driven by chemotaxis \citep{Painter1999}. Numerical studies include those by \cite{Barrio1999} and hybrid approaches that couple stochastic and deterministic reaction-diffusion equations \citep{Spill2015}. The question of robustness of patterns in the face of perturbations has been considered by \cite{MainiByrne2012} and the so-called Turing instability by \cite{Korvasova2015}. The interesting case of robustness of patterns on uniformly growing systems was taken up by \cite{Crampin1999}. The effects of mixed and uniform boundary conditions on the uniqueness, stability and sensitivity of solutions to domain changes were studied by \cite{Dillon1994}. An early mechanochemical model by \cite{MurrayMaini1988} was one of the more comprehensive in its treatment of diffusing and reacting morphogens as well as migrating cell populations modelled by random walks (diffusion), chemotaxis and advection by tissue deformation.

A somewhat different use of reaction-diffusion equations has become central to explaining the development of the imaginal wing of \emph{Drosophila melanogaster}, as reviewed by \cite{Wartlick2011}. It is commonly accepted that decapentaplegic (DPP), a morphogen, is secreted by cells in the compartment boundary, and that the concentration in a given cell controls growth there. The theories proposed have included scaling of the DPP gradient by an  expander molecule that acts to repress the former's action \citep{BenZvi2011,Restrepo2011}, control of the morphogen's spatial gradient by communication between cells \citep{Day2000}, as well as complementary inhibition of the morphogen \citep{Campbell1999}. Other models include a role for mechanics via the stress induced by growth: The growing disc induces tension in the periphery, stimulating further growth there as well as compression in the center. Above a threshold, the compressive stress shuts off growth in the center, and consequently in the periphery \citep{Aegerter2007,Hufnagel2007}. 

Transport equations also describe the motion of cell populations; in the continuum limit cells are not individually tracked, but are represented by concentration fields. In the treatment advanced by \cite{MurrayMaini1988} diffusion represented short range cell migration, advection modelled chemo- and haptotaxis, and reaction terms were used for cell division and death. The authors also included what they termed as long-range diffusion to account for non-dilute cell concentrations, and the nonlocal interactions of cells bearing filopodia. They represented this effect by fourth-order diffusion, recognizing its potential to stabilize the transport equation. The reaction-advection-diffusion treatment of the evolution of cell populations has been used in modelling tumor dynamics; see for example {\cite{Jackson2002,Byrne2003,Byrne12003}}, as well as \cite{Narayanan2010} and \cite{Rudraraju2013}, of which the latter work also considered stress-driven migration as an advection term. A number of authors have included phase field methods to represent the progression of the tumor wall in models of tumor dynamics \citep{Wise2008,Cristini2009,Lowengrub2010,Oden2010,CristiniLowengrub2010,Chatelain2011}. The development of more complex morphologies has also been modelled, such as of angiogenesis \citep{Vilanova2013,Vilanova2014,Xu2016}.

\subsection{Morphogenesis by elasticity; size and position}
\label{sec:sizefield}

During development, the patterns on an organism and its morphological form evolve on the time scales of cell division or migration, which are much greater than those of elastic wave propagation in soft tissue. Morphogenesis can driven by differential growth---a subject of classical studies by  \cite{Thompson1917} and  \cite{Huxley1932}---and governed by quasistatic nonlinear elasticity that determines the placements of material points. {While the mechanisms of morphogenesis during early development can include cell intercalation and contraction, the main emphasis here is on differential or inhomogeneous, growth, which leads to local morphological form by elastic buckling followed by folding, wrinkling or creasing \citep{Cai2010}.} The onset of these instabilities, and the nonlinear evolution of the surface post-instability have been applied to study the formation of intestinal villi \citep{Freddo2016} and crypts \citep{Hannezo2011}, gut looping \citep{Savin2011}, intestinal tissues \citep{BenAmar2013}, the morphology of petals and leaves {\citep{Dervaux2008,BenAmar2012}}, seashells \citep{Chirat2013}, of gels viewed as a surrogate for soft tissue \citep{Hong2009,Jin2011,BenAmar2010,Li2012,Prost2015}, as well as extensively to the development of folds (sulci and gyri) in the brain \citep{Richman1975,Xu2010,Bayly2013,Tallinen2013,Budday2014,Goriely2015,Tallinen2016}.

\begin{figure}[hbt]
  \centering
 \includegraphics[width=0.5\textwidth]{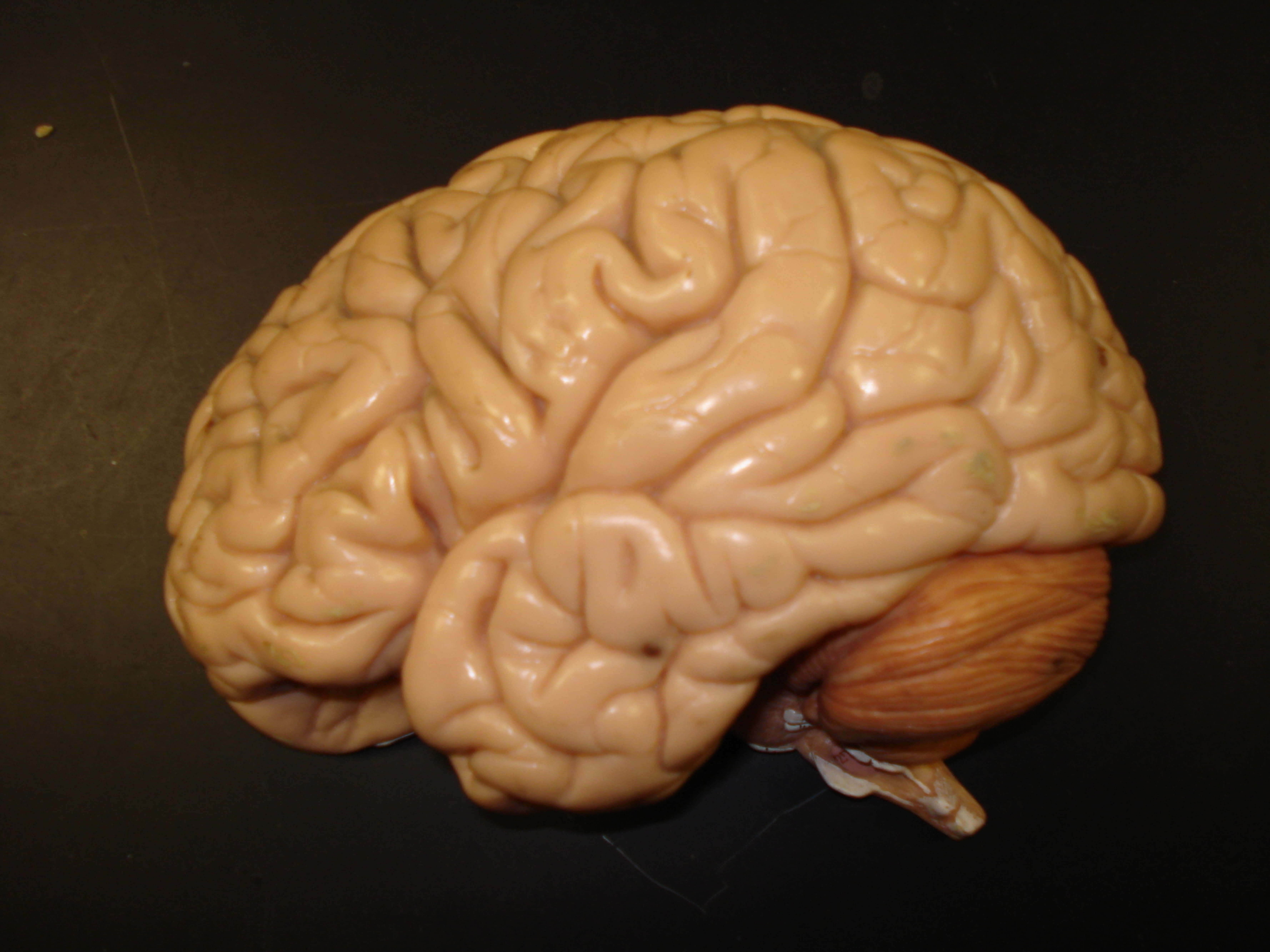}
\caption{The brain may appear to have morphological form with features of fixed size but not fixed field values. However, certain sulci and gyri define centers controlling specific neurological functions, and it is important for normal function that they occur in fixed locations and have fixed sizes (image by Shannan Muskopf, used under a Creative Commons license).}
\label{fig:brain}
\end{figure}

\begin{figure}[hbt]
  \centering
 \includegraphics[width=0.5\textwidth]{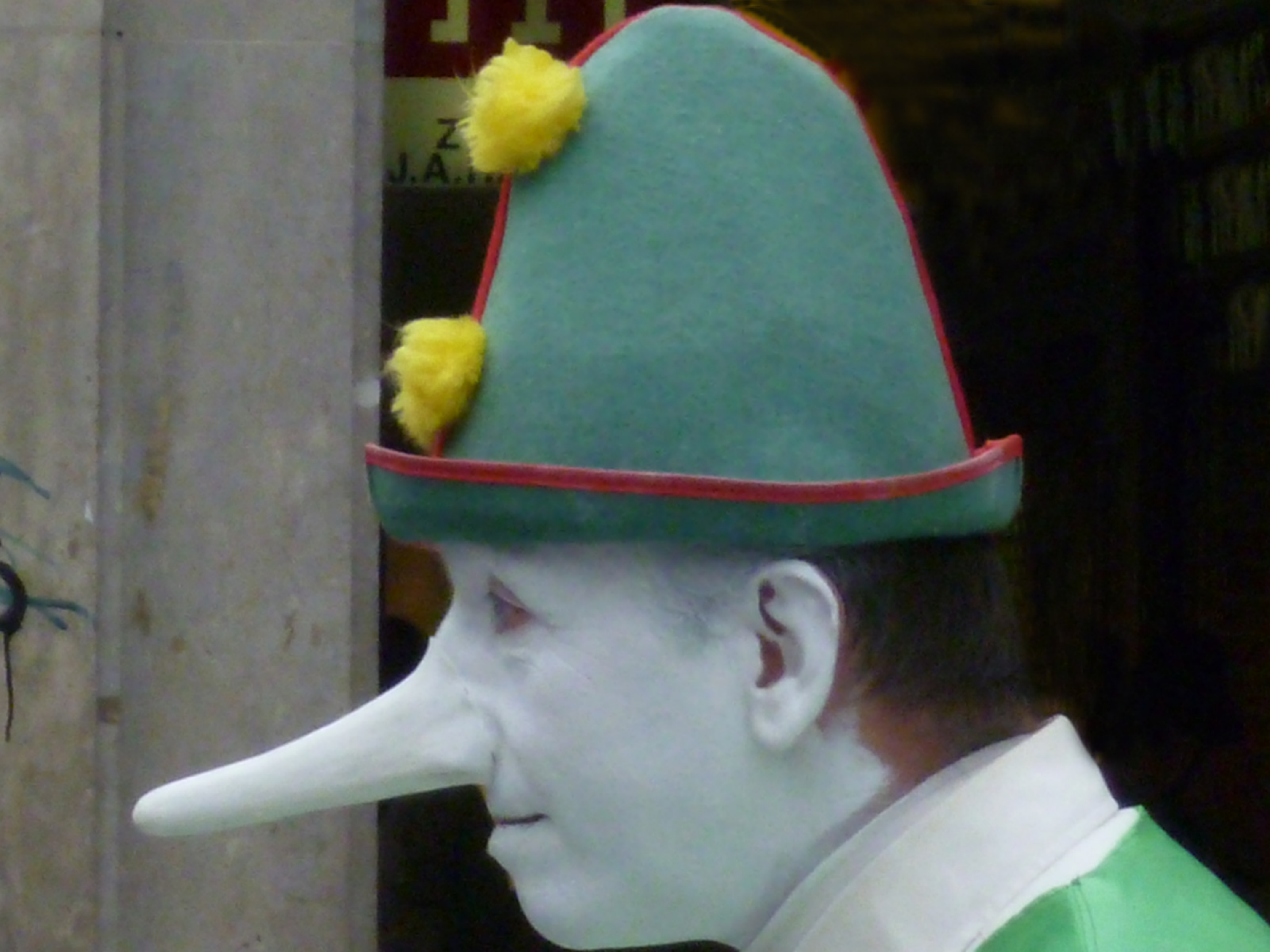}
\caption{An illustrative example of morphological form to emphasize, via exaggeration, that its field values as well as position are fixed (image by Rosmarie Voegtli, used under a Creative Commons license).}
\label{fig:nose}
\end{figure}

As in the case of patterns, the morphological forms appearing in some organs or systems, such as intestinal villi, could show robustness of size but not of position. The spacing between folds, wrinkles and creases are reproduced between individuals, while their positions of themselves are of no consequence. Thus, normally developing animals may or may not have a villus at a given position on the inner wall of their intestines (Figure \ref{fig:patternSizePosition}). We recall the case of human fingerprints discussed in Section \ref{sec:sizepattern} (Figure \ref{fig:patternSizepartPosition}): The spacing of the pattern of ridges would be the same on identically-sized index fingers, and the whorl occurs in a precise position; however, the whorl may be open or closed, and field values of the pattern away from the whorl are unique to the individual. The long history of the use of fingerprints to establish identity is based on this uniqueness. There is some analogy with the brain. To the untrained eye, it may appear that the many sulci and gyri have prescribed sizes between individuals, but arbitrary positions (Figure \ref{fig:brain}). However, neurologists will point out that for healthy brain function, certain centers, demarcated by sulci and gyri, must have positions within fairly tight bounds. The extreme case of the consequence of both size and position of morphological form is at the organ scale. Pinocchhio-like proportions remain the stuff of fantasy (Figure \ref{fig:nose}), and healthy individuals tend to have facial features that obey reasonably tight limits on size and position (eyes, ears and nose, each in their places). It is evident then that gene expression is controlled by feedback on size as well as position.


\subsection{The coupled progression of patterning and morphogenesis}
\label{sec:patternmorph}

It bears note that in the literature of mathematical biology and theoretical biophysics, the treatments of patterning and of morphogenesis have been developed almost independently of each other. Reaction-transport-based treatments of patterning via scalar fields have remained uncoupled from those of elasticity for the modelling of morphological form, except perhaps in the area of tumor growth. (The robustness of the size of a pattern during growth of the underlying domain has been treated by \cite{Crampin1999}, as noted above. However, this is quite different from a progressive pattern of a scalar field and and a growing morphological form influencing each other.) This is curious when one considers the control of size and position of morphological form: The most obvious approach to such control would be to lay down a pattern by a scalar field, followed by differential growth enslaved to this pattern. This was clearly presaged by \cite{Turing1952}, who stated as much in his seminal work on the chemical basis of morphogenesis. The fields of mathematical biology and theoretical biophysics appear to have not yet followed through on thus combining patterning and morphogenesis. This survey of the partial differential equations that have been proposed to govern patterning and morphogenesis addresses this unification by first dwelling on reaction-transport equations for patterning, followed by nonlinear elasticity for morphogenesis, before concerning itself with the coupling of these equation systems. The aim is not to reproduce specific examples of patterning and morphogenesis in great detail. It is, instead, to lay down the relevant systems of equations, outline how they control size and position, and offer a few, possibly novel, insights on replicating some patterning and morphogenetic phenomena. The next three sections in turn consider patterning by reaction-diffusion equations, phase segregation, and morphogenesis by nonlinear elastic instabilities. The coupling of patterning and morphogenesis also is taken up in the fourth section. Concluding thoughts are summarized in the fifth section.

\section{Reaction-diffusion equations and patterning; Turing instabilities}

Before reviewing the role of reaction-diffusion equations in biological pattern generation, it helps to consider the simpler case of Fickian diffusion. For a scalar field whose concentration is $c$, the canonical diffusion problem can be posed over a domain $\Omega\in\mathbb{R}^3$ that represents an organ or developing system with a combination of concentration boundary conditions, and zero flux (vanishing concentration gradient for homogeneous, isotropic diffusion) boundary conditions being appropriate:
\begin{align}
\frac{\partial c}{\partial t} &= D\nabla^2 c &\mathrm{in}\;\Omega\times[0,T]\nonumber\\
c &= \overline{c} &\mathrm{on}\;\partial\Omega_c\nonumber\\
\nabla c\cdot\bn &= 0 &\mathrm{on}\;\partial\Omega_j\nonumber\\
c(\bx,t) &= c^0(\bx) &\mathrm{at}\;t=0\nonumber
\end{align}

\noindent Because diffusional driving forces smooth out gradients, it follows that steady state or equilibrium patterns cannot be sustained by this equation. However, a time-dependent length scale can be defined via the diffusion length, for instance as $l_\mathrm{D} = \sqrt{D T}$.  It is therefore plausible that if there exists an internal clock to trigger genes at programmed times, this diffusion length scale could control organ size. Unarguably, it requires the existence of a complex timing machinery, but the well-established sequence of events during embryo development would seem to obey such a finely programmed timing \citep{Alberts2008}.

Moving on to reaction-diffusion, consider a system of two morphogens, $c_1$ and $c_2$, diffusing and reacting with each other, while satisfying zero flux boundary conditions.
\begin{subequations}
\begin{align}
\frac{\partial c_1}{\partial t} &= D_1\nabla^2 c_1 + R_{11}c_1 + R_{12}c_2, &\frac{\partial c_2}{\partial t} &= D_2\nabla^2 c_2 + R_{21}c_1 + R_{22}c_2 &&\mathrm{in}\;\Omega\times[0,T]\label{eq:diffreacpdes}\\
\nabla c_1\cdot\bn &= 0, &\nabla c_2\cdot\bn &=0 &&\mathrm{on}\;\partial\Omega\label{eq:diffreacbcs}\\
c_1(\bx,t) &= c_1^0(\bx), &c_2(\bx,t) &= c_2^0(\bx) &&\mathrm{at}\;t=0\label{eq:diffreacics}
\end{align}
\end{subequations}

\noindent For this linear reaction-diffusion system, $R_{11}, R_{22} >  0$ represents auto-activation, $R_{11}, R_{22} <  0$ represents auto-inhibition, $R_{12}, R_{21} >  0$ represents cross-activation and $R_{12}, R_{21} <  0$ represents cross-inhibition. General solutions take the form
\begin{equation}
c_\alpha(\bx,t) = \sum\limits_{k} T_{\alpha k}^0 e^{i\bk\cdot\bx} e^{\omega_{k} t},\quad \alpha = 1,2.
\end{equation}
Substitution in Equation (\ref{eq:diffreacpdes}) yields the conditions:

\begin{subequations}
\begin{align}
T_{1k}^0 \omega_{k} &= -D_1 \vert\bk\vert^2 T_{1k}^0 + R_{11}T_{1k}^0  + R_{12} T_{2k}^0\\
T_{2k}^0 \omega_{k} &= -D_2 \vert\bk\vert^2 T_{1k}^0 + R_{21}T_{1k}^0  + R_{22} T_{2k}^0
\end{align}
\label{eq:linearstabilityeqs}
\end{subequations}
For non-trivial $T_{1k}^0$ and $T_{2k}^0$ the frequency and wave numbers must satisfy
\begin{equation}
\omega_k = \frac{-\left((D_1+D_2)\vert\bk\vert^2-R_{11}-R_{22} \right)\pm\sqrt{\left((D_1+D_2)\vert\bk\vert^2 - R_{11} - R_{22} \right)^2 + 4R_{12}R_{21}}}{2},
\label{eq:diffreaclinstability}
\end{equation}
showing that the stabilizing effect of diffusion can be lost for certain combinations of reaction coefficients, leading to $\omega_k > 0$ and growth in time of the $\bk$ mode. On this basis, Turing considered a number of oscillatory-in-time cases, where the imaginary component, $\mathrm{Im}(\omega_k) \neq 0$ as well as ``stationary'' cases that are stable with the real component $\mathrm{Re}(\omega_k) \le 0$ or unstable with $\mathrm{Re}(\omega_k) > 0$.

\begin{figure}
\begin{minipage}[]{0.24\textwidth}
\centering
\subfloat[$c_1\;\mbox{at}\;t = 0$]{\includegraphics[width=\textwidth]{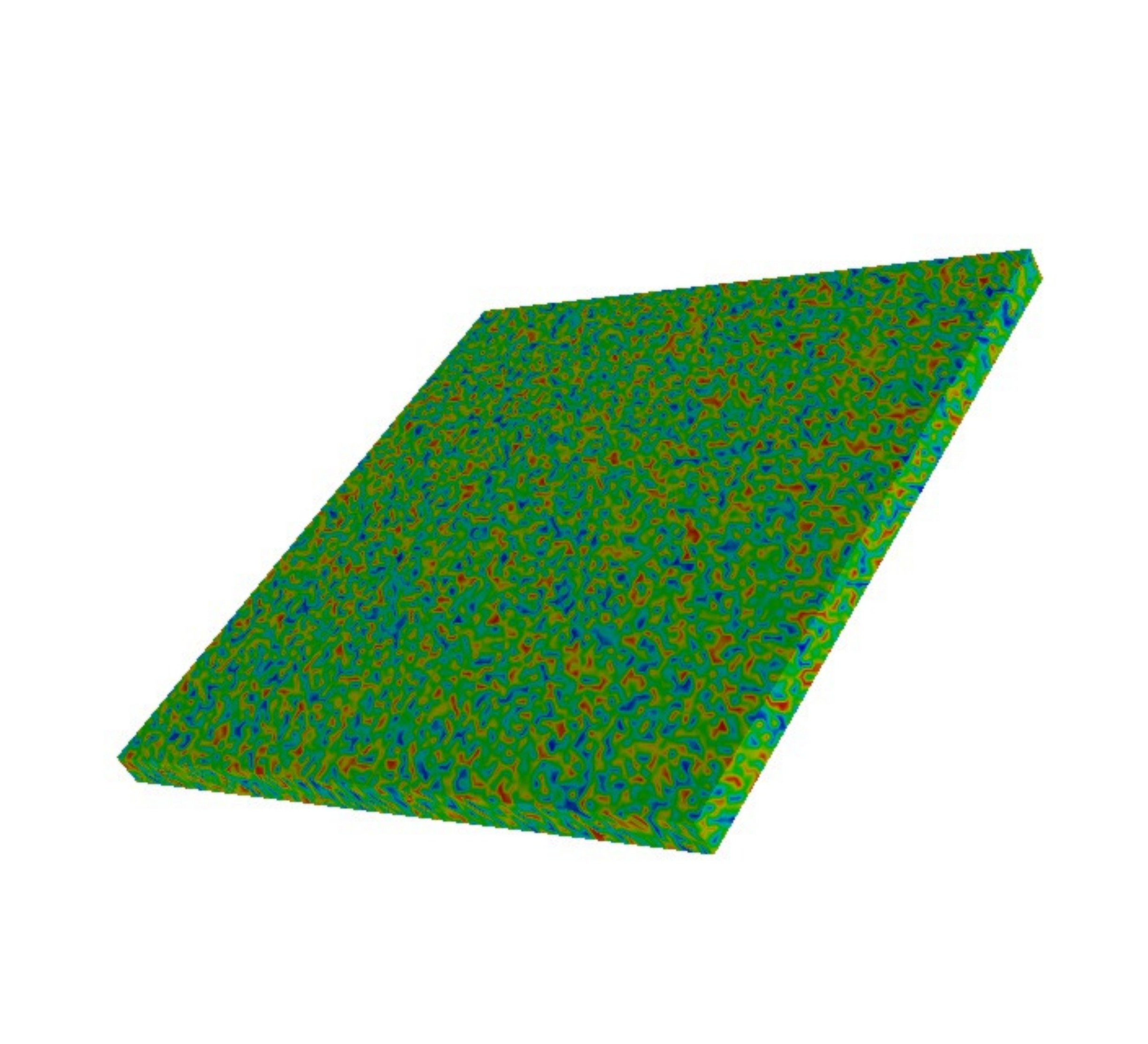}}
\end{minipage}
\begin{minipage}[]{0.24\textwidth}
\centering
\subfloat[$c_1\;\mbox{at}\;t = 4.4$]{\includegraphics[width=\textwidth]{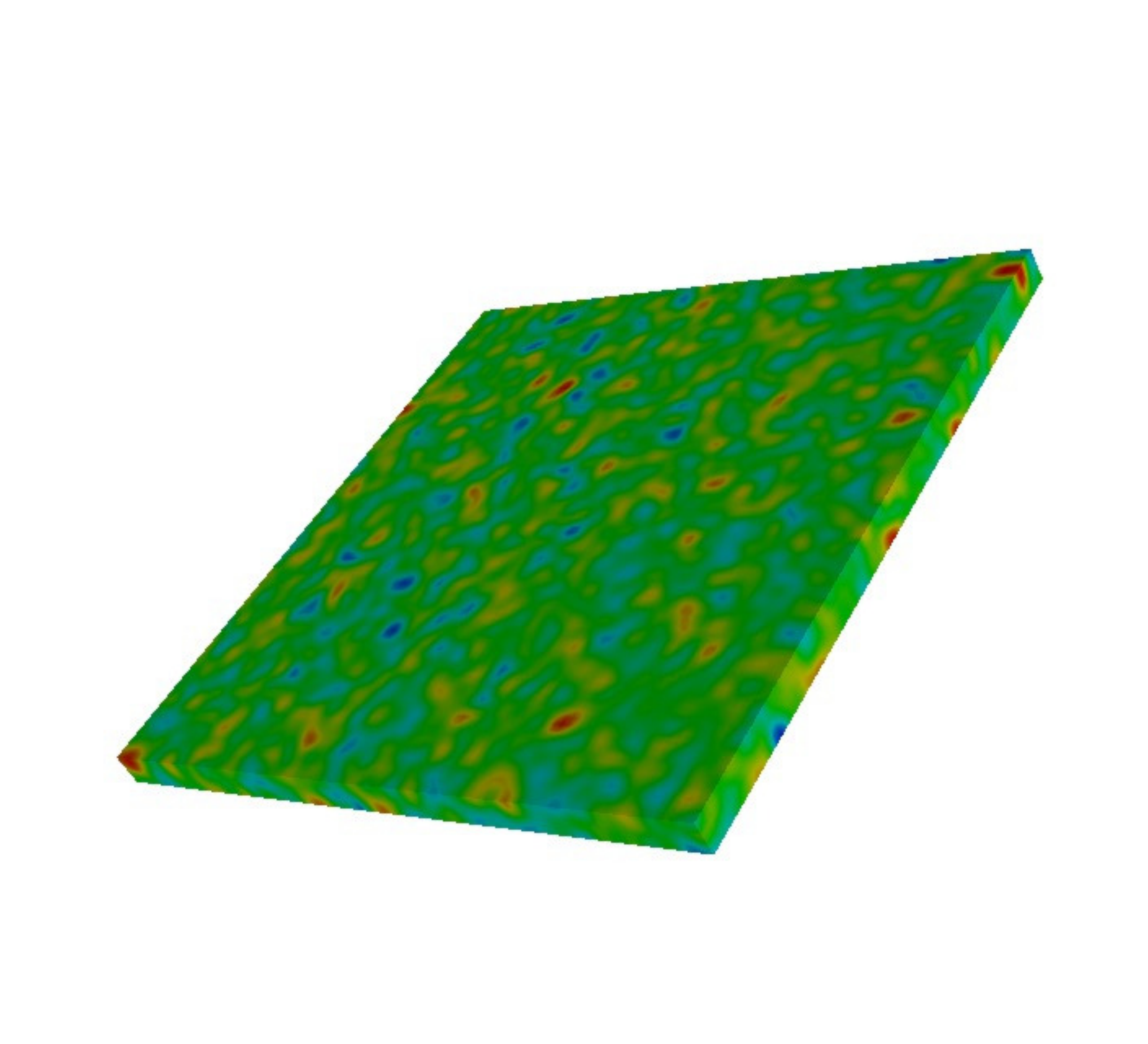}}
\end{minipage}
\begin{minipage}[]{0.24\textwidth}
\centering
\subfloat[$c_1\;\mbox{at}\;t = 68$]{\includegraphics[width=\textwidth]{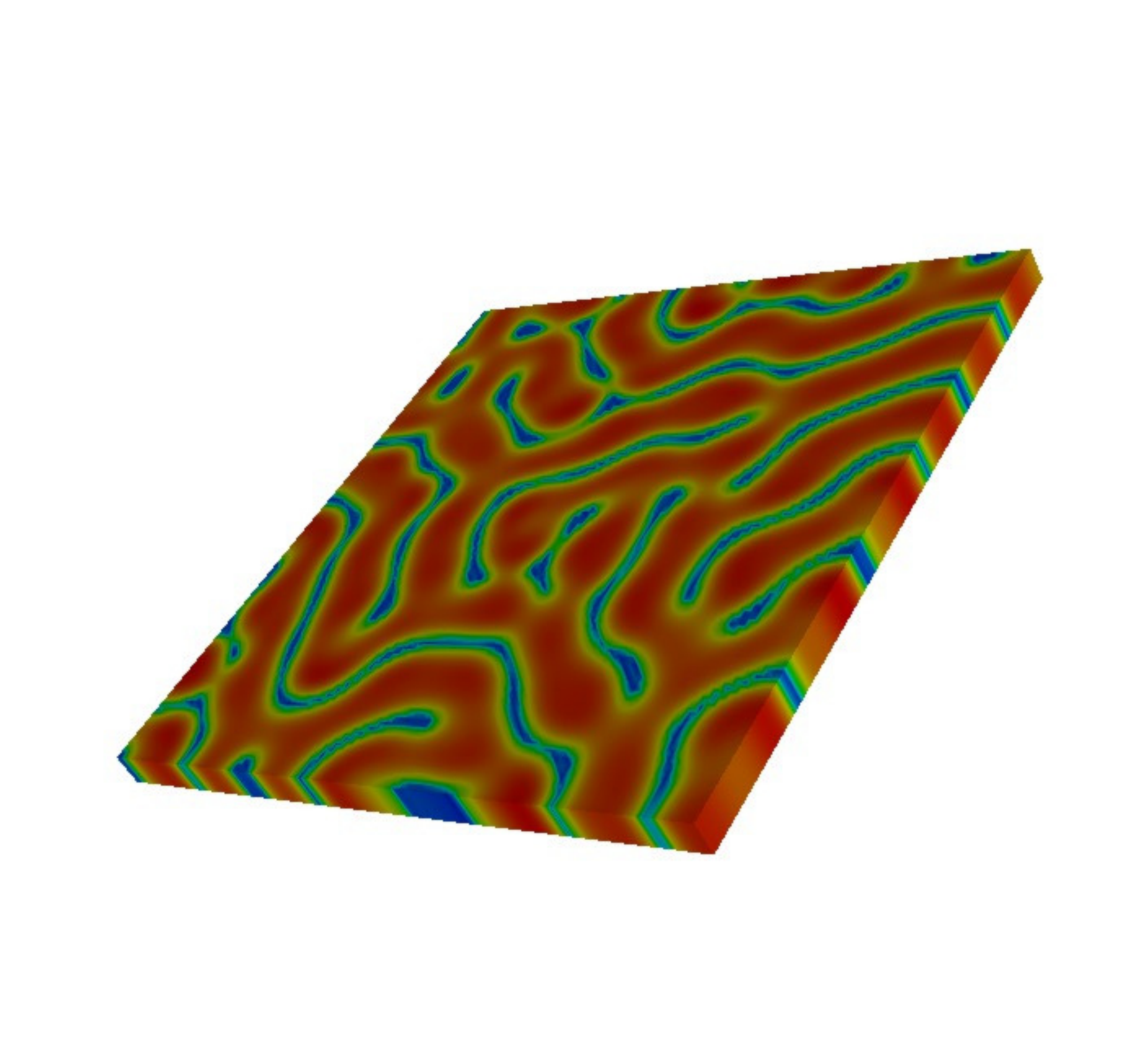}}
\end{minipage}
\begin{minipage}[]{0.24\textwidth}
\centering
\subfloat[$c_2\;\mbox{at}\;t = 625$]{\includegraphics[width=\textwidth]{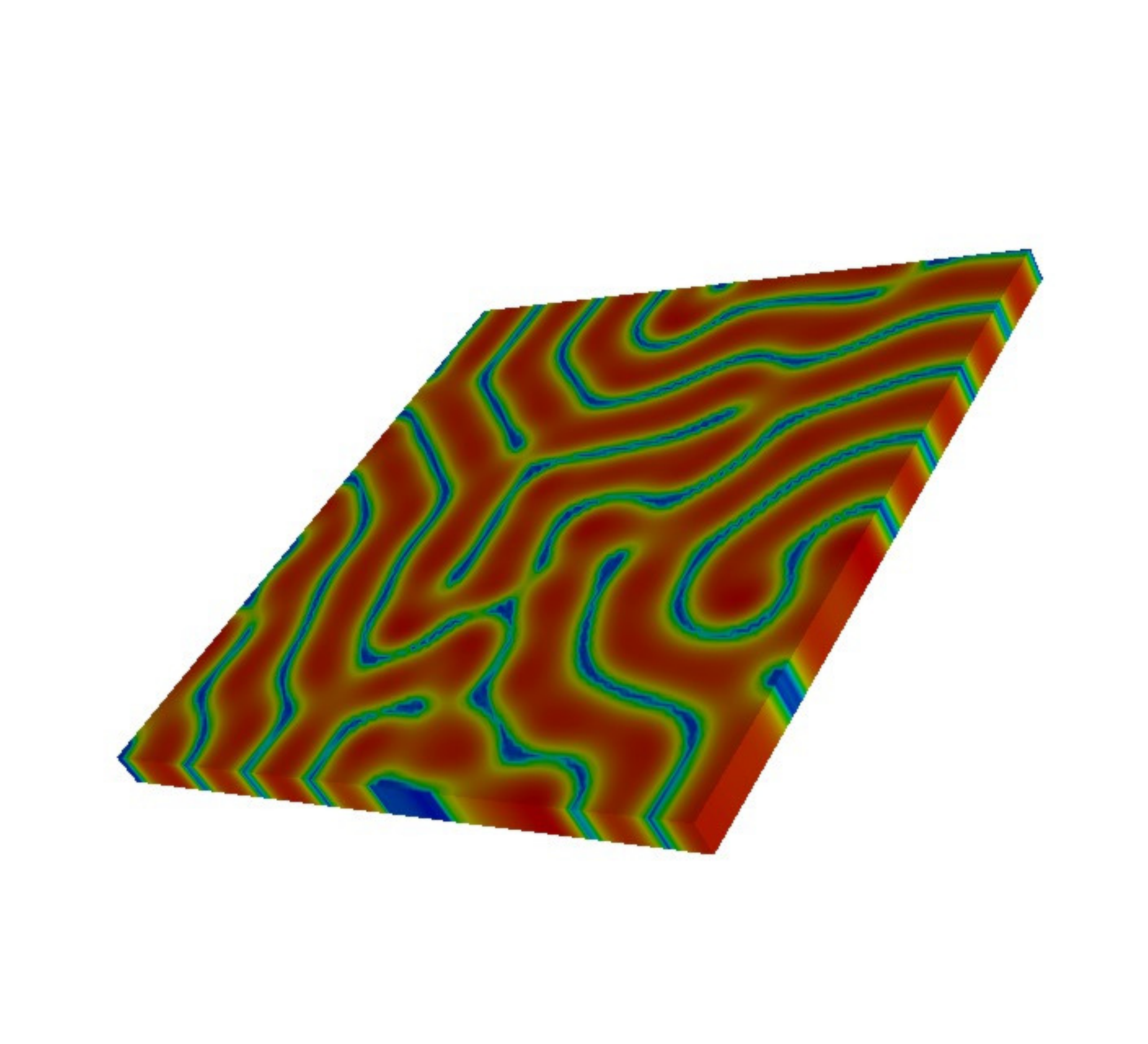}}
\end{minipage}\\
\begin{minipage}[]{0.24\textwidth}
\centering
\subfloat[$c_2\;\mbox{at}\;t = 0$]{\includegraphics[width=\textwidth]{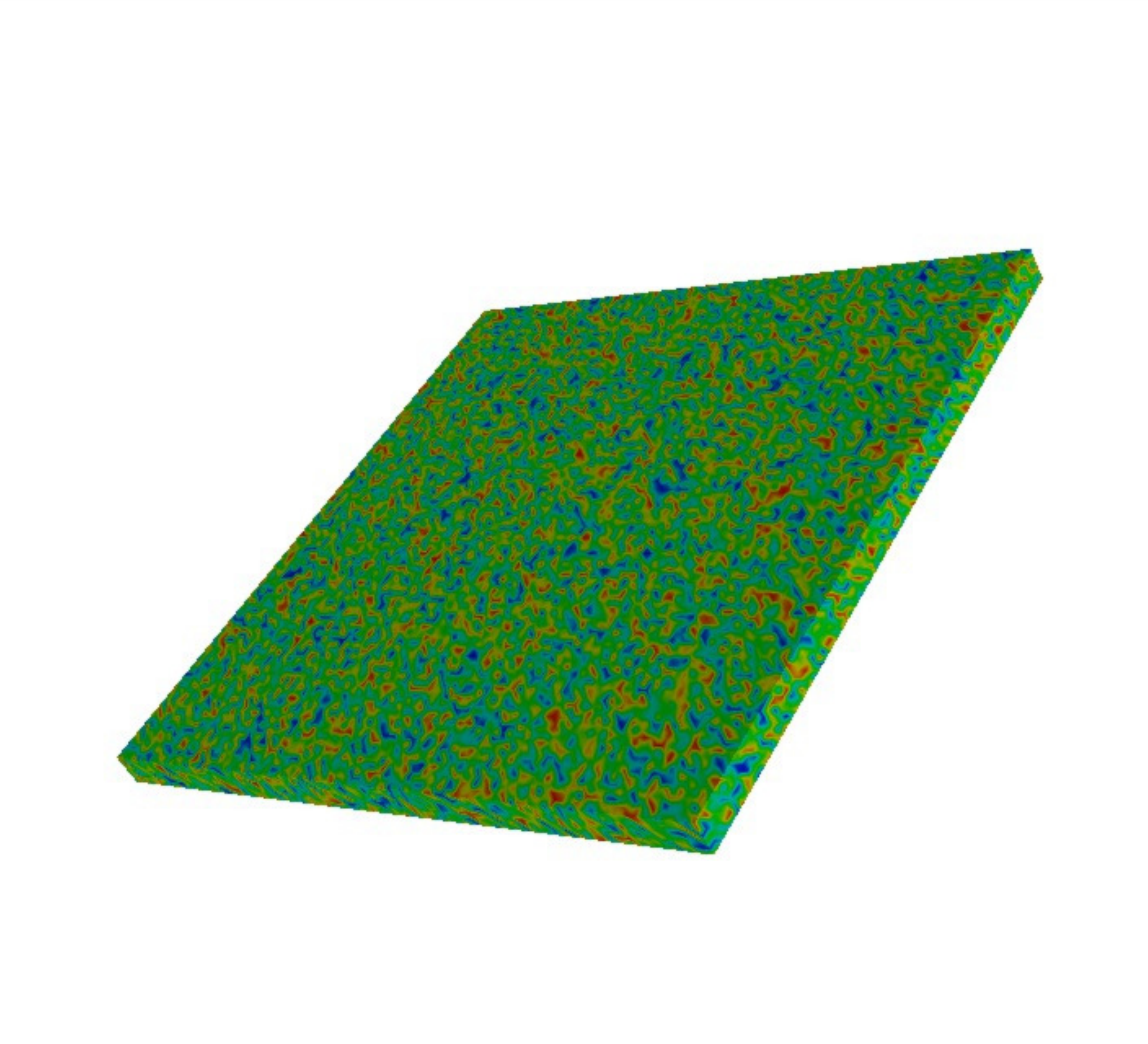}}
\end{minipage}
\begin{minipage}[]{0.24\textwidth}
\centering
\subfloat[$c_2\;\mbox{at}\;t = 4.4$]{\includegraphics[width=\textwidth]{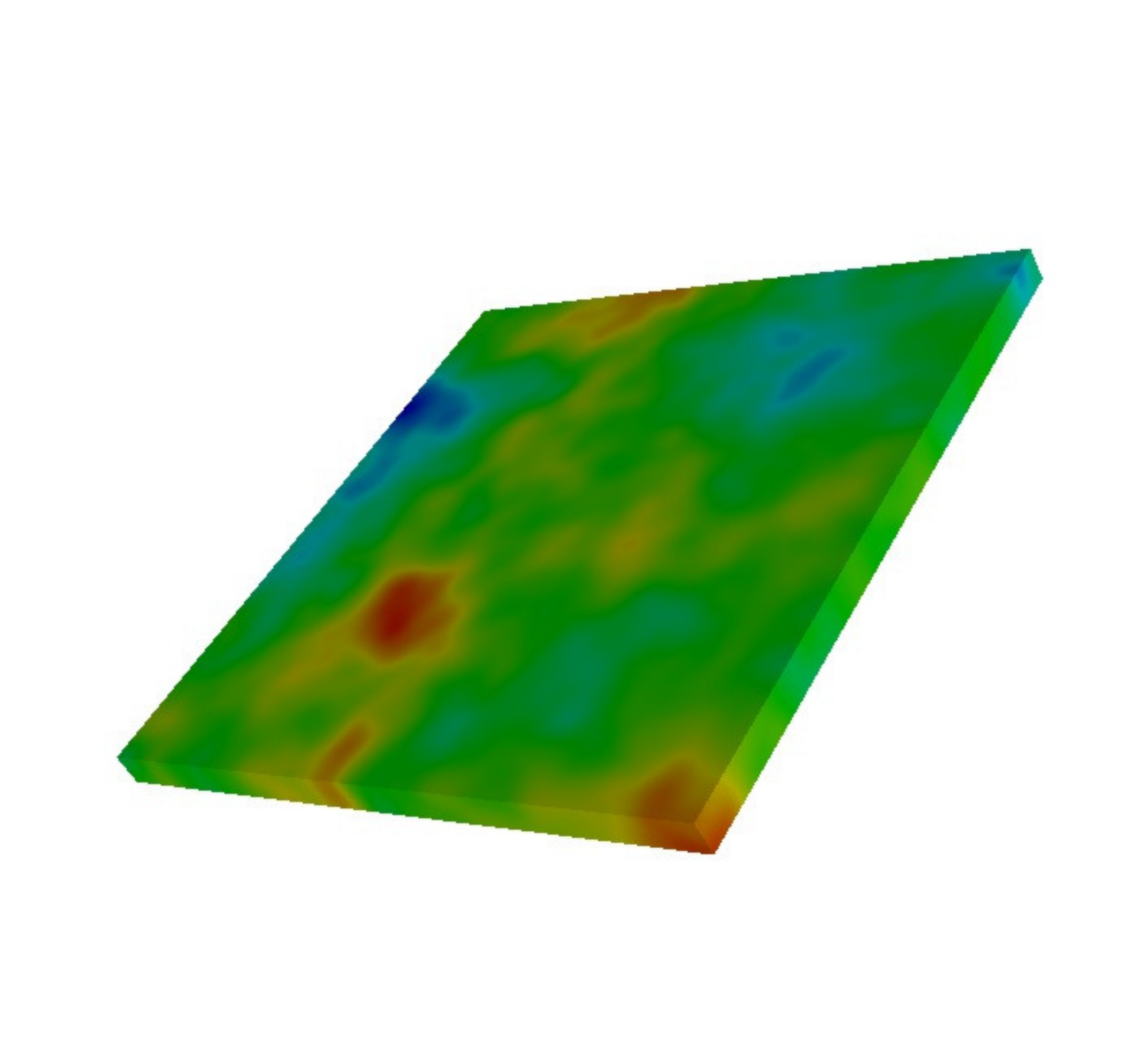}}
\end{minipage}
\begin{minipage}[]{0.24\textwidth}
\centering
\subfloat[$c_2\;\mbox{at}\;t = 68$]{\includegraphics[width=\textwidth]{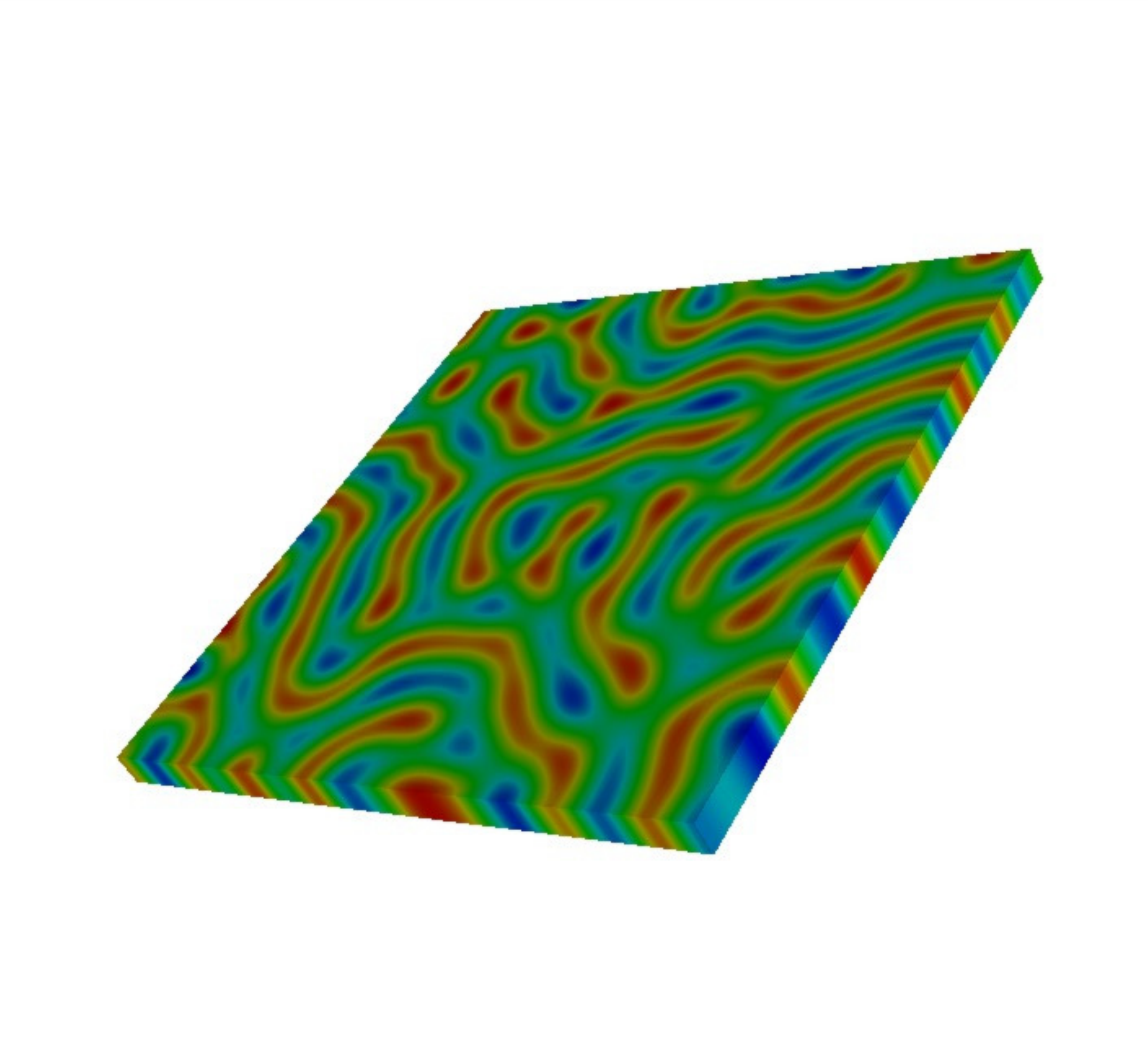}}
\end{minipage}
\begin{minipage}[]{0.24\textwidth}
\centering
\subfloat[$c_1\;\mbox{at}\;t = 625$]{\includegraphics[width=\textwidth]{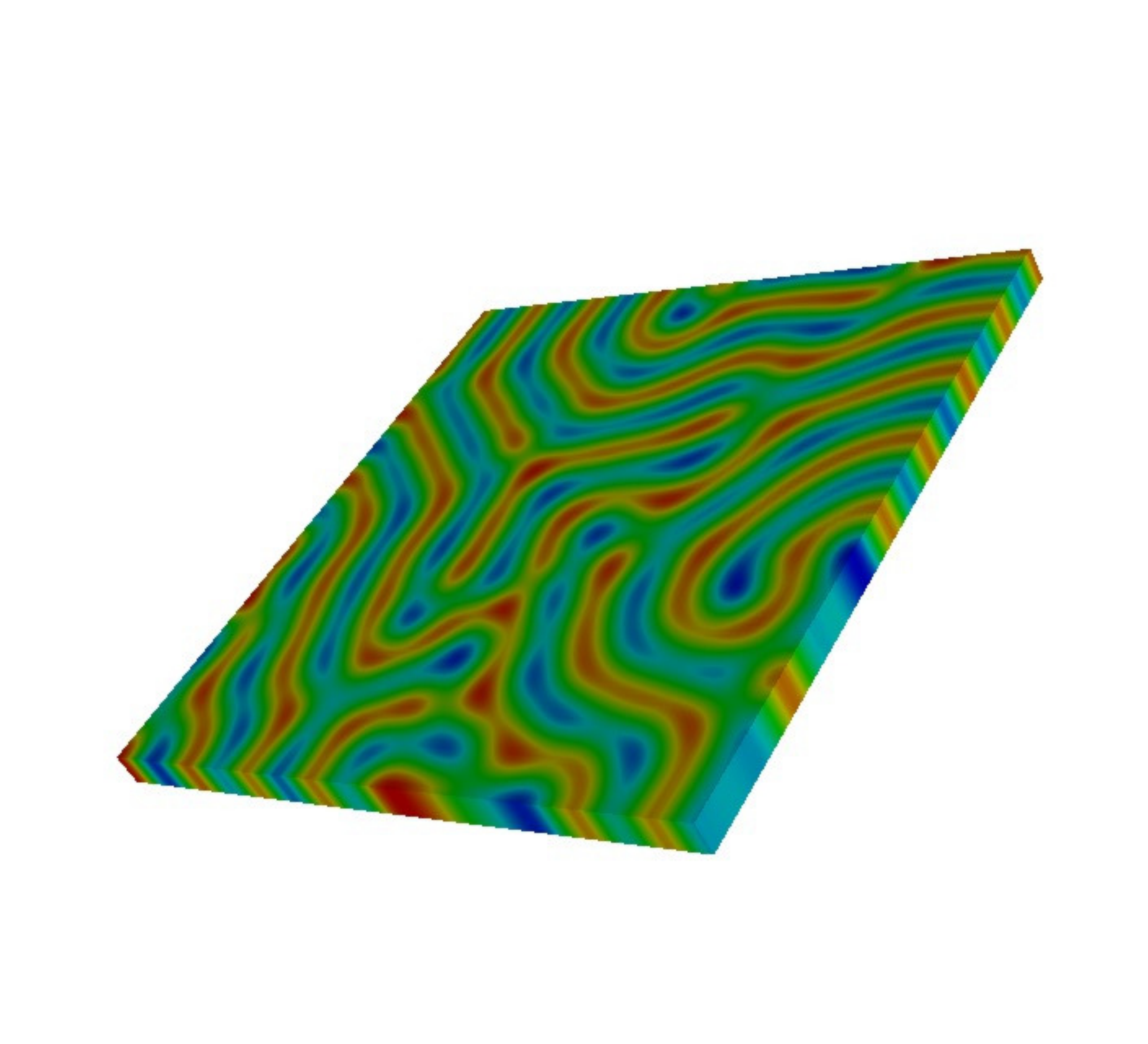}}
\end{minipage}
\caption{Schnakenberg patterns evolving in the two interacting fields. Also see Supplementary Movies S1 and S2.}
\label{fig:schnakenberg}
\end{figure}

The above linearized stability analysis serves only to indicate the possibility of growth in modes. {With nonlinear reaction terms, of course, the stability analysis becomes slightly more complicated, but remains tractable.} The application of reaction-diffusion equations to studies of patterns in the animal and plant kingdoms is based entirely on nonlinear reaction terms \citep{Murray1981,Meinhardt1987,MurrayMaini1988,Meinhardt2009,MainiByrne2012,Gong2012}. \cite{Badugu2012} used nonlinear reaction-diffusion equations in the form of Schnakenberg kinetics \citep{Schnakenberg1976} as a putative explanation for digit patterning during limb development in mouse embryos. Equation (\ref{eq:nonlindiffreacpdes}) summarizes the Schnakenberg reaction-diffusion equations:

\begin{align}
\frac{\partial c_1}{\partial t} &= D_1\nabla^2 c_1 + R_{10} + R_{11}c_1 + R_{12}c_1^2  c_2, \nonumber\\
\frac{\partial c_2}{\partial t} &= D_2\nabla^2 c_2  + R_{20} + R_{21}c_1 + R_{22}c_1^2 c_2\quad\mathrm{in}\;\Omega\times[0,T]
\label{eq:nonlindiffreacpdes}
\end{align}

\noindent with parameters in Table \ref{tbl:schnakenberg} and the same boundary conditions as (\ref{eq:diffreacbcs}). Figure \ref{fig:schnakenberg} is an example of initially random concentration fields evolving to a steady state, { which was shown to exist by \cite{Schnakenberg1976}. The Schnakenberg model induces a rich dynamics including limit cycles and fixed points \citep{Vellela2009}.} Also see Supplementary Movies S1 and S2.

\begin{table}[h]
\centering
\caption{Schnakenberg kinetics' parameters.}
\begin{tabular}{ |c|c|c|c|c|c|c|c|c|  }
\hline
 Parameter & $D_1$ & $D_2$ & $R_{10}$ & $R_{11}$ & $R_{12}$ & $R_{20}$ & $R_{21}$ & $R_{22}$ \\
 \hline
 Value & $1$ & $40$  & $0.1$ & $-1$ & $1$ & $0.9$ & $0$ & $-1$\\
 \hline
\end{tabular}
 \label{tbl:schnakenberg}
\end{table}

The numerical examples illustrated in Figure \ref{fig:schnakenberg}, and those that follow in this communication have been posed and solved in three dimensions by the finite element method programmed using the open source library {\tt deal.ii} \citep{BangerthHartmannKanschat2007,deal284}. The code is parallelized with {\tt Message Passing Interface (MPI)}, and uses the direct SuperLU and iterative GMRES solvers from {\tt PETSc} (\url{https://www.mcs.anl.gov/petsc/}). Algorithnmic differentiation as implemented in the {\tt Sacado} package of the {\tt Trilinos} project (\url{trilinos.org}) has been used to generate Jacobian matrices of the nonlinear residual equations that also reflect the coupling of several fields. The code for all numerical examples presented here is available at \url{https://github.com/mechanoChem/patternMorph}.

\subsection{Control of size and position by reaction-diffusion}
\label{sec:reacDiffSizePos}

Reaction-diffusion equations yield a length scale $l_\mathrm{RD} = \sqrt{D/R}$ (where $D$ and $R$ are generic diffusion and linear reaction constants), which sets the wavelength of patterns. This setting of partial differential equations furnishes a framework to make more precise the notion of a pattern with fixed size, but not position of field: The surface of an animal can be approximated as a 2-manifold on a periodic domain on which such patterns develop (Figure \ref{fig:patternSize}a). The patterns can be approximated as periodic, or at least quasi-periodic, and in the latter case arise far from boundaries. In contrast, butterfly wings and leafs also are $2-$manifolds, but with well-defined boundaries, and develop patterns with robust size and field position (Figure \ref{fig:patternSizePosition}) under influence of the boundary. In this case of patterning by reaction-diffusion phenomena, size is set by the partial differential equation coefficients. In the example of Figure \ref{fig:schnakenberg}, zero flux boundary conditions (\ref{eq:diffreacbcs}) align the stripes perpendicular to the boundary, thus controlling the pattern to some degree. With concentration boundary conditions, of course, precise control of the pattern is attained at the boundaries. With both types of boundary conditions, the pattern propagates into the domain under control of the wavelengths.

\section{Patterning by cell segregation in tissues}
\label{sec:phaseseg}
While reaction-diffusion models do give rise to patterning, stable or steady state patterns only arise for special ranges of coefficients predicted by (\ref{eq:diffreaclinstability}) in the linear case, and examples such as Schnakenberg kinetics in the nonlinear case. The thesis that Turing-like biological patterns can be generated by reaction-diffusion equations relies on the identification of morphogens subject to Turing patterns that then trigger cell differentiation. Morphogen distributions controlled by Fickian diffusion are well-established in the developmental biology of \emph{D. melanogaster} \citep{Campbell1999,Day2000,Aegerter2007,Hufnagel2007,Wartlick2011,BenZvi2011,Restrepo2011} as well as the mouse \citep{Suwinska2008,Nishioka2009}, humans \citep{Warmflash2014} and other mammals \citep{Yu2015}. In chemical systems, such as the chlorite-iodide-malonic acid-starch reaction, the role of Turing patterns appears well established \citep{Maini1997}. However, while there is some evidence in its favor \citep{Raspopovic2014}, the thesis that morphogen fields  form Turing patterns and promote cell differentiation is not yet central to developmental biology. This has left room for consideration of other models for patterning of spots and stripes on insect wings, animal skins, seashells, leaves and flowers. 

\subsection{Continuum phase segregation as a model for differential intercellular adhesion}
\label{sec:segdiffadhesn}
Any reasonably complex biological organism displays patterning of cells. The differentiation and segregation of cell types is a common feature of tissues. The difficulties with Turing patterns notwithstanding, models for cell segregation by type remain of importance. In a now classical experiment performed more than 60 years ago, \cite{Townes1955} showed that when mesoderm cells, neural plate cells and epidermal cells were disaggregated from an embryo and then reaggregated in a random mixture, they underwent segregation into an arrangement reminiscent of a normal embryo with a neural tube internally and epidermis externally, separated by mesoderm. Similar results are seen when cells from different organs, such as the retina and liver are mixed. Over time, the cells segregate into clusters originating from the same organ. Clearly, cells have preferential adhesion for their own type, and in more complex tissues with several cell types, there is a hierarchy of adhesion preferences, { with a biophysical basis in the type and number of cadherin molecules expressed on the surface of each sell type \citep{Alberts2008}.} A well-known continuum model describes such cell sorting, and we outline it below.

\begin{figure}[hbt]
  \centering
  \includegraphics[width=0.6\textwidth]{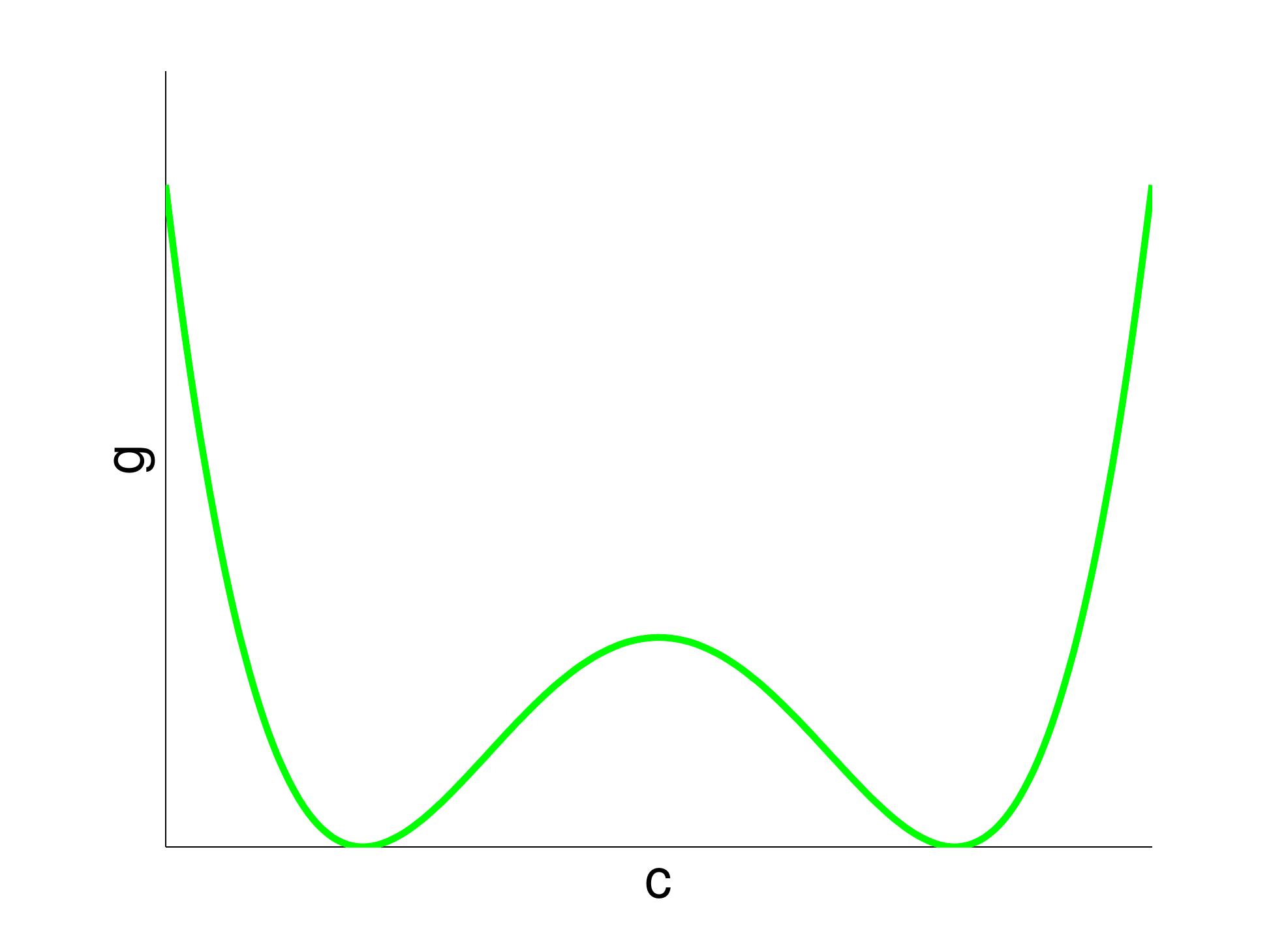}
\caption{A schematic of the non-convex tissue energy density function that drives segregation of a tissue into two distinct cell types $\alpha$ and $\beta$, with cell concentration values $c_\alpha$ and $c_\beta$, corresponding to the minima of $g$.}
\label{fig:nonconvex}
\end{figure}

Consider a continuum mixture of two cell types, $\alpha$ and $\beta$. Cells of each type have higher adhesion strengths for their own type over their adhesion strength with dissimilar cells. In terms of energy, the adhesion energy of $\alpha$-$\alpha$ and $\beta$-$\beta$ bonds is lower than that of $\alpha$-$\beta$ bonds. Without loss of generality we assume also that the $\alpha$-$\alpha$ bond energy is equal to that of the $\beta$-$\beta$ bond. Let $c$ be a cell concentration field with $c = c_\alpha$ and $c = c_\beta$ being its values corresponding to the two cell types.\footnote{It may seem more natural to have $c$ represent cell type, with $c = 0$ representing a population of $\alpha$ cells and $c = 1$ representing a population of $\beta$ cells. However, a concentration so-defined can always be rescaled to the more general form used here.} The cell adhesion energies also can be extended to the continuum setting to define a tissue adhesion energy density, $g(c)$, with equal minima at $c_\alpha$ and $c_\beta$. The tissue adhesion energy density has a smooth local maximum at $c = c_\gamma$, for $c_\alpha <  c_\gamma < c_\beta$. This concentration, $c_\gamma$ and the corresponding maximum in energy density represent the least preferred arrangement of a mixture of type $\alpha$ and type $\beta$ cells, because the $\alpha$-$\beta$ bonds are of the highest adhesion energy. A simple tissue adhesion energy density function of this form is
\begin{equation}
g(c) = \omega (c - c_\alpha)^2(c - c_\beta)^2,\quad \omega > 0,
\label{eq:homogenergy}
\end{equation}
shown schematically in Figure \ref{fig:nonconvex} with the minima corresponding to $c_\alpha$ and $c_\beta$ ($c_\alpha < c_\beta$), and the maximum to $c_\gamma = \frac{1}{2}(c_\alpha + c_\beta)$. Non-convex functions of this form correspond to the so-called \emph{homogeneous} free energy density used in phase segregation models. However, they suffer from a well-known drawback in representing the sought-after tissues with heterogeneous cell types: They do not differentiate between cases which have the same volumes, say $V_\alpha$ and $V_\beta$ of type $\alpha$ and type $\beta$ cells, respectively, but different spatial distributions. This is understood as an instability in $g(c)$ stemming from its non-convex form, and needs regularization by penalizing the concentration gradients between type $\alpha$ and $\beta$ cell clusters. It has motivated the inclusion of an inhomogeneous, or concentration gradient energy density in the mathematical model of phase segregation \citep{CahnHilliard1958,Wise2008,Cristini2009,Lowengrub2010,Oden2010,Chatelain2011,Vilanova2013,Vilanova2014,Xu2016}. The total tissue energy density is then written as
\begin{equation}
f(c,\nabla c) = {\underbrace{\omega (c - c_\alpha)^2(c - c_\beta)^2}_{g(c)}} + \frac{\kappa}{2}\vert\nabla c\vert^2.
\label{eq:totalchenergy}
\end{equation}
for gradient parameter $\kappa$. The gradient energy penalizes interfaces between the different cell clusters, thus representing a tissue interface energy density.

The total tissue energy functional is $\mathscr{F}[c] = \int_\Omega f \mathrm{d}V$. From variational considerations, the chemical potential is obtained as a Gateaux variation:
\begin{equation}
\mu = \frac{\delta \mathscr{F}}{\delta c} = g^\prime - \kappa\nabla^2 c
\label{eq:chempot}
\end{equation}
provided equilibrium is assumed at the boundaries, in which case the additional condition is $\nabla c\cdot\bn = 0$ at $\partial\Omega$ for unit outward normal, $\bn$. With these conditions, we consider the following mass transport equations for cell migration:
\begin{displaymath}
\frac{\partial c}{\partial t} = \nabla\cdot\left(M\nabla \mu\right) \quad\mathrm{in}\;\Omega,
\end{displaymath}
which on substituting (\ref{eq:chempot}), and adding initial and boundary conditions, leads to the Cahn-Hilliard equation \citep{CahnHilliard1958}:
\begin{subequations}
\begin{align}
\frac{\partial c}{\partial t} &= \nabla\cdot\left(M\nabla g^\prime\right)  - \nabla\cdot\left(M\nabla(\kappa\nabla^2 c) \right)\quad\mathrm{in}\;\Omega\label{eq:chA}\\
-M\nabla\mu\cdot\bn &= 0 \quad\mathrm{on}\;\partial\Omega\label{eq:chB}\\
\nabla c\cdot\bn &= 0 \quad\mathrm{on}\;\partial\Omega\label{eq:chC}\\
c(\bx,t) &= c^0(\bx) \quad\mathrm{at}\;t = 0\label{eq:chD}
\end{align}
\end{subequations}

\begin{figure}[hbt]
  \centering
 \includegraphics[width=0.6\textwidth]{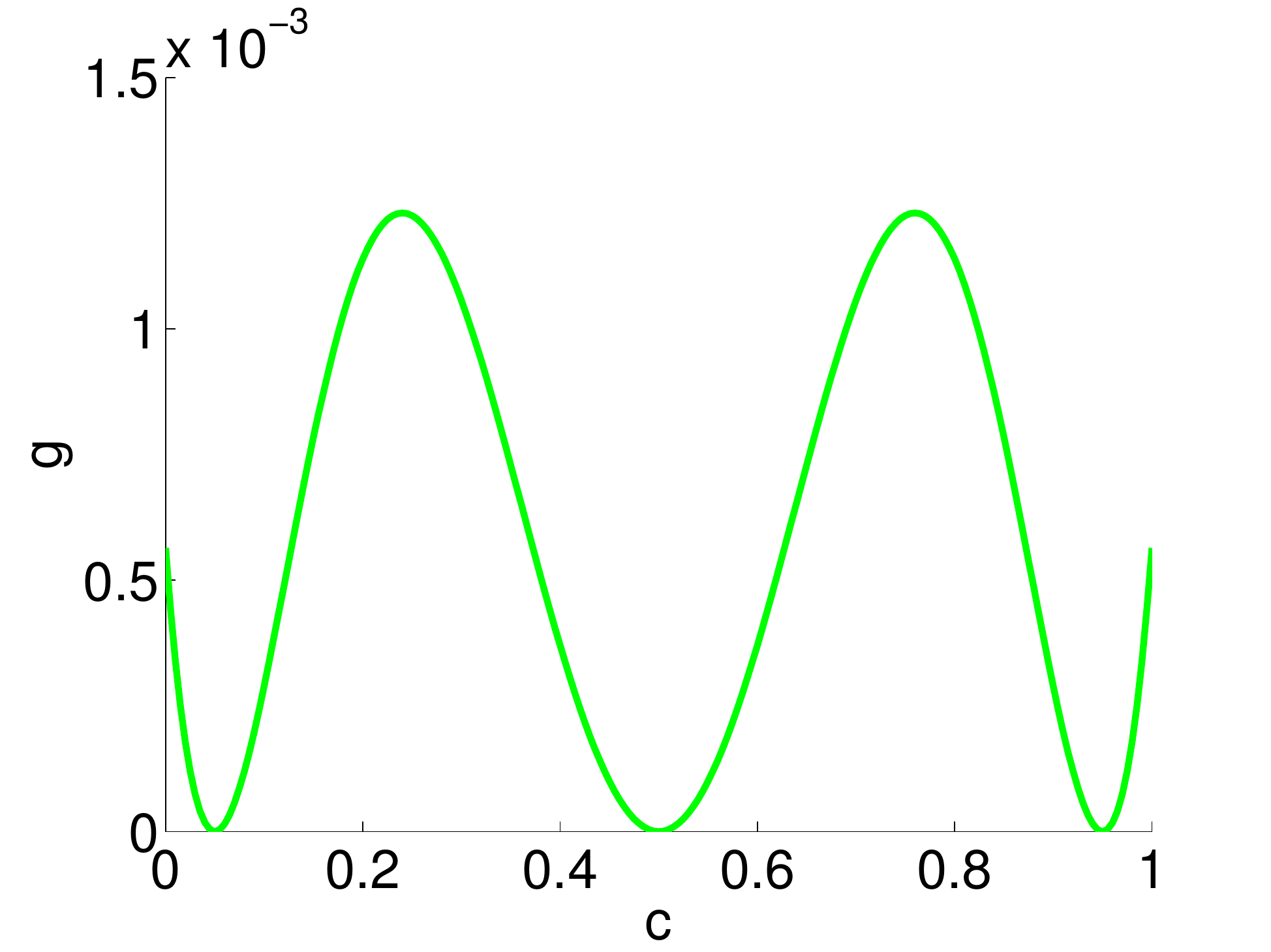}
\caption{A non-convex tissue energy density function parameterized by a single field, $c$, representing cell concentration, with wells $c_\alpha = 0.05, \; c_\beta = 0.5, \; c_\gamma 0.95$ for three distinct cell types. }
\label{fig:ch3well1field}
\end{figure}

Attention is called to the well-known fourth-order nature of this partial differential equation in the concentration $c$.\footnote{{The long-range diffusion by which \cite{MurrayMaini1988} modelled the cell tractions imposed via filopodia leads to a fourth-order diffusion term with the same form as the last term on the right hand-side in Equation (\ref{eq:chA}). }} A natural length scale residing in this equation is $l_\mathrm{CH} = \sqrt{\kappa/\omega}$. If the initial condition (\ref{eq:chD}) places the concentration over some neighborhood in the spinodal, defined as the concentration regime where $g^{\prime\prime} < 0$, then the tissue segregates into cell clusters with concentrations $c_\alpha$ and $c_\beta$ representing each distinct cell type. The initial dynamics modelled by this equation are very fast, and referred to as spinodal decomposition. A slower dynamic stage follows, called Ostwald ripening in the materials physics literature, in which larger regions of both cell clusters grow at the expense of smaller ones. The equilibrium state is determined by the average concentration, $\bar{c}$, over $\Omega$. If $\bar{c}$ lies outside the spinodal the equilibrium cell concentration field is uniform at $c = \bar{c}$. If $\bar{c}$ lies within the spinodal, there remains a single, connected sub-domain of each phase, $\Omega_\alpha$ and $\Omega_\beta$ such that $\bar{c}\cdot\mathrm{m}(\Omega) = c_\alpha\cdot\mathrm{m}(\Omega_\alpha) + c_\beta\cdot\mathrm{m}(\Omega_\beta)$. This presents an interesting model, and can replicate the results of an experiment along the lines of \cite{Townes1955} for tissues with two cell types.\footnote{For discrete systems, the Cellular Potts Model \citep{Graner1992}, presents a comparable approach.} A natural question is whether it can be extended to more than two cell types, for instance with a homogeneous tissue adhesion energy density of the form $g(c) = \omega(c - c_\alpha)^2(c - c_\beta)^2(c - c_\gamma)^2$, for three cell types, where $c_\alpha < c_\beta < c_\gamma$ are the three minima in Figure \ref{fig:ch3well1field} with parameters in Table \ref{tbl:ch3well1field}.

\begin{figure}
\begin{minipage}[]{0.3\textwidth}
\centering
\subfloat[$t = 0$]{\includegraphics[width=1.25\textwidth]{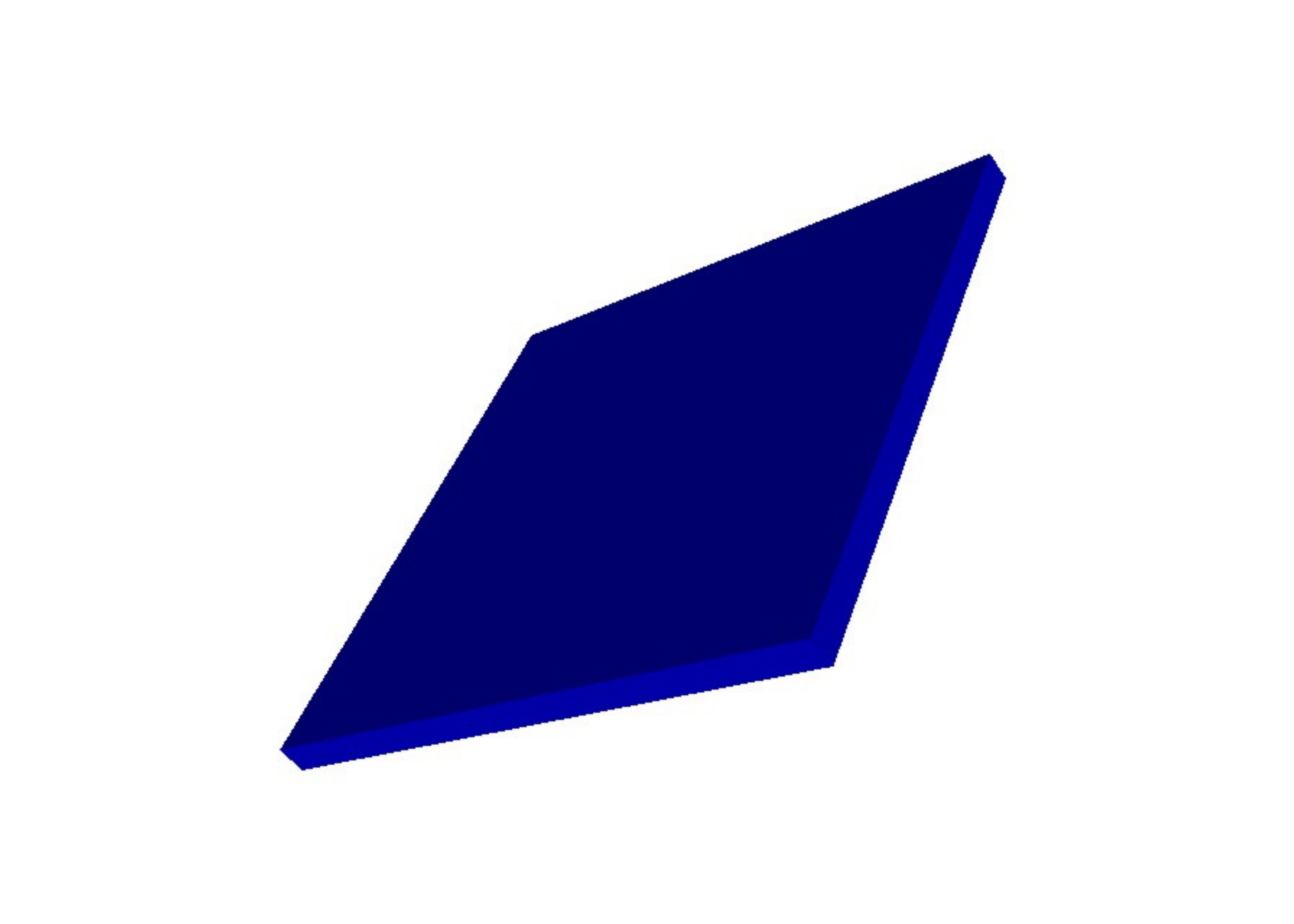}}
\end{minipage}
\begin{minipage}[]{0.3\textwidth}
\centering
\subfloat[$t = 0.9$]{\includegraphics[width=1.25\textwidth]{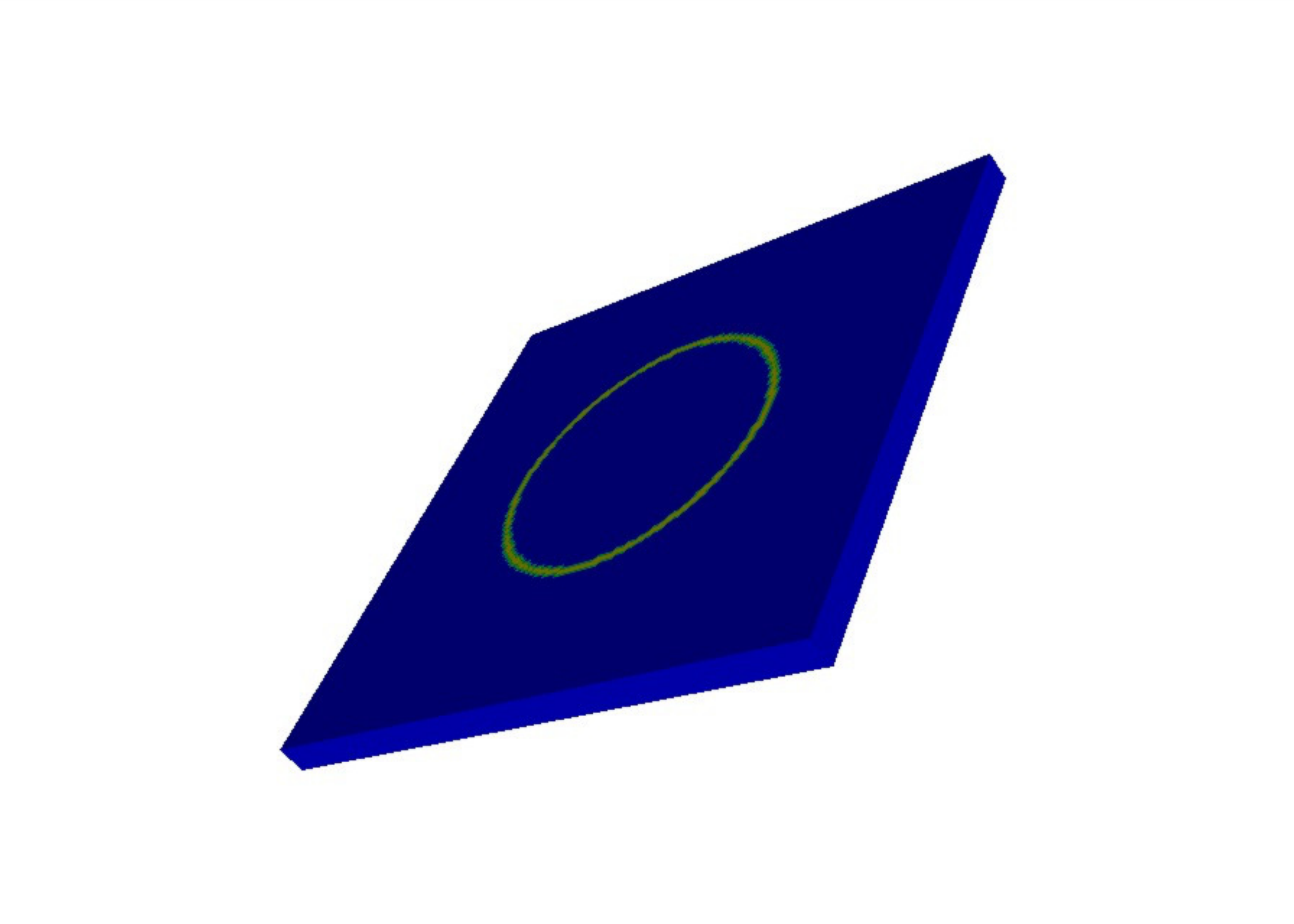}}
\end{minipage}
\begin{minipage}[]{0.3\textwidth}
\centering
\subfloat[$t = 220$]{\includegraphics[width=1.25\textwidth]{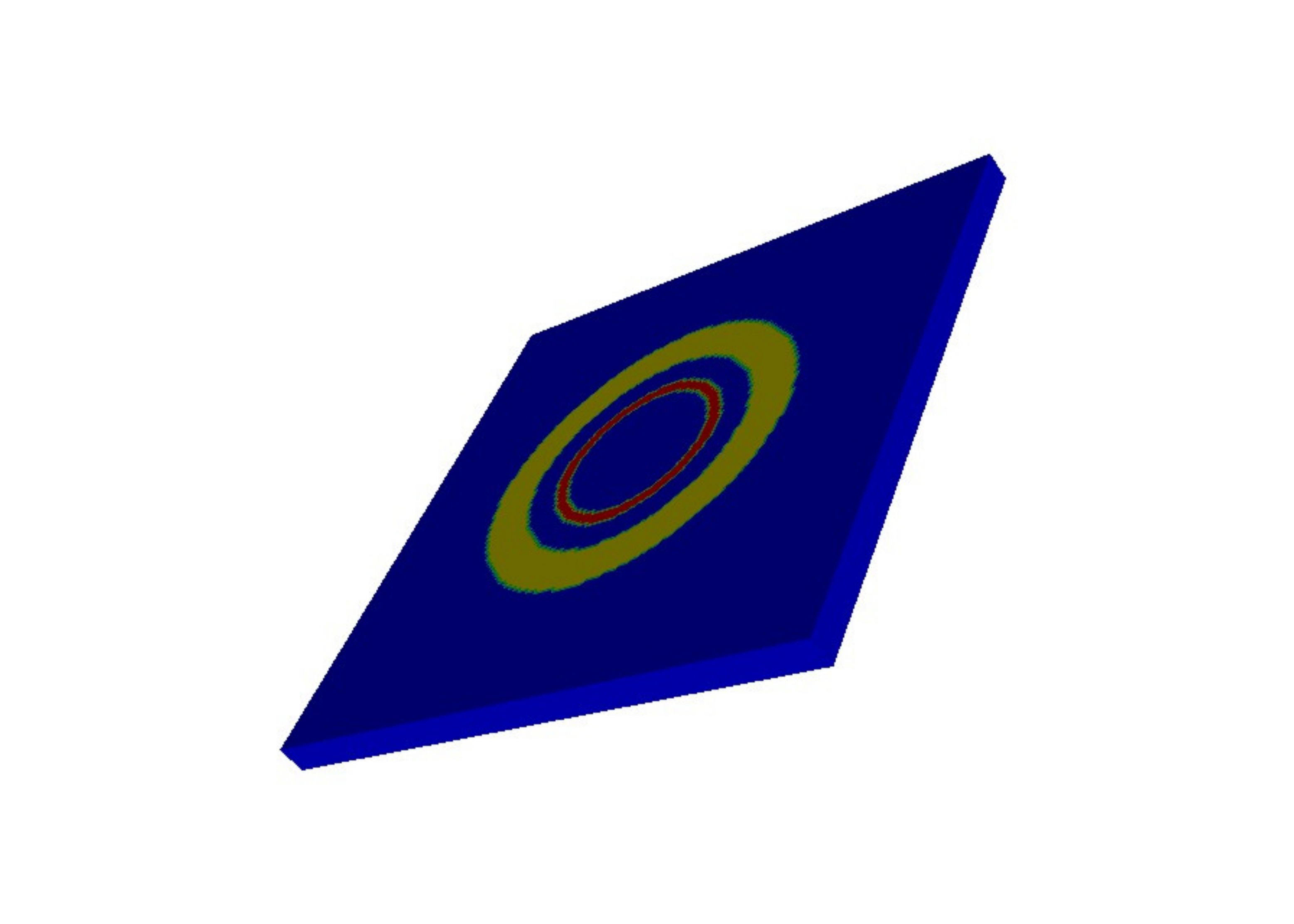}}
\end{minipage}\\
\begin{minipage}[]{0.3\textwidth}
\centering
\subfloat[$t = 420$]{\includegraphics[width=1.25\textwidth]{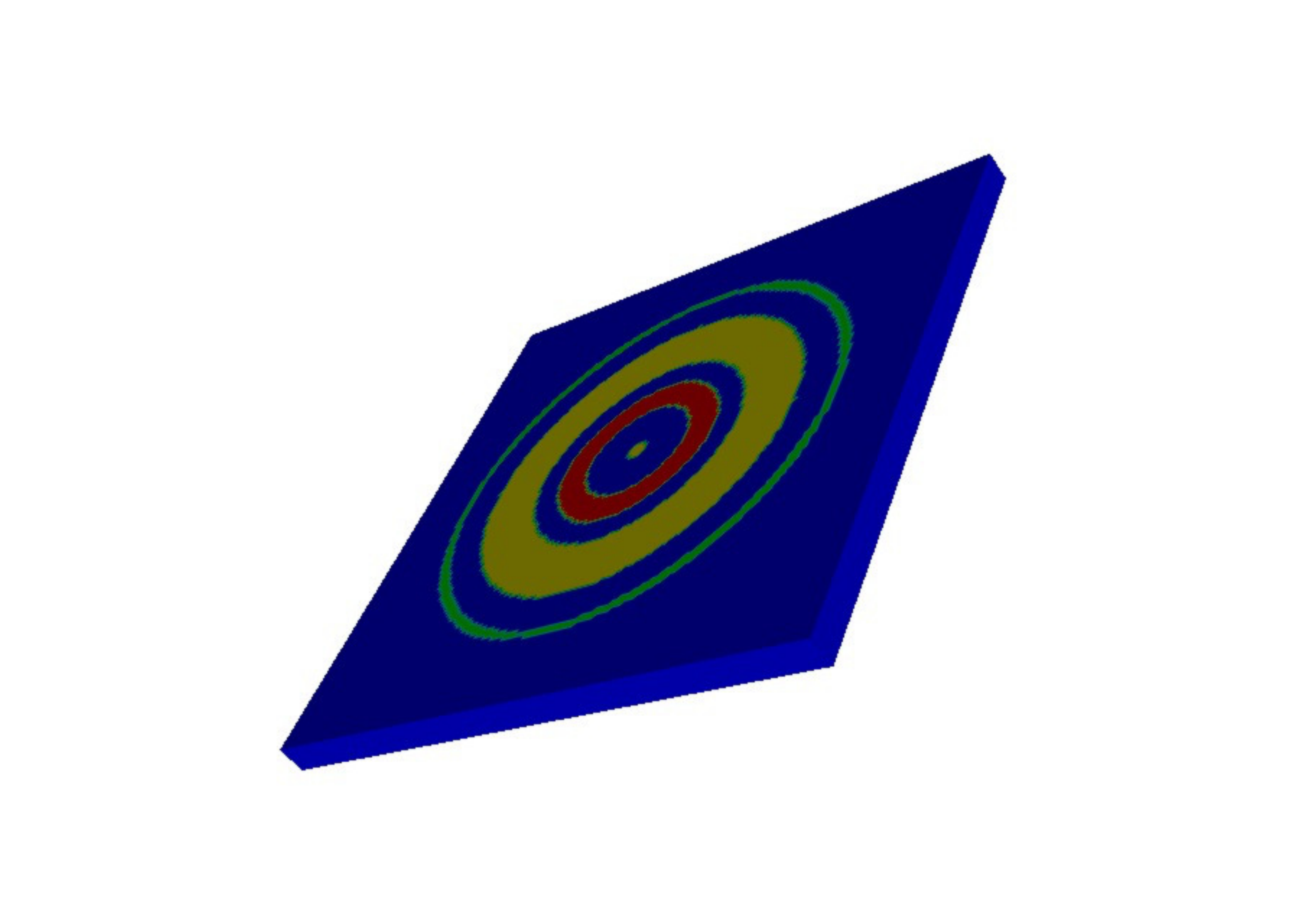}}
\end{minipage}
\begin{minipage}[]{0.3\textwidth}
\centering
\subfloat[$t = 7120$]{\includegraphics[width=1.25\textwidth]{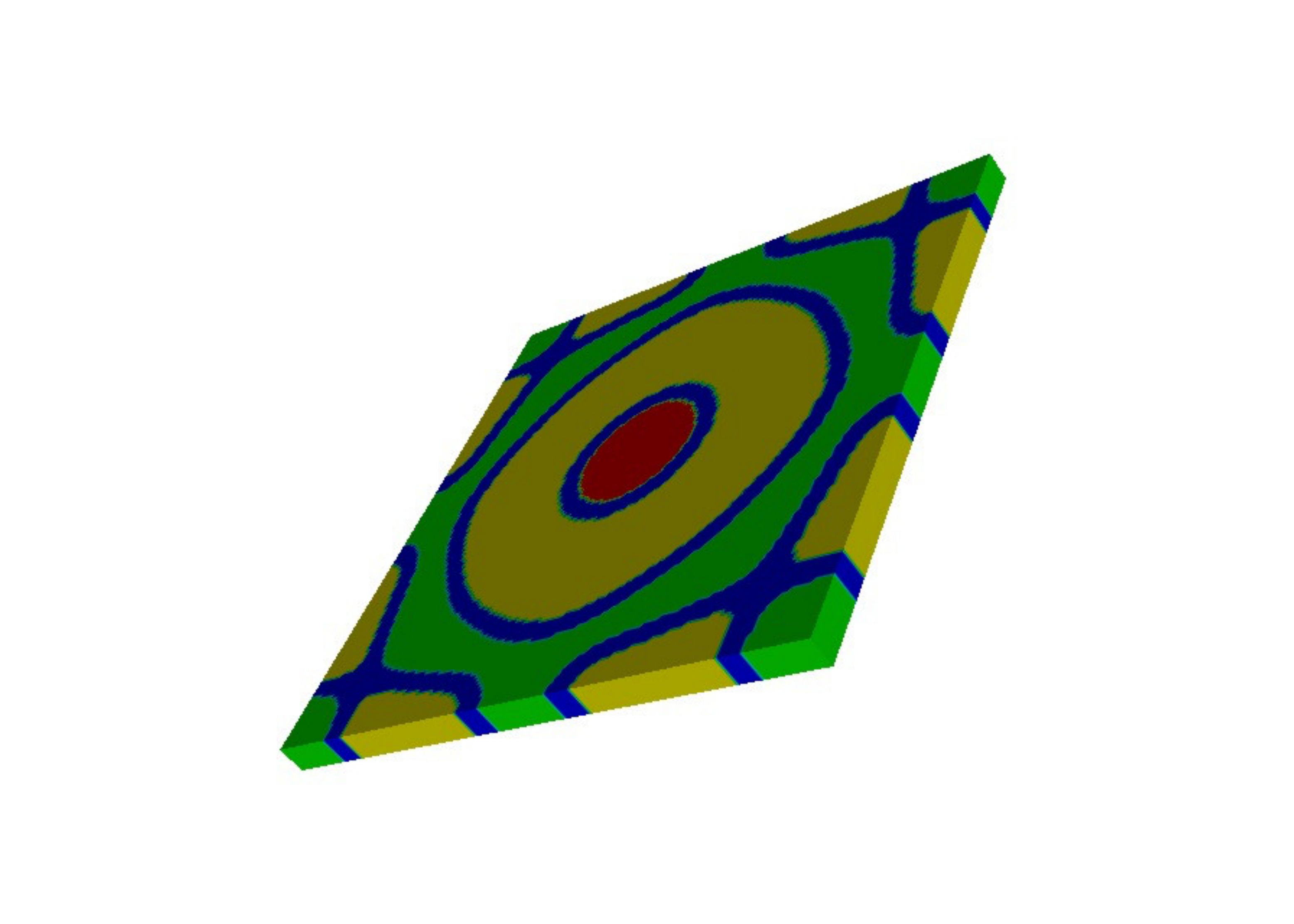}}
\end{minipage}
\begin{minipage}[]{0.3\textwidth}
\centering
\subfloat[$t = 15000$]{\includegraphics[width=1.25\textwidth]{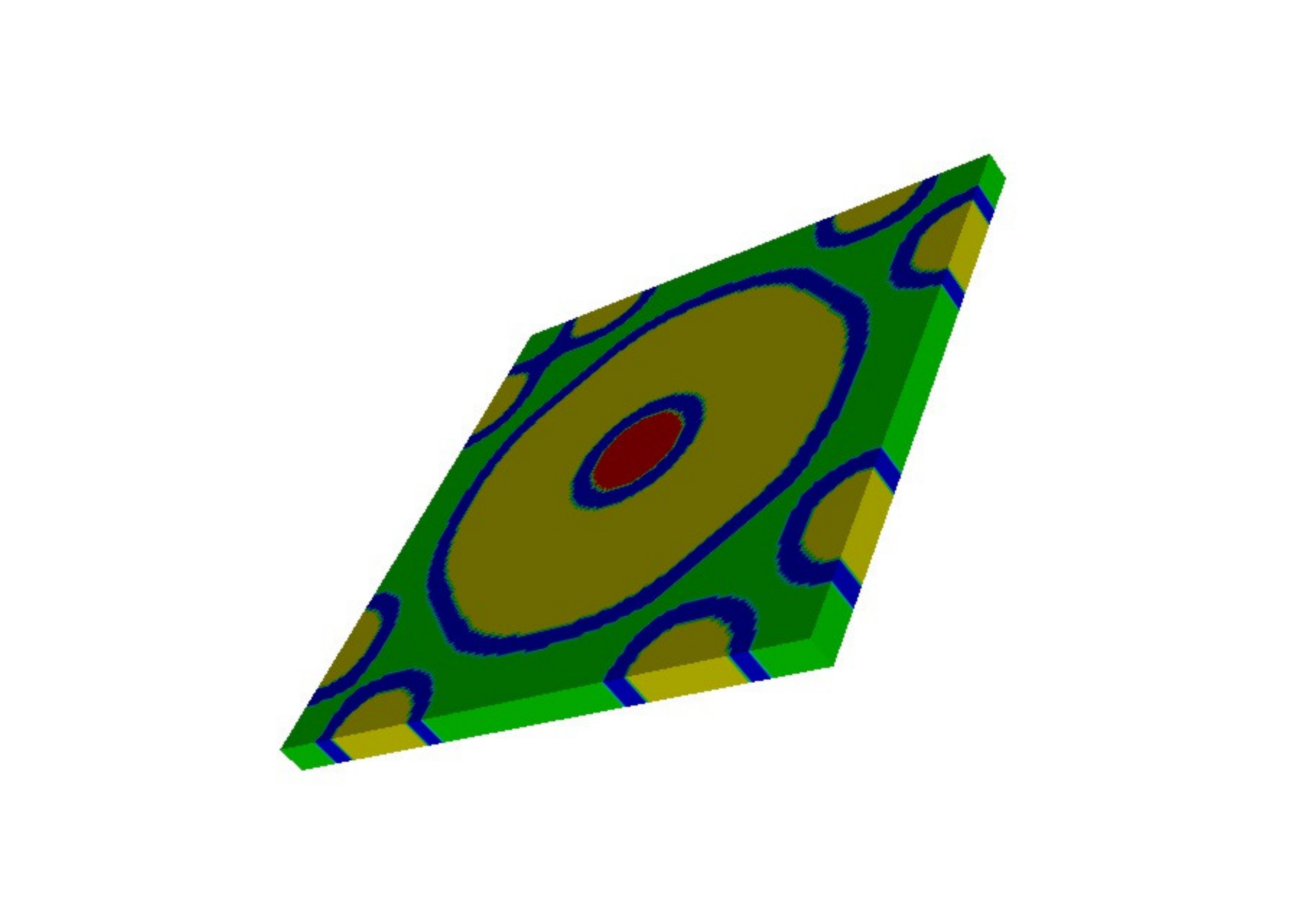}}
\end{minipage}
\caption{A slowly segregating tissue with cell concentration values in each of three wells in Figure \ref{fig:ch3well1field} and Table \ref{tbl:ch3well1field}. These values represent distinct cell types: red ($\alpha$), yellow ($\beta$) and green ($\gamma$). Also see Supplementary Movie S3.}
\label{fig:ch3well1fieldcontours}
\end{figure}

\begin{table}[h]
\centering
\caption{Parameters for segregation of a tissue into three distinct cell types from a single cell concentration field parameterizing a tissue energy density function with three wells.}
\begin{tabular}{ |c|c|c|c|c|c|c|  }
\hline
 Parameter & $\omega$ & $c_\alpha$ & $c_\beta$ & $c_\gamma$ & $\kappa$ & $M$ \\
 \hline
 Value & $1$ & $0.05$  & $0.5$ & $0.95$ & $1$ & $0.1$\\
 \hline
\end{tabular}
 \label{tbl:ch3well1field}
\end{table}

However, a straightforward analysis negates such a result. The best that can be achieved is an equilibrium state of two cell clusters as argued above, with either cell types $\alpha$ and $\beta$ if $\bar{c}$ lies in the spinodal region between the corresponding minima in Figure \ref{fig:ch3well1field}, or $\beta$ and $\gamma$ if $\bar{c}$ lies in the spinodal region between the $\beta$ and $\gamma$ minima. If $\bar{c}$ lies outside of these spinodals, the equilibrium state achieved by the tissue will lie in the well around the corresponding minimum: $c_\alpha, c_\beta$ or $c_\gamma$. There remains the possibility, however, that if time-dependent kinetics are included, a non-equilibrium state of the tissue could be frozen in if the mobility $M \to 0$ at large times. An example of such a slowly evolving state appears in Figure \ref{fig:ch3well1fieldcontours}. Also see Supplementary Movie S3. Initial conditions for this computation are
\begin{equation}
c^0(\bx) = 
\begin{cases}
0.24 \pm \delta,\quad\vert\bx\vert \le 1\\
0.76 \pm \delta,\quad\vert\bx\vert > 1
\end{cases}
\end{equation}
where $\delta \in [-0.01,0.01]$ is a random real number. The initially sharp interface gets smeared out into a rapidly fluctuating field $c(\bx,t)$ that takes on values in each of the three wells. Red for $\alpha$, yellow for $\beta$ and green for $\gamma$. The average concentration at equilibrium would be $c = 0.658$ which lies outside both spinodals, in the $\beta$ well. This is seen in the predominance of the yellow cell cluster in Figure \ref{fig:ch3well1fieldcontours}. However, the dynamics are extremely slow, and the stage shown may be mistaken for a steady state with rings of the different cell clusters.

From a biological viewpoint, such slowly evolving dynamics may be relevant, especially if the steady states are reached in asymptotic time. But, it is important to note that the topology of the free energy function allows only certain cell clusters to neighbor each other. Specifically, the $\beta$ (yellow) cell type can be bordered by $\alpha$  (red) and $\gamma$ (green), but the latter two cell types cannot be neighbors. Of course, this is not a very satisfying representation of a tissue with three cell types. However, a reformulation of the cell segregation problem with two concentration fields $c_1, c_2$, is possible with a homogeneous tissue energy density function of the form
\begin{equation}
g(c_1,c_2) =  \frac{3 d}{2 s^4} (c_1^2+c_2^2)^2 + \frac{d}{s^3} c_2 (c_2^2-3c_1^2) - \frac{3 d}{2 s^2} (c_1^2+c_2^2), \quad d,s > 0
\label{eq:homogenergy3well2field}
\end{equation}
\begin{figure}[hbt]
  \centering
  \includegraphics[width=0.7\textwidth]{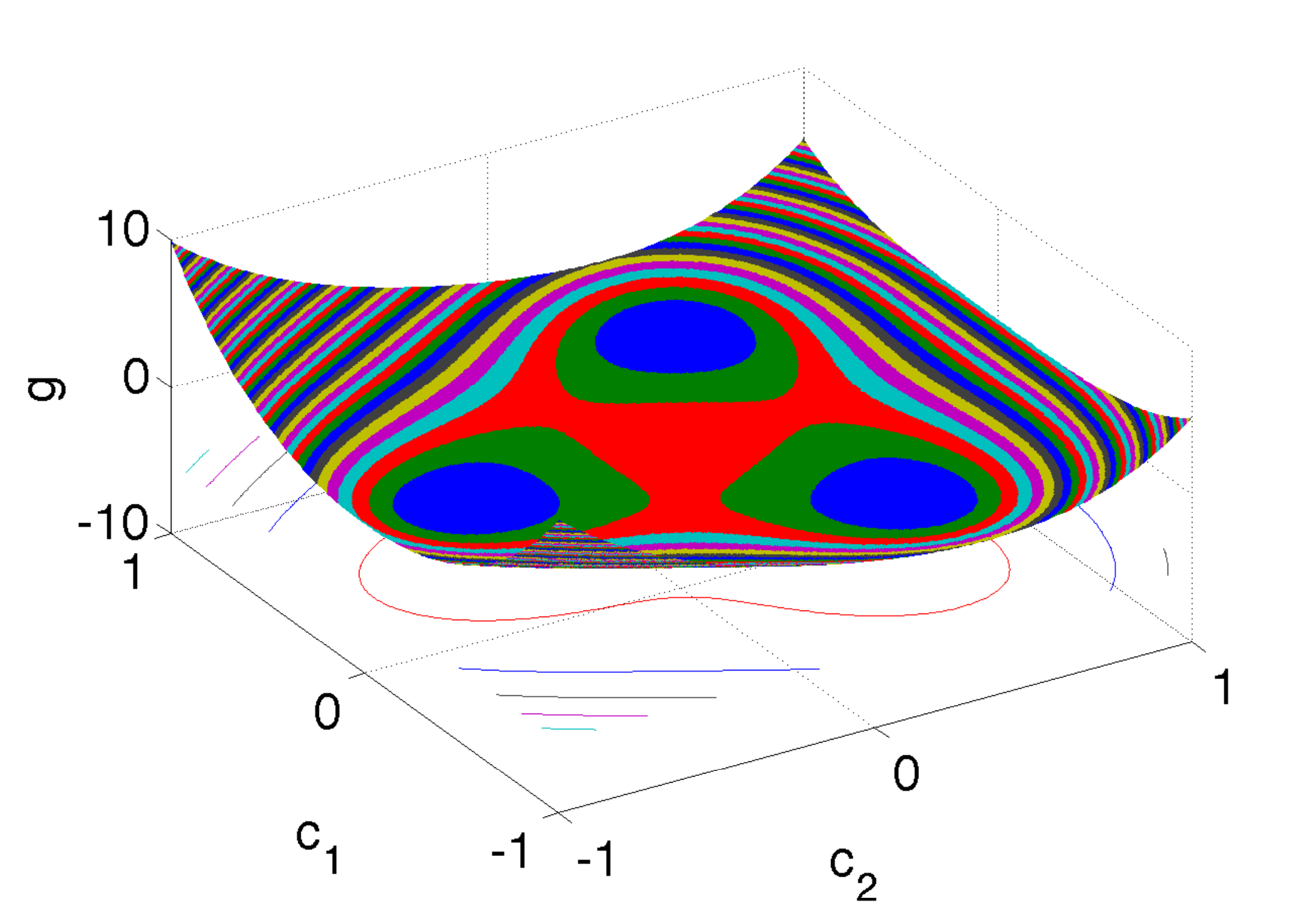}
\caption{A non-convex tissue energy function parameterized by two concentration fields, $c_1$ and $c_2$ with wells for three distinct cell types.}
\label{fig:ch3well2field}
\end{figure}
In the $c_1-c_2$ plane, this function has three minima as shown in Figure \ref{fig:ch3well2field}, where the concentrations have been rescaled to reach over into the negative half planes. These minima represent three cell types. Here, $d$ and $s$ control the common depth of the three wells (minima), and $s$ determines their radial location in the $c_1-c_2$ plane. Equation (\ref{eq:homogenergy3well2field})  replaces Equation (\ref{eq:homogenergy}). Because of the non-convexities it harbors in between these valleys, it too must be regularized by gradient energy densities, representing the energy at tissue interfaces. The simplest forms penalize $\vert\nabla c_1\vert$ and $\vert\nabla c_2\vert$:
\begin{equation}
f(c_1,c_2,\nabla c_1, \nabla c_2) =  \frac{3 d}{2 s^4} (c_1^2+c_2^2)^2 + \frac{d}{s^3} c_2 (c_2^2-3c_1^2) - \frac{3 d}{2 s^2} (c_1^2+c_2^2) + \frac{\kappa_1}{2}\vert\nabla c_1\vert^2 + \frac{\kappa_2}{2}\vert\nabla c_2\vert^2.
\label{eq:totalchenergy3well2field}
\end{equation}
With the total free energy $\mathscr{F} = \int_\Omega f\mathrm{d}V$, chemical potentials are then defined as

\begin{equation}
\mu _i= \frac{\delta \mathscr{F}}{\delta c_i} = \frac{\partial g}{\partial c_i} - \kappa_i\nabla^2 c_i, \quad i = 1,2,\label{eq:chempot3well2field}
\end{equation}
and governed by the equations
\begin{subequations}
\begin{align}
\frac{\partial c_i}{\partial t} &= \nabla\cdot\left(M_i\nabla \mu_i\right) \quad\mathrm{in}\;\Omega\label{eq:ch3well2fieldA}\\
-M_i\nabla\mu_i\cdot\bn &= 0 \quad\mathrm{on}\;\partial\Omega\label{eq:ch3well2fieldB}\\
\nabla c_i\cdot\bn &= 0 \quad\mathrm{on}\;\partial\Omega\label{eq:ch3well2fieldC}\\
c_i(\bx,t) &= c_{i}^0(\bx)\quad\mathrm{at}\; t = 0,\quad \mathrm{for}\; i = 1,2.\label{eq:ch3well2fieldD}
\end{align}
\end{subequations}
Starting from randomized initial conditions for $c_1,c_2 \in [-1,1]\times[-1,1]$, this formulation does indeed evolve through spinodal decomposition in regions where $(c_1,c_2)$ lie in regions of non-convexity of $g(c_1,c_2)$ (Figure \ref{fig:ch3well2field}), to Ostwald ripening and tend towards a genuine equilibrium with three cell clusters. Figure \ref{fig:threephaseequil} shows a quasi-equilibrium state with clusters of the three cell types, which will persist until equilibrium. Also see Supplementary Movie S4. Parameters for this computation appear in Table \ref{tbl:ch3well2field}. This is a viable model for tissues with more than two cell types without relying on kinetically frozen regimes of slow dynamics as an approximation of equilibrium tissue structures. It also can be extended to more than three cell types by constructing free energy functions with the corresponding number of minima in the $c_1-c_2$ plane. 

\begin{figure}
\begin{minipage}[]{0.3\textwidth}
\centering
\subfloat[$t = 0$]{\includegraphics[width=\textwidth]{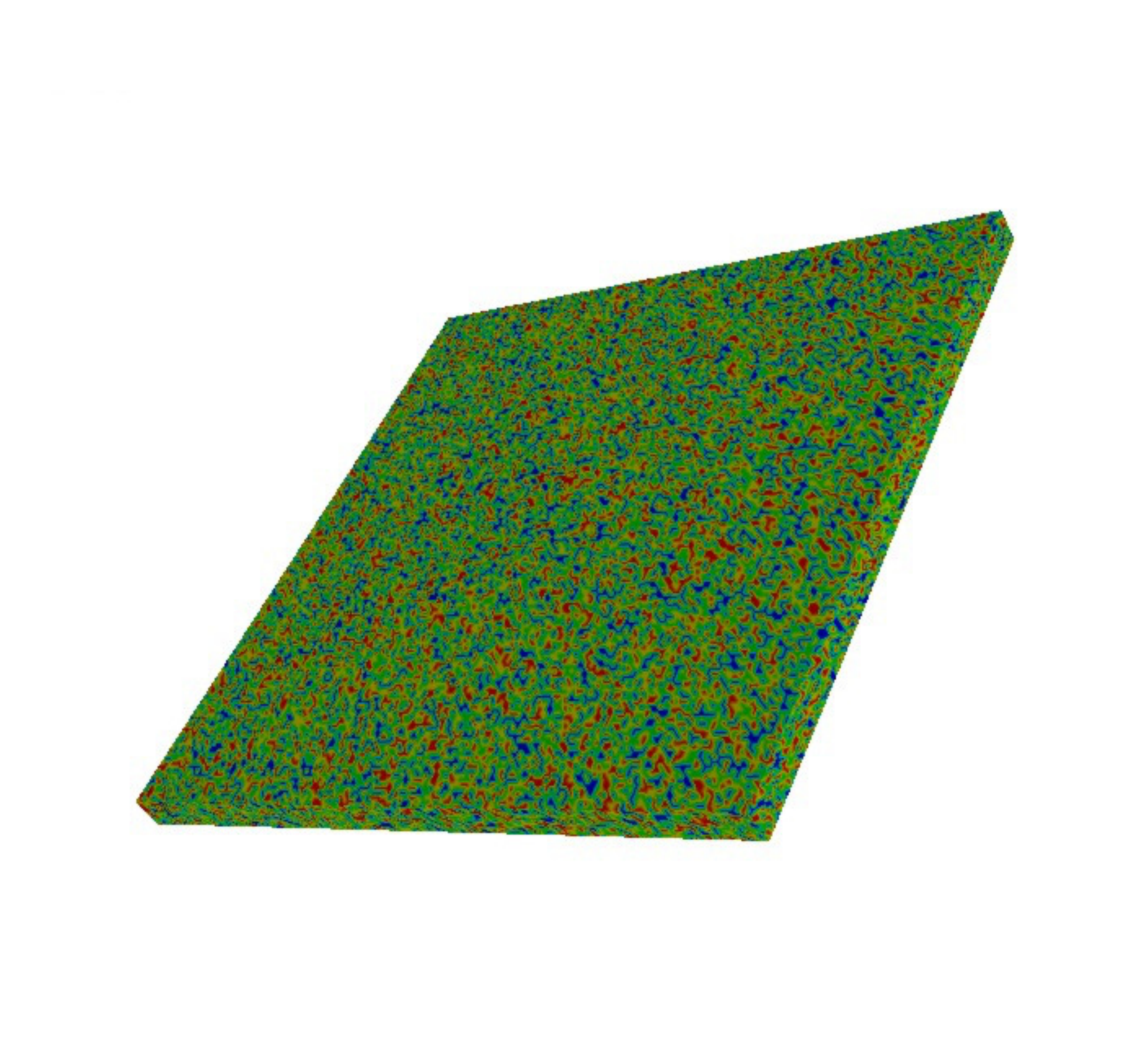}}
\end{minipage}
\begin{minipage}[]{0.3\textwidth}
\centering
\subfloat[$t = 22$]{\includegraphics[width=\textwidth]{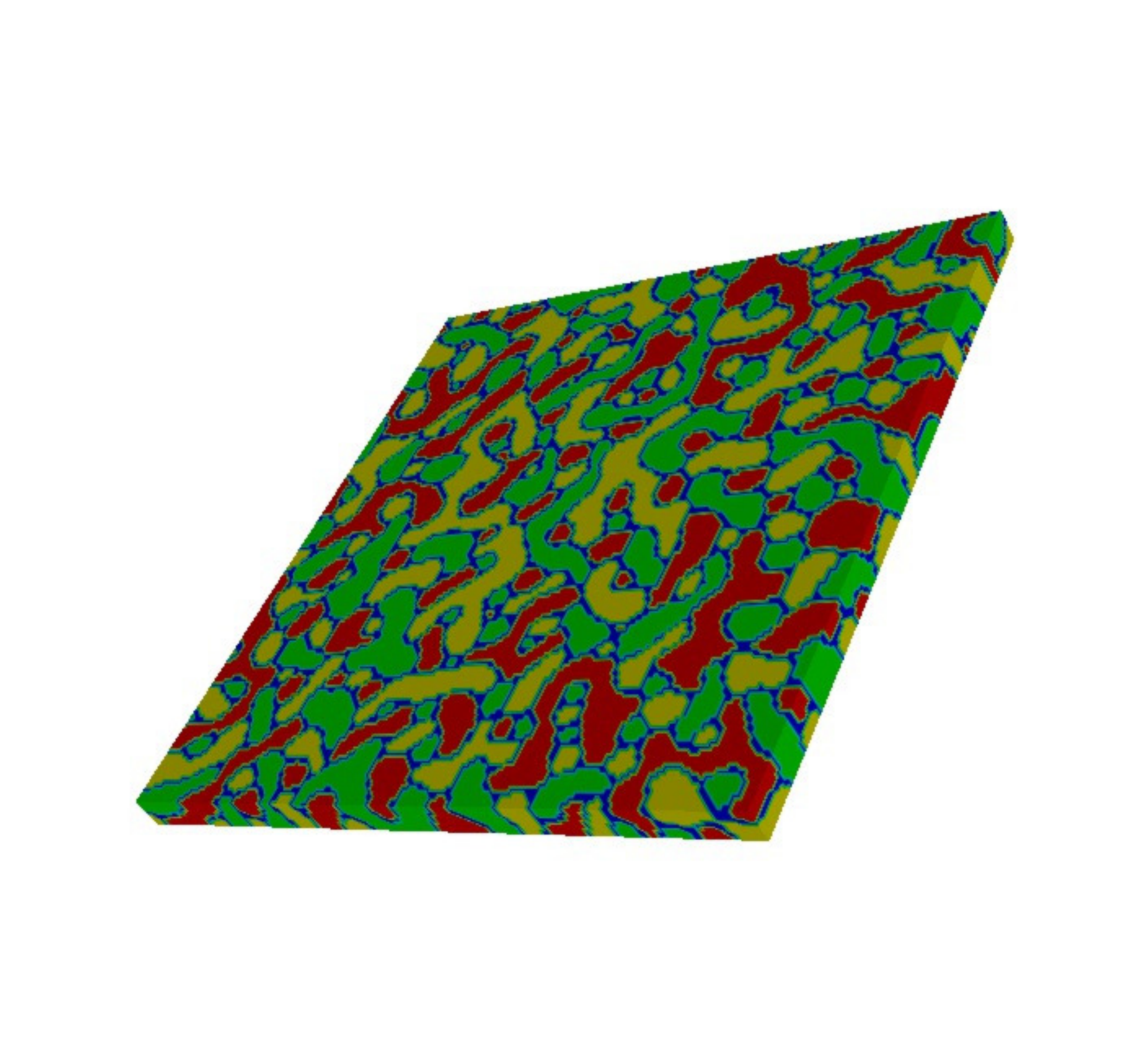}}
\end{minipage}
\begin{minipage}[]{0.3\textwidth}
\centering
\subfloat[$t = 114$]{\includegraphics[width=\textwidth]{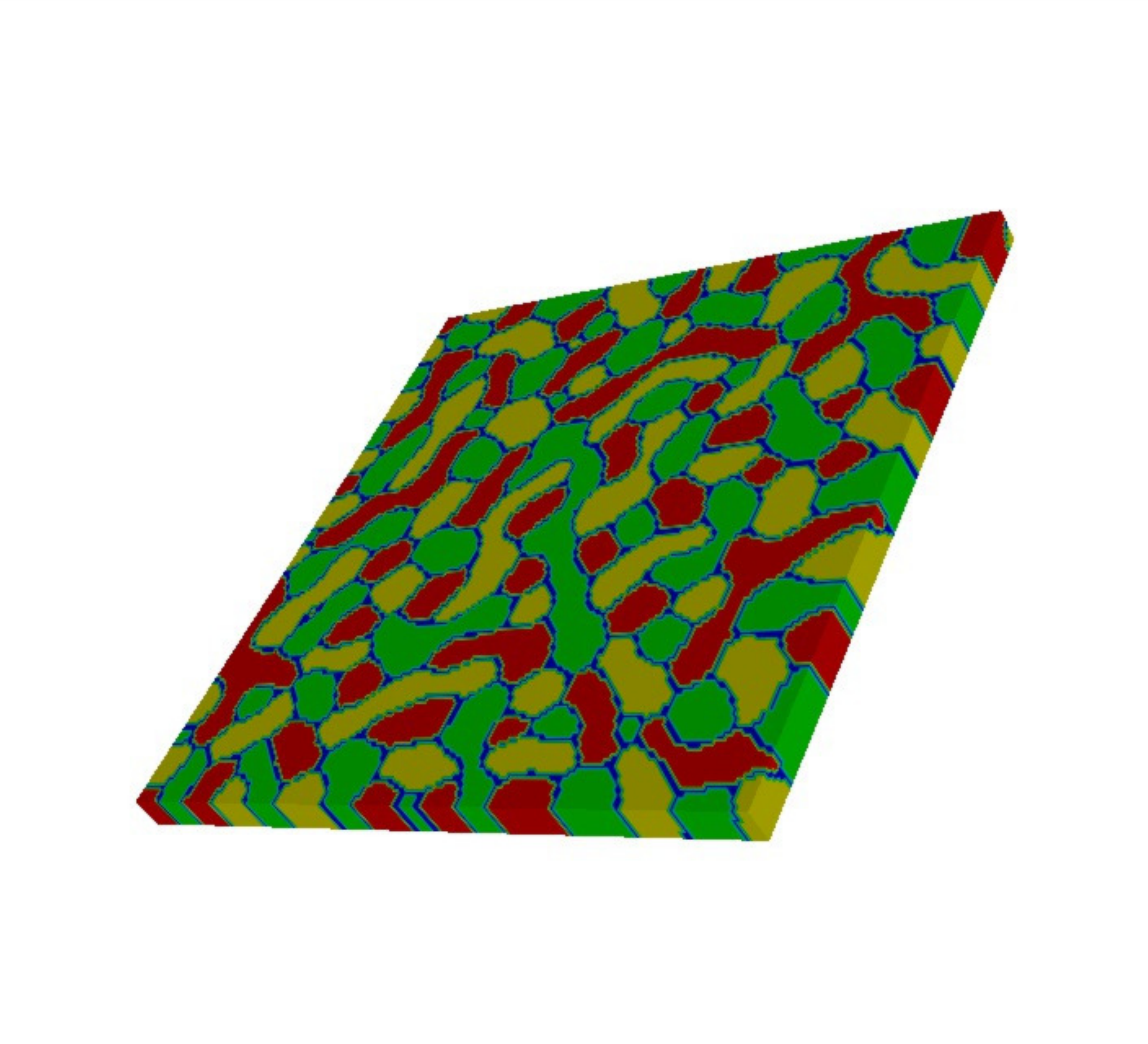}}
\end{minipage}\\
\begin{minipage}[]{0.3\textwidth}
\centering
\subfloat[$t = 714$]{\includegraphics[width=\textwidth]{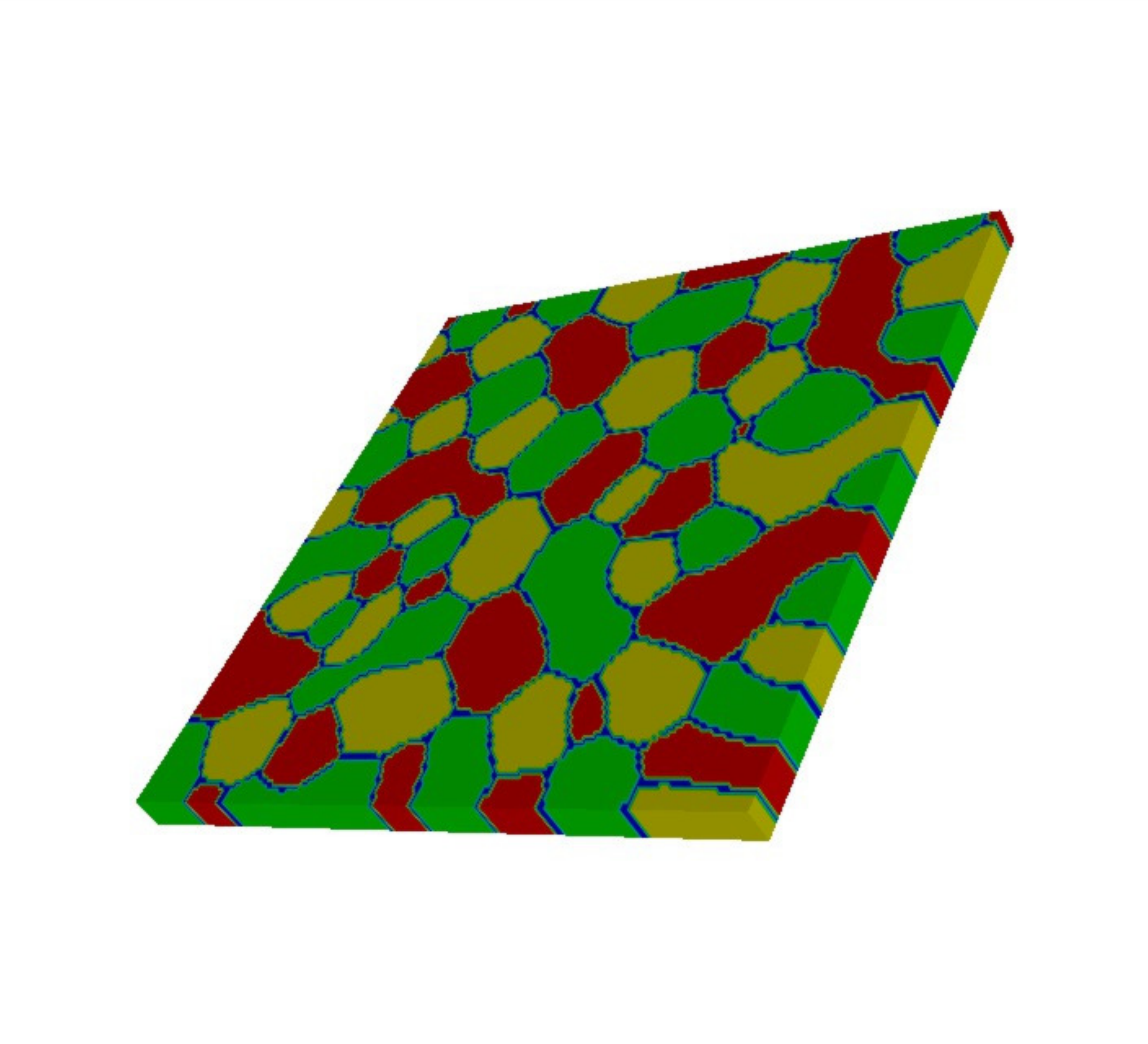}}
\end{minipage}
\begin{minipage}[]{0.3\textwidth}
\centering
\subfloat[$t = 1514$]{\includegraphics[width=\textwidth]{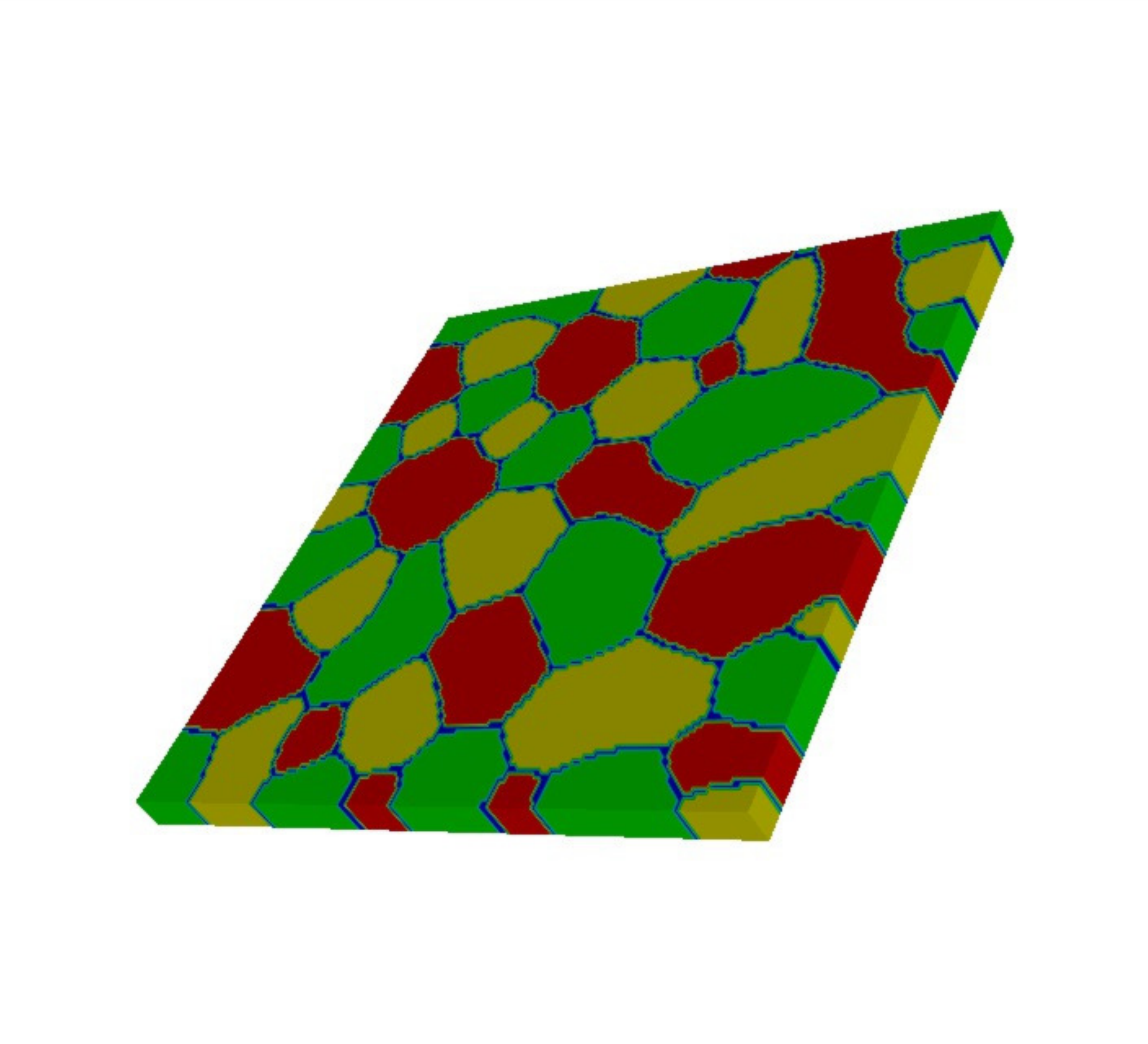}}
\end{minipage}
\begin{minipage}[]{0.3\textwidth}
\centering
\subfloat[$t = 2154$]{\includegraphics[width=\textwidth]{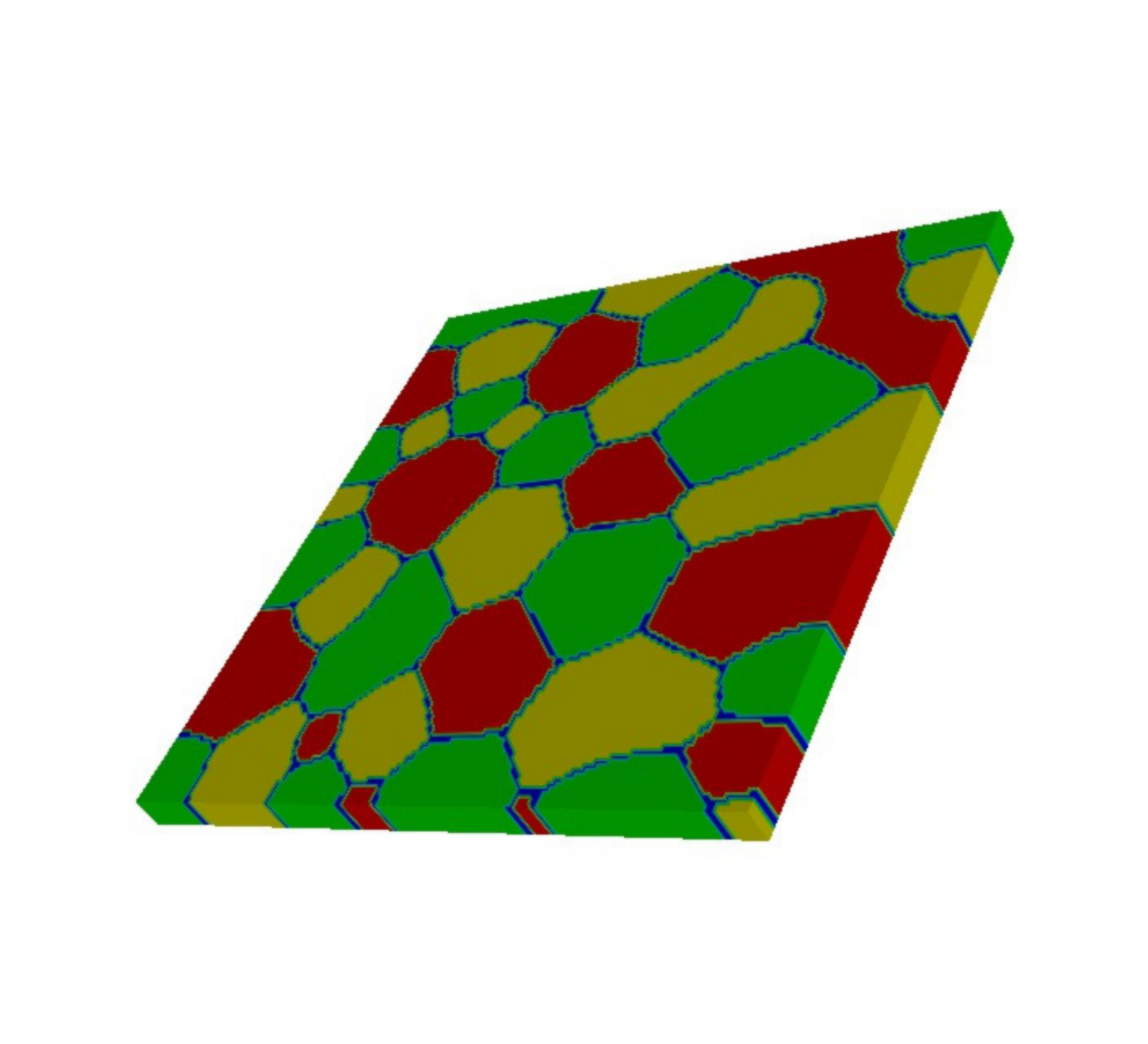}}
\end{minipage}
\caption{Clusters of three cell types emerging from the three-well, non-convex tissue energy density function in two concentration fields that appears in Figure \ref{fig:ch3well2field}. Also see Supplementary Movie S4.}
\label{fig:threephaseequil}
\end{figure}

\begin{table}[h]
\centering
\caption{Parameters for segregation of a tissue into three distinct cell types from two concentration fields parameterizing a tissue adhesion energy density function with three wells.}
\begin{tabular}{ |c|c|c|c|c|c|c|  }
\hline
 Parameter & $d$ & $s$ & $\kappa_1$ & $\kappa_2$ & $M_1$ & $M_2$ \\
 \hline
 Value & $0.4$ & $0.7$  & $1$ & $1$ & $0.1$ & $0.1$\\
 \hline
\end{tabular}
 \label{tbl:ch3well2field}
\end{table}

\begin{figure}
\begin{minipage}[]{0.3\textwidth}
\centering
\subfloat[$t = 0$]{\includegraphics[width=\textwidth]{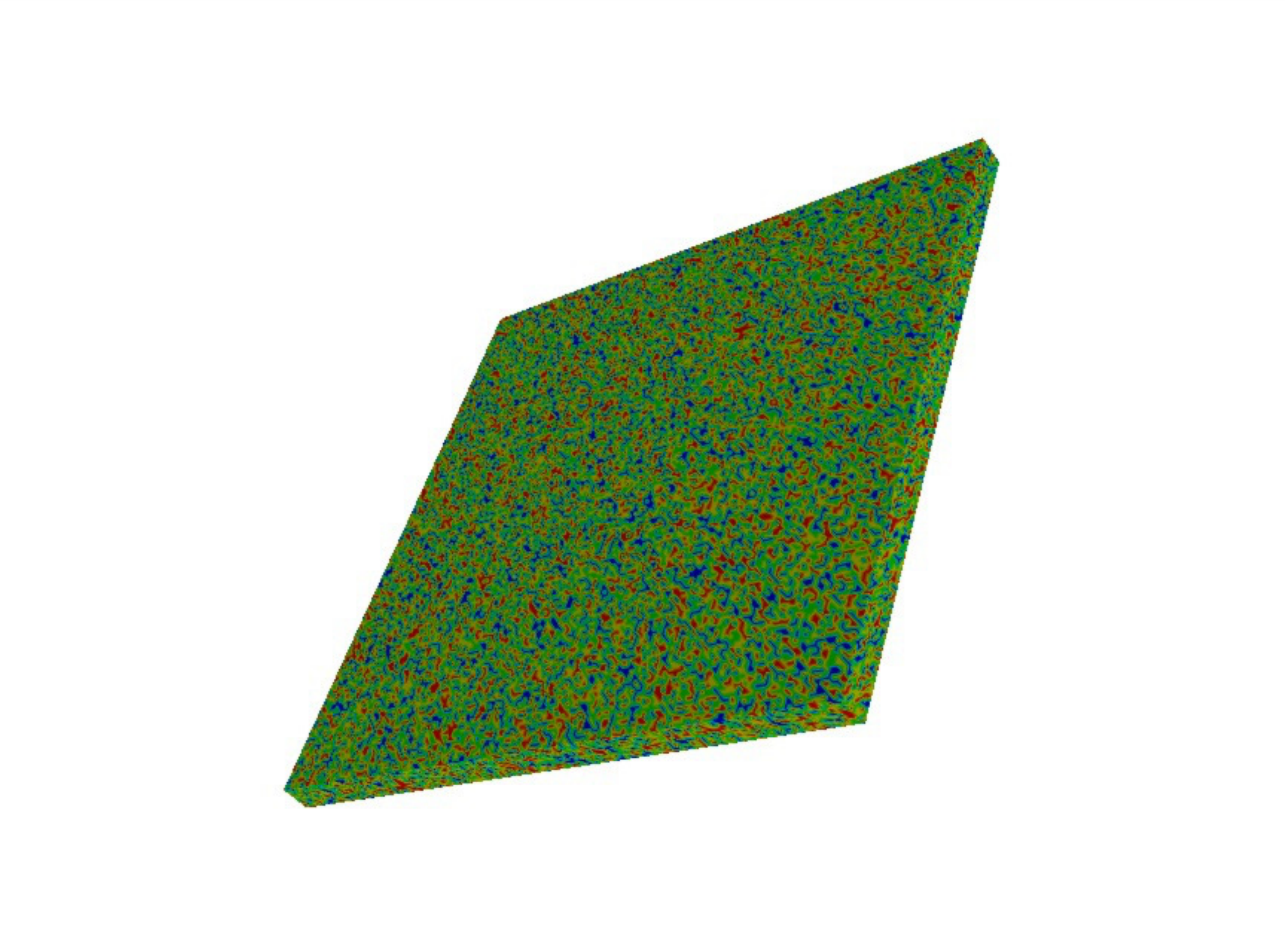}}
\end{minipage}
\begin{minipage}[]{0.3\textwidth}
\centering
\subfloat[$t = 22$]{\includegraphics[width=\textwidth]{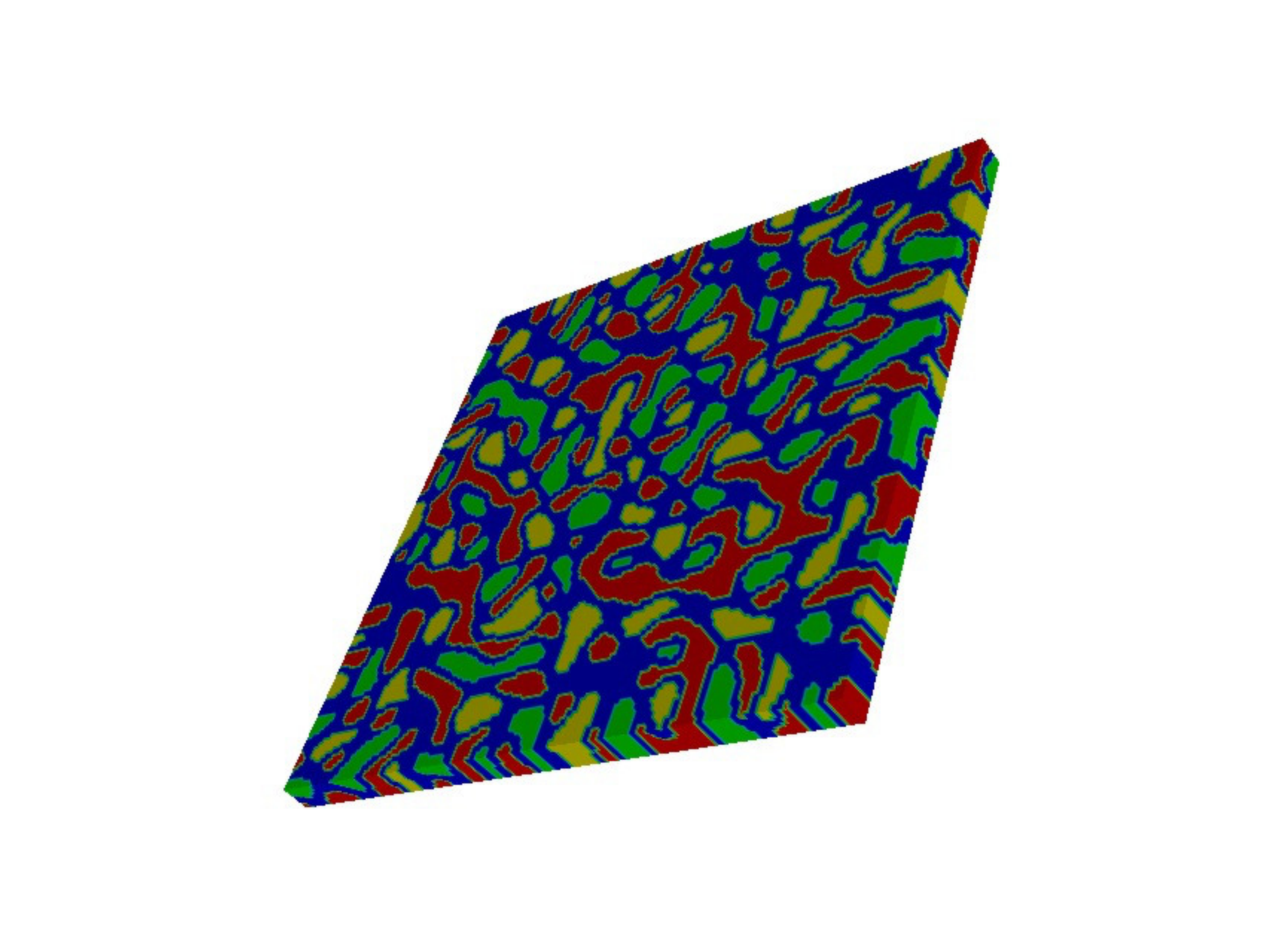}}
\end{minipage}
\begin{minipage}[]{0.3\textwidth}
\centering
\subfloat[$t = 114$]{\includegraphics[width=\textwidth]{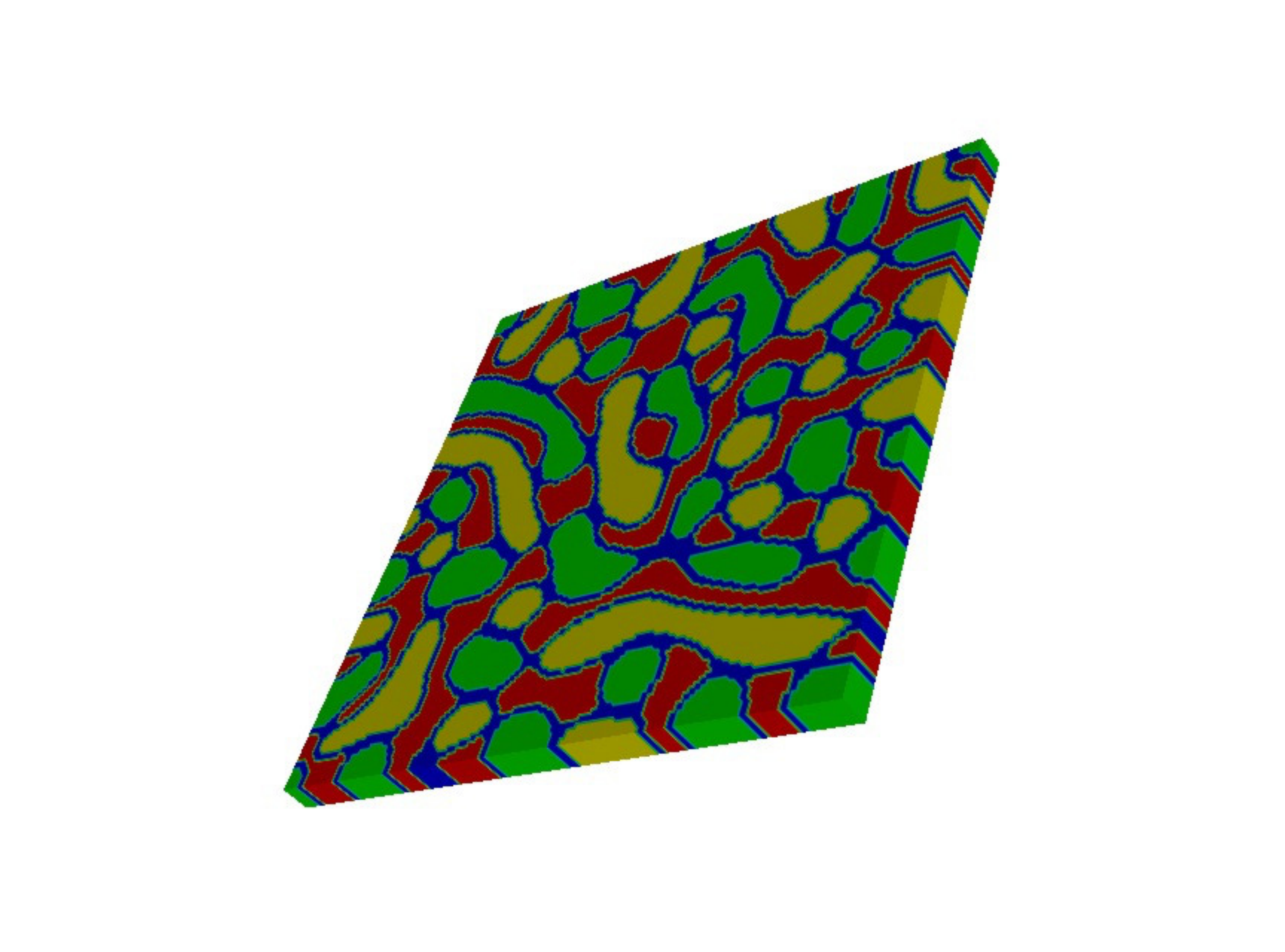}}
\end{minipage}\\
\begin{minipage}[]{0.3\textwidth}
\centering
\subfloat[$t = 714$]{\includegraphics[width=\textwidth]{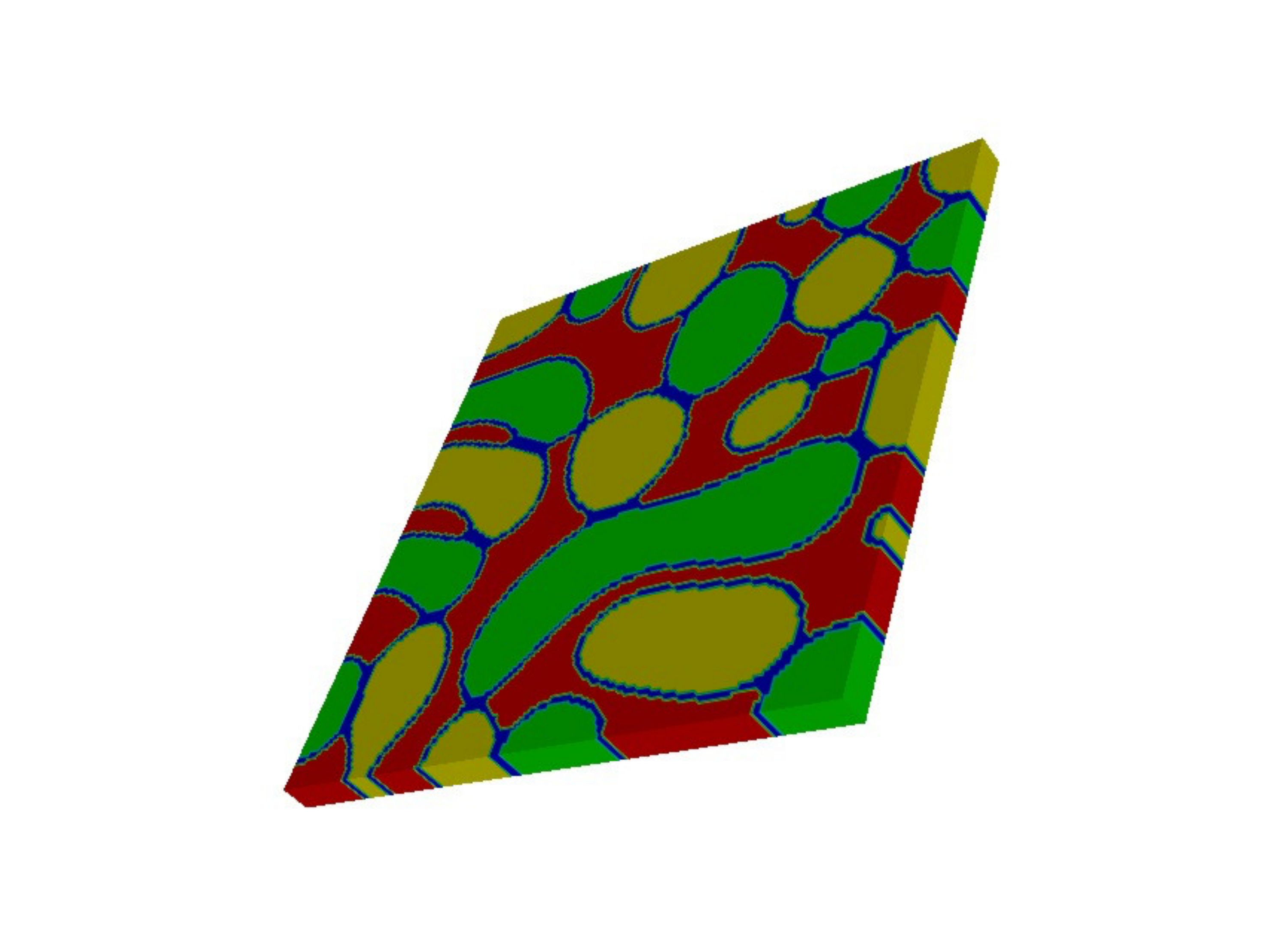}}
\end{minipage}
\begin{minipage}[]{0.3\textwidth}
\centering
\subfloat[$t = 1514$]{\includegraphics[width=\textwidth]{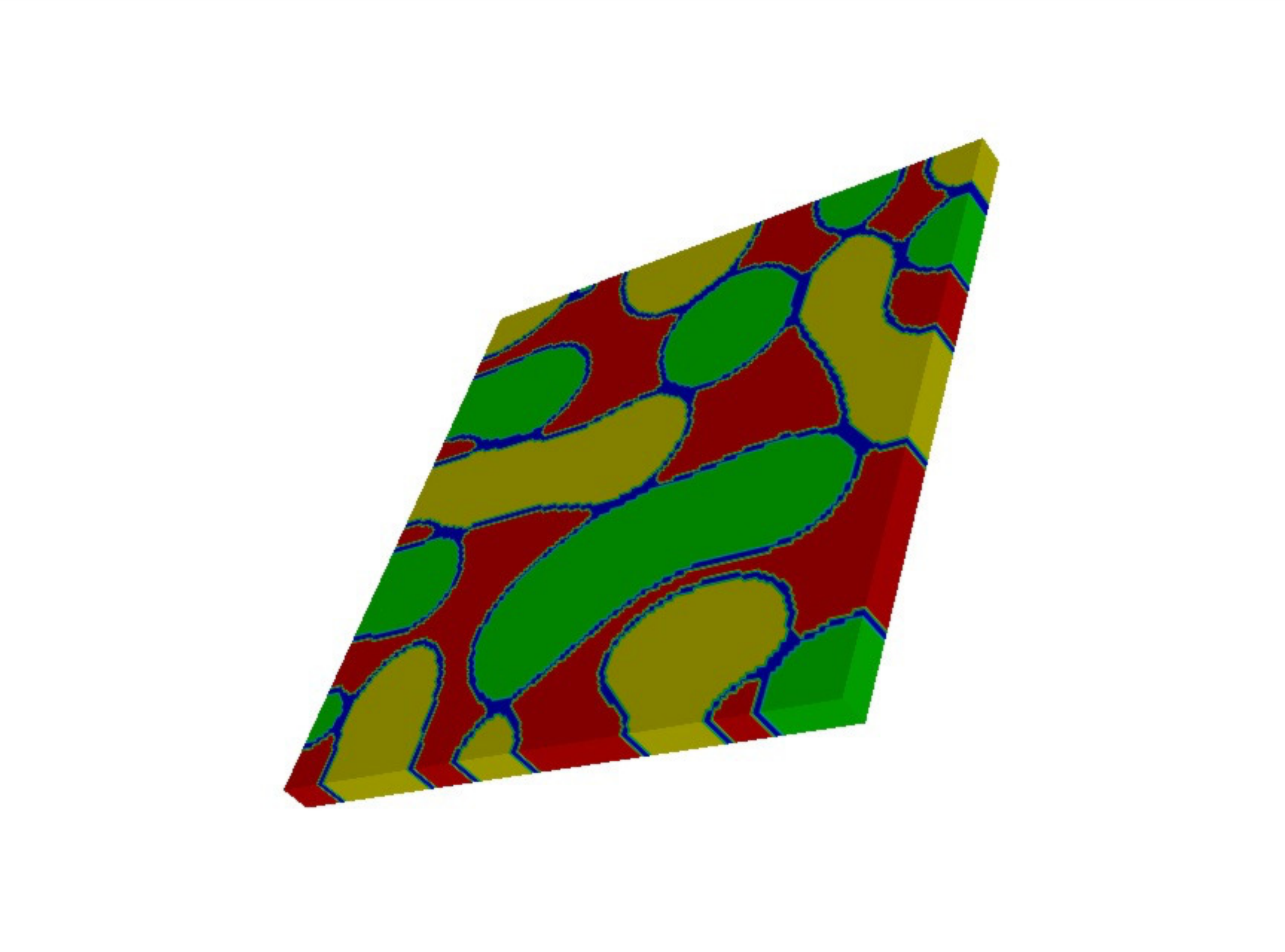}}
\end{minipage}
\begin{minipage}[]{0.3\textwidth}
\centering
\subfloat[$t = 2154$]{\includegraphics[width=\textwidth]{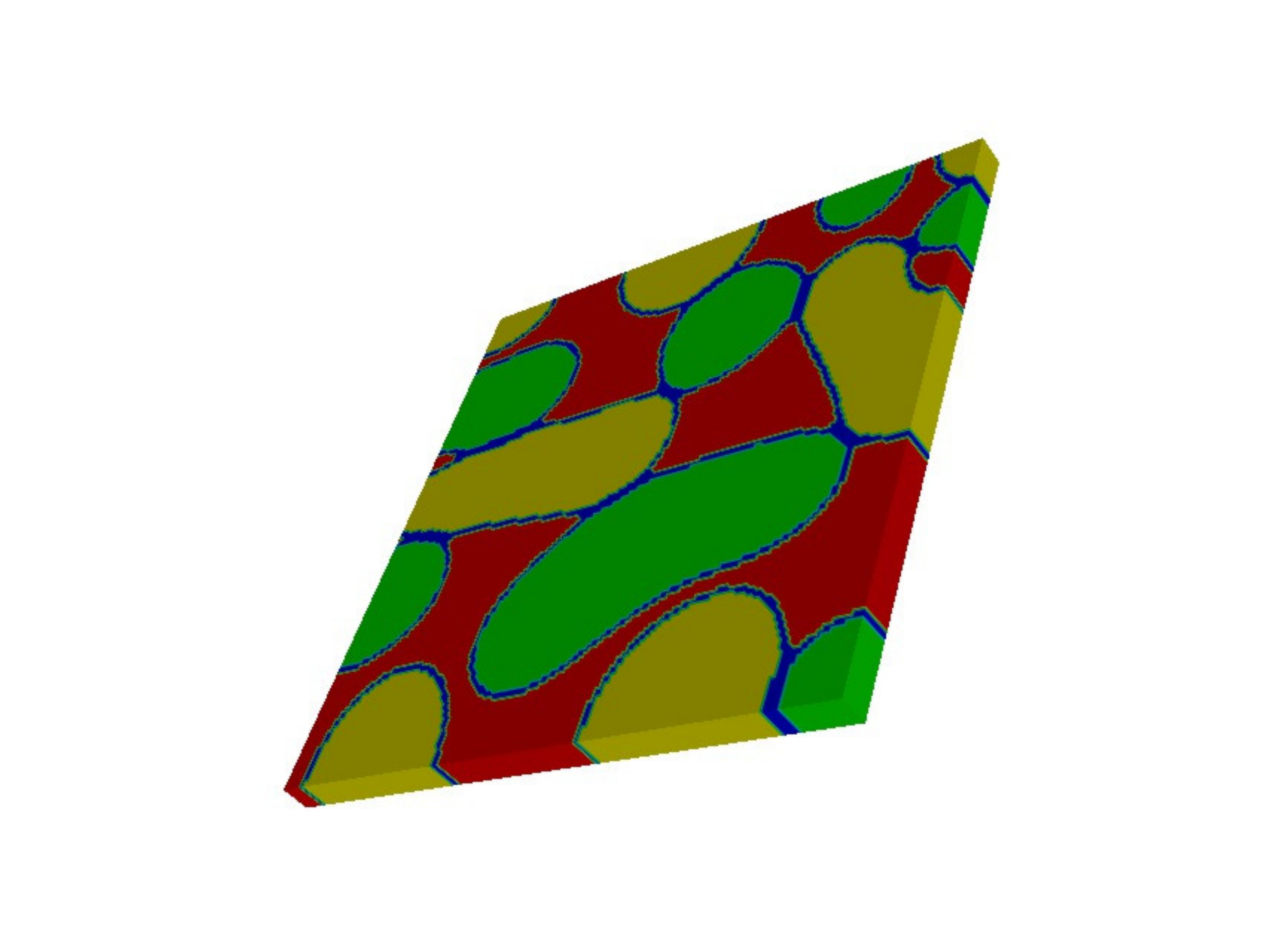}}
\end{minipage}
\caption{Preferential adhesion between the red and yellow, and green and yellow cell types, over the red and green cell types. Compare with Figure \ref{fig:threephaseequil}, and note the reduced length of yellow-green interfaces in favor of red-yellow and red-green interfaces. Also see Supplementary Movie S5.}
\label{fig:threephaseequilpref}
\end{figure}

Furthermore, the interfaces between equilibrium cell clusters are defined by gradients in certain linear combinations of $c_1$ and $c_2$. For the three-well tissue energy density function in Figure \ref{fig:ch3well2field} these linear combinations are $c_1$ and $c_2\pm c_1/\sqrt{3}$. Preferential adhesion of distinct cell types leads to unequal tissue interface energy densities, and can be modelled by different penalties applied to the gradients $\nabla c_1$, $\nabla(c_2 \pm c_1/\sqrt{3})$. For instance, a higher penalty, $\kappa_1 = 10$, applied to $\nabla c_1$  penalizes the yellow-green phase interfaces, allowing the red-yellow and red-green interfaces to form in preference. See Figure \ref{fig:threephaseequilpref}, where the yellow-green interfaces have decreased, and the red-yellow and red-green interfaces have increased in length relative to the computation in Figure \ref{fig:threephaseequil}. The yellow-green interfaces are also wider, suggesting increased matrix material (blue) that is distinct from the cell types represented by the red, yellow and green clusters. Also see Supplementary Movie S5. {The functional forms of such tissue energy density functions can be biophysically motivated from the existence of different types and numbers of cadherins molecules on surfaces of distinct cell types \citep{Alberts2008}.}

\subsection{A model for the epithelial-mesenchymal transition in cancer}
\label{sec:epMes}
\begin{figure}[hbt]
  \centering
  \includegraphics[width=0.6\textwidth]{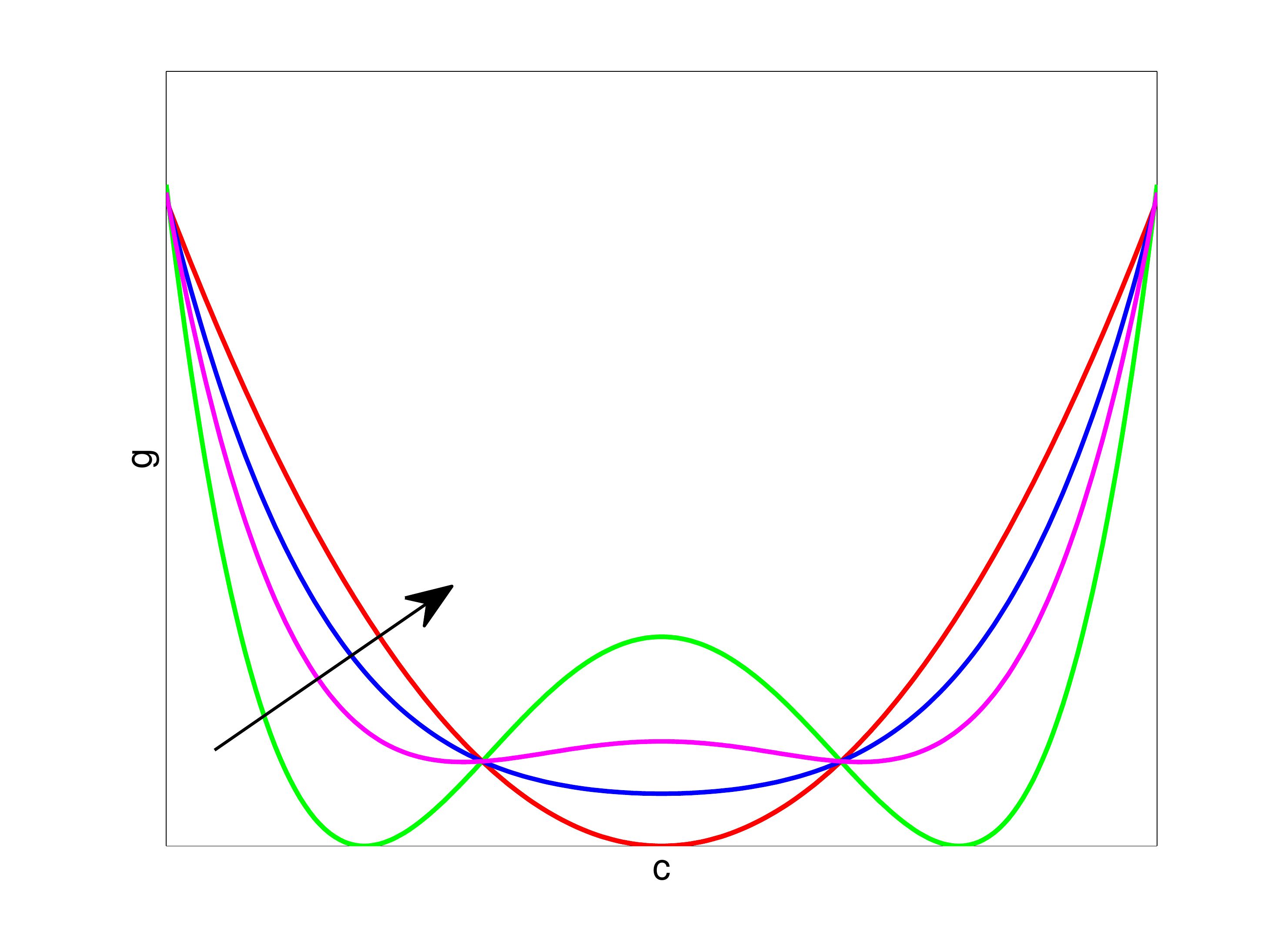}
\caption{The transition of the free energy from non-convex to convex form.}
\label{fig:convex-nonconvex}
\end{figure}

\begin{figure}
\begin{minipage}[]{0.3\textwidth}
\centering
  \includegraphics[width=0.4\textwidth]{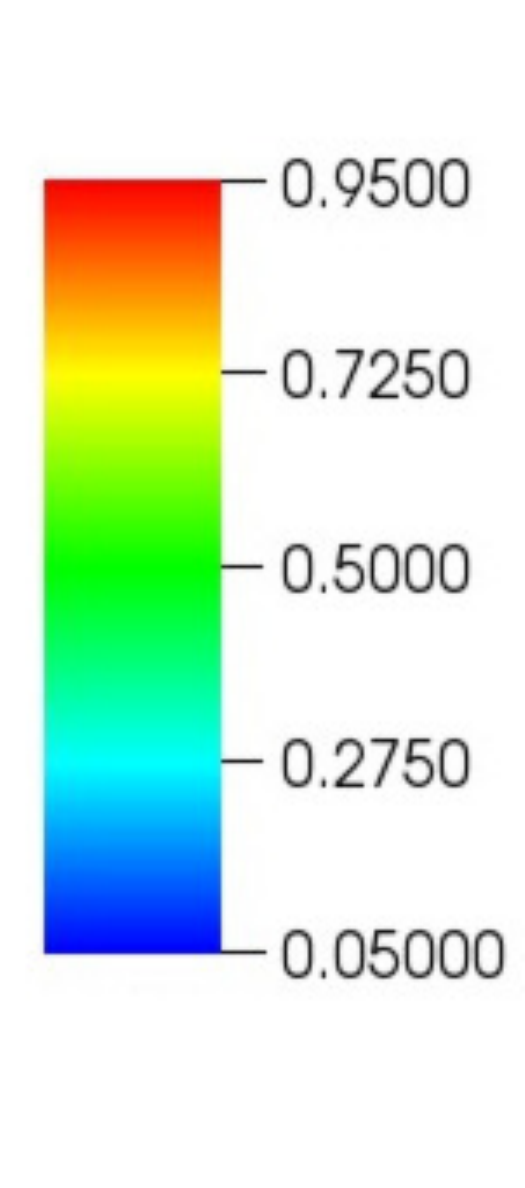}
\end{minipage}
\begin{minipage}[]{0.3\textwidth}
\centering
\subfloat[$t = 0$]{\includegraphics[width=\textwidth]{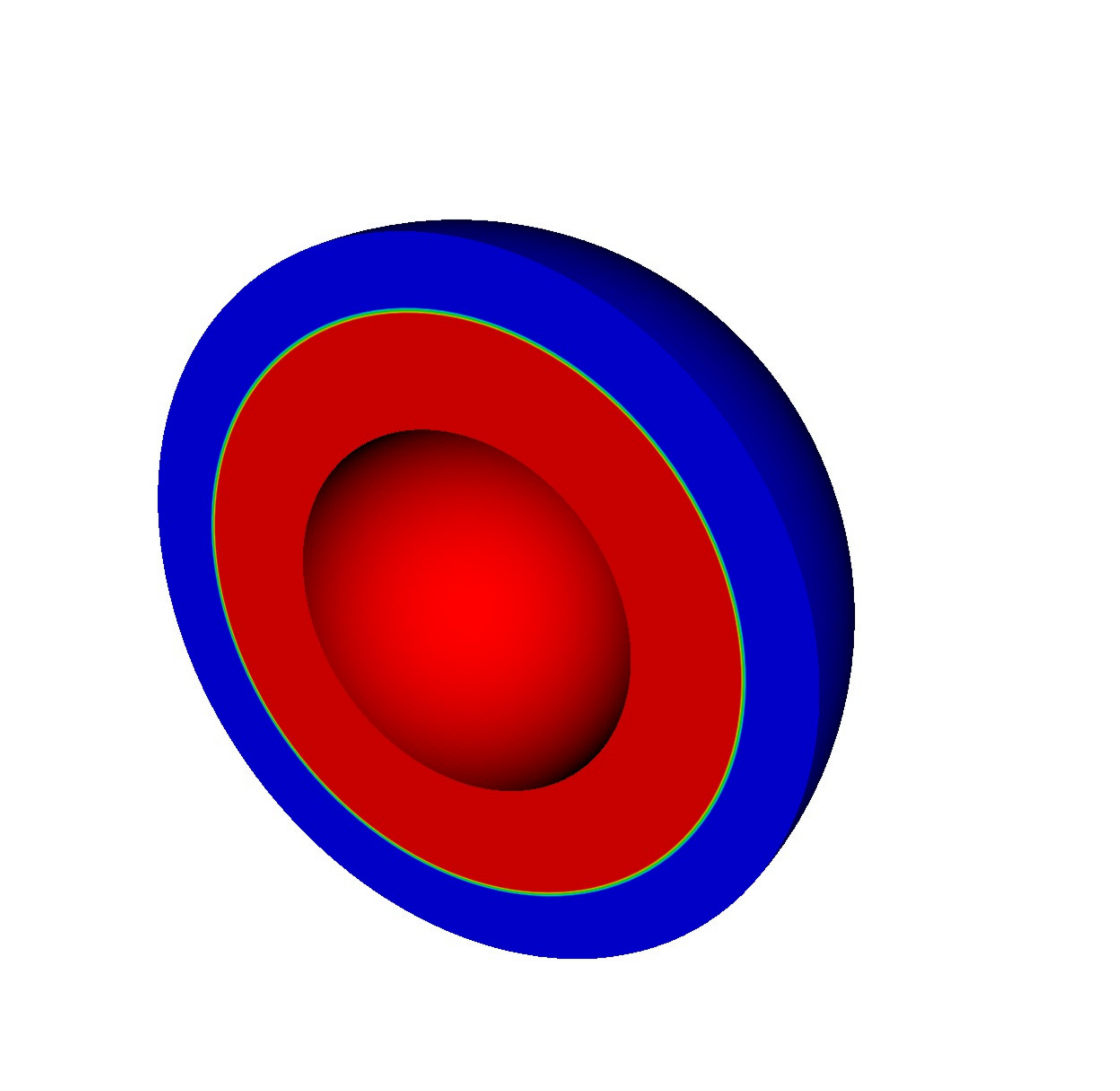}}
\end{minipage}
\begin{minipage}[]{0.3\textwidth}
\centering
\subfloat[$t = 175$]{\includegraphics[width=\textwidth]{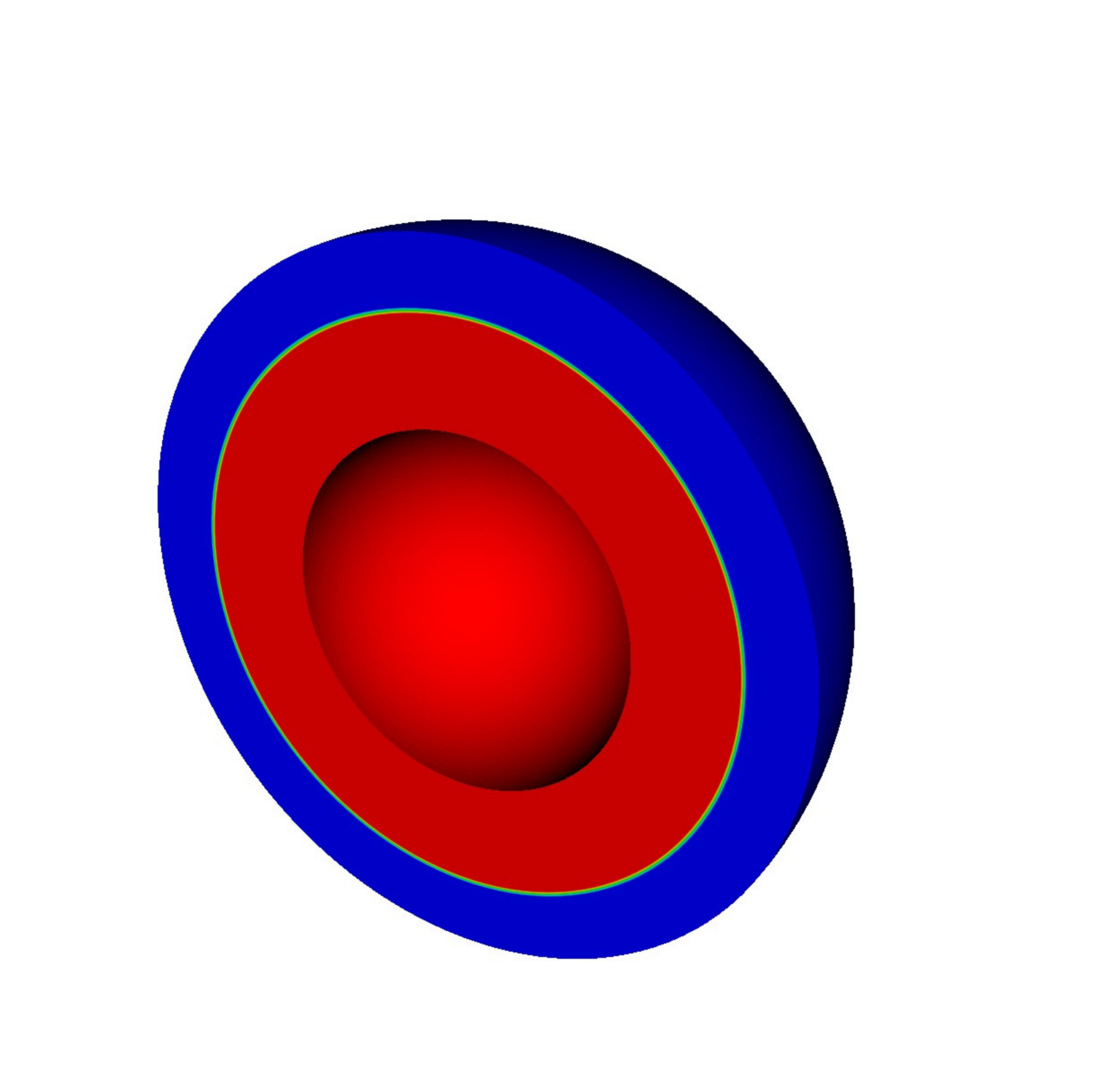}}
\end{minipage}\\
\begin{minipage}[]{0.3\textwidth}
\centering
\subfloat[$t = 275$]{\includegraphics[width=\textwidth]{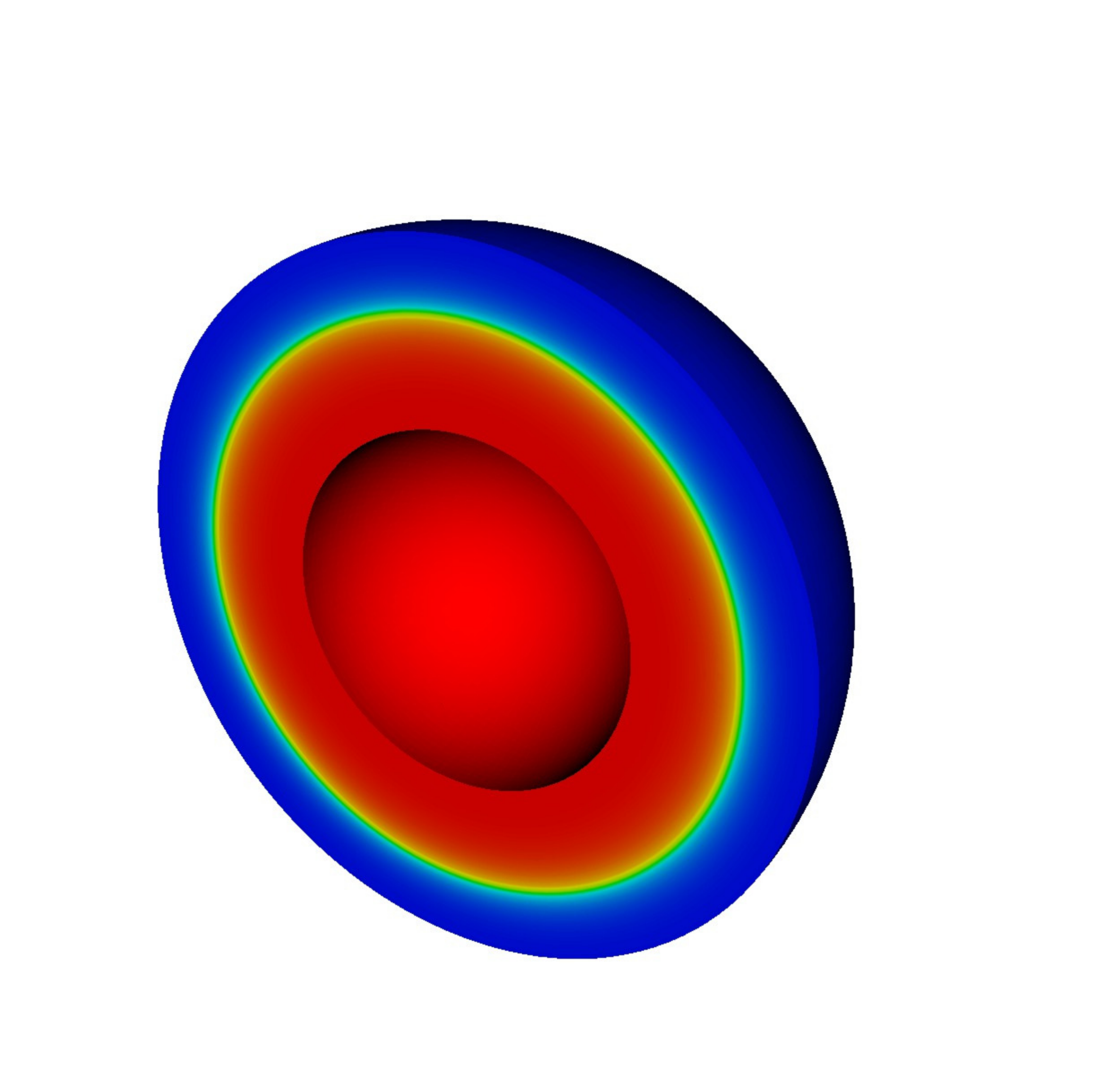}}
\end{minipage}
\begin{minipage}[]{0.3\textwidth}
\centering
\subfloat[$t = 575$]{\includegraphics[width=\textwidth]{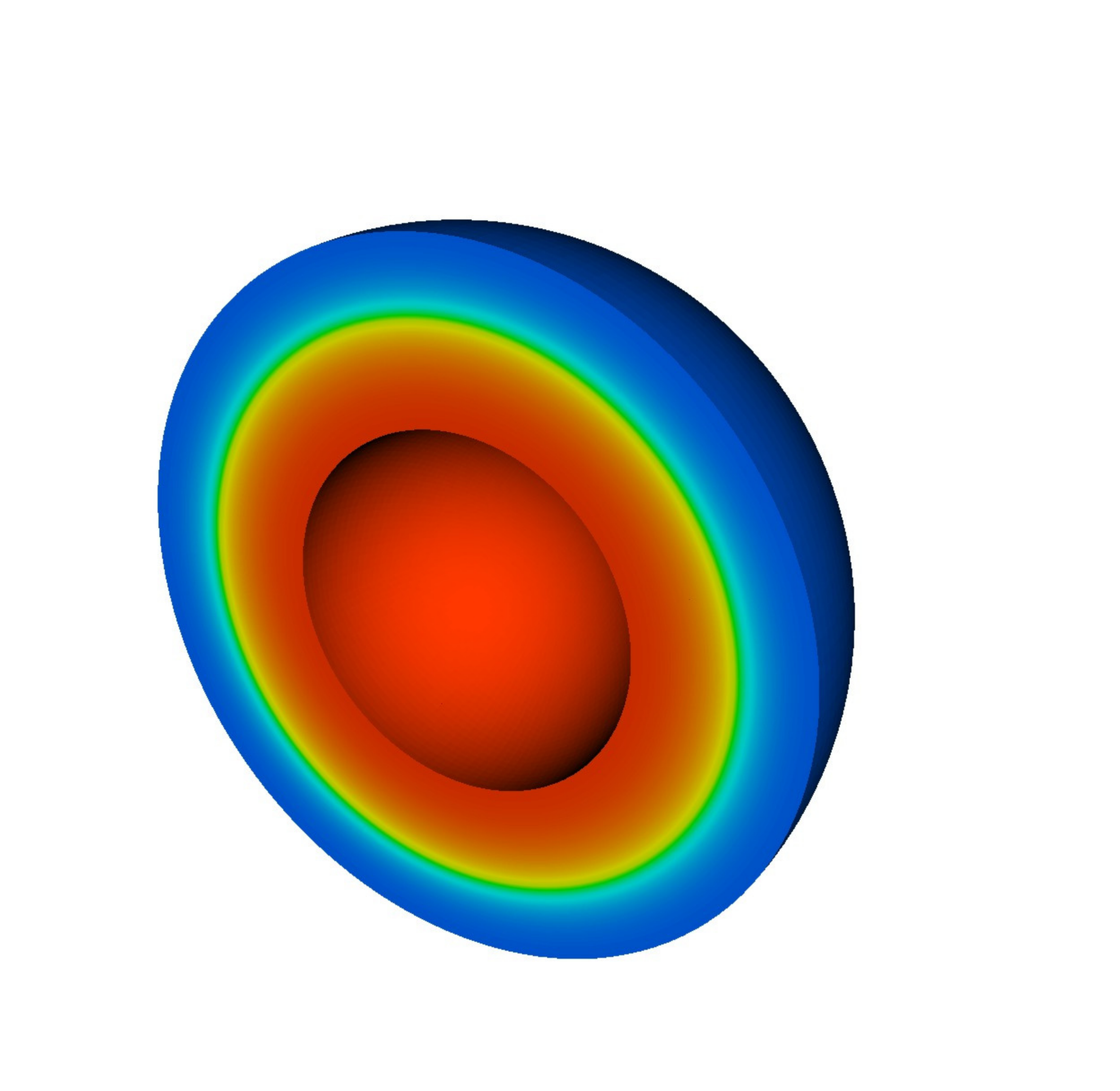}}
\end{minipage}
\begin{minipage}[]{0.3\textwidth}
\centering
\subfloat[$t = 875$]{\includegraphics[width=\textwidth]{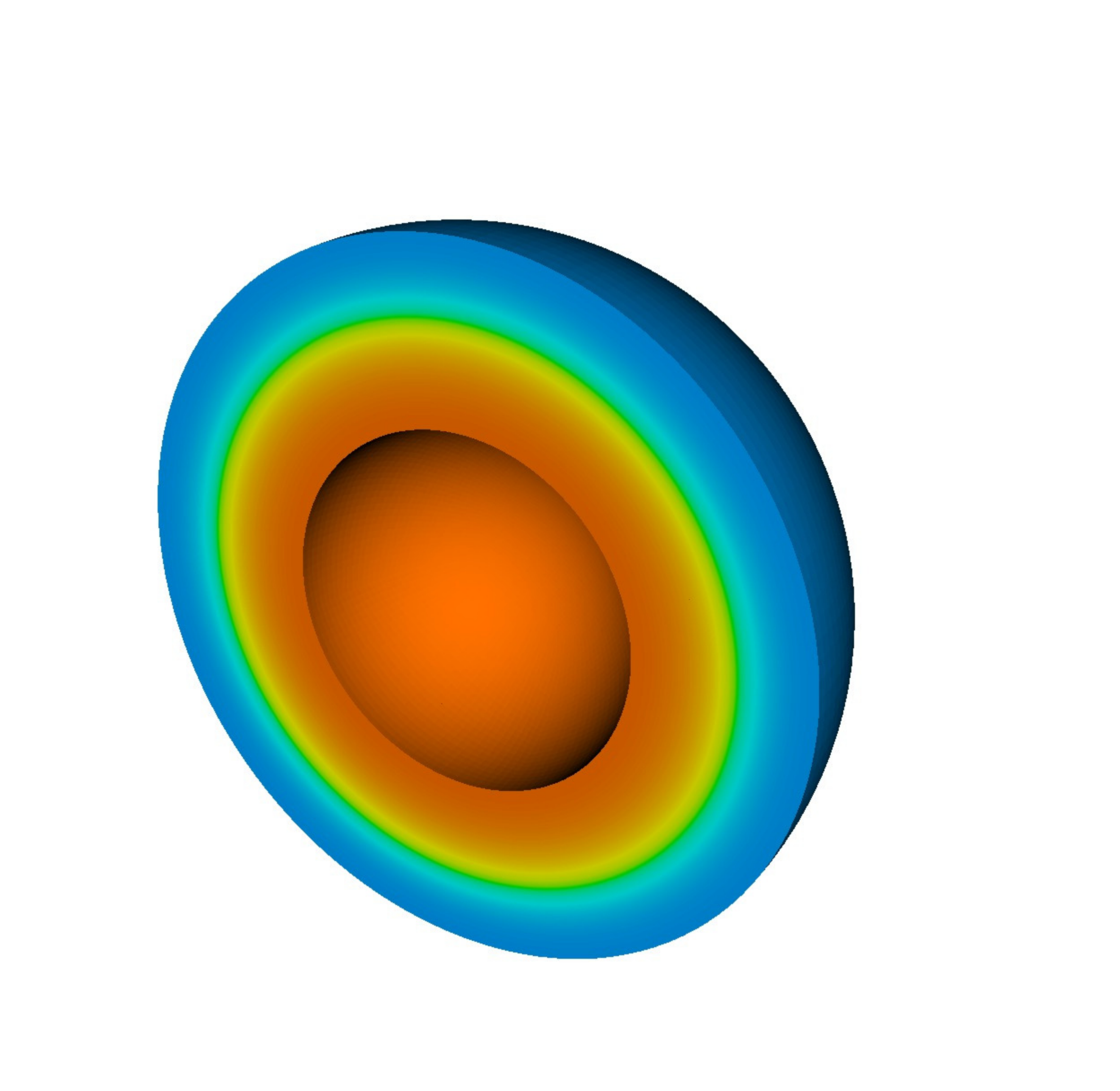}}
\end{minipage}
\caption{The two-well to single-well tissue energy density transition of Figure \ref{fig:convex-nonconvex} forms a cell segregation-based model for the epithelial (a,b) to mesenchymal (c-e) transition. Also see Supplementary Movie S6.}
\label{fig:epmes}
\end{figure}

A rather facile application of segregation of a tissue into two cell types also serves as a model for the epithelial to mesenchymal transition in cancer. Recall that this transformation of cells in a tumor recapitulates the motile, mesenchymal state of otherwise sessile, epithelial cells. Cell migration becomes possible, and with it, cell escape from a tumor mass as an early step towards malignancy \citep{Weinberg2007}. A distribution of cells, spatially segregated into tumor and non-tumor cells will remain so separated by a two-well tissue energy density function of the form in Equation (\ref{eq:totalchenergy}). In this case $c_\alpha$ and $c_\beta$ would represent the tumor and non-tumor cell concentrations. The system of equations (\ref{eq:chA}--\ref{eq:chD}) would maintain this equilibrium state. However, a time-parameterized tissue energy density function with
\begin{subequations}
\begin{align}
g(c;t) &= \omega (c - c_\alpha)^2(c - c_\beta)^2\left( \frac{1-\tanh{(t-t_0)/\tau}}{2}\right) + \omega (c - c_\gamma)^2\left( \frac{1+\tanh{(t-t_0)/\tau}}{2}\right)\label{eq:epmesA}\\
\kappa(t) &= \bar{\kappa}\left( \frac{1-\tanh{(t-t_0)/\tau}}{2}\right)\label{eq:epmesB}\\
f(c, \nabla c;t) &= g(c;t) + \frac{\kappa(t)}{2}\vert \nabla c\vert^2 \label{eq:epmesC}
\end{align}
\end{subequations}
enforces a transition from segregated equilibrium of two cell types to diffusive transport modelling random migration of the tumor cells. This transition in tissue energy density happens over a time interval $(t_0-\tau/2, t_0+\tau/2)$, and is illustrated in Figure \ref{fig:convex-nonconvex} for $c_\alpha < c_\gamma < c_\beta$. The resulting sequence of cell concentration fields appears in Figure \ref{fig:epmes}, evolving from the epithelial state in Figures \ref{fig:epmes}a, b to the mesenchymal state in Figures \ref{fig:epmes}c-e. Also see Supplementary Movie S6. Parameters for this computation appear in Table \ref{tbl:epmes}

\begin{table}[h]
\centering
\caption{Parameters for the epithelial to mesenchymal transition modelled by a transition of the tissue energy density from double-welled to single-welled.}
\begin{tabular}{ |c|c|c|c|c|c|c|c|c|  }
\hline
 Parameter & $\omega$ & $c_\alpha$ & $c_\beta$ & $c_\gamma$ & $t_0$ & $\tau$ & $\bar{\kappa}$ & $M$ \\
 \hline
 Value & $1$ & $0.05$  & $0.95$ & $0.5$ & $200$ & $0.01$ & $0.1$ & $0.1$\\
 \hline
\end{tabular}
 \label{tbl:epmes}
\end{table}
\subsection{Control of size and position by phase segregation}
\label{sec:reacDiffSizePos}

The length scale, $l_\mathrm{CH} = \sqrt{\kappa/\omega}$, of the type of phase segregation phenomena considered in this section determines the interface width between phases, and for this reason is fundamentally different from the reaction-diffusion length scale $l_\mathrm{RD}$ that determines wavelengths of the pattern. In phase segregation phenomena with flux-free boundaries, which are appropriate to patterning on biological systems, the  relative sizes of sub-domains of each phase are determined by the initial conditions. In kinetically frozen states there is the further question of how far from equilibrium the system is, which determines how many unconnected sub-domains the total mass of each phase is distributed over. At equilibrium, the minimal number of interphase interfaces is achieved, and together with the initial conditions, determines the sizes of the distinct phase sub-domains. These are the factors that control size in phase segregation-driven patterning. As with reaction-diffusion systems, and indeed any partial differential equation the boundary conditions control position in the pattern and have relevance to the examples illustrated in Figures \ref{fig:patternSizePosition} and \ref{fig:patternSizepartPosition}. In the preceding phase segregation examples, the gradient boundary condition $\nabla c\cdot\bn = 0$ aligns interfaces perpendicular to the boundary, thus controlling position of the pattern to some degree. As with reaction-diffusion systems, concentration boundary conditions would precisely control the pattern at the boundary, while its propagation into the domain would depend on the factors discussed above.

\section{Morphogenesis by elastic buckling, and the post-bifurcated shape}
{ Many morphogenetic phenomena in three dimensions can be explained by inhomogeneous growth.} For the case of sulcification and gyrification of the brain, a competing theory held that axonal tension played a role \citep{VanEssen1997}. However, that hypothesis has largely been ruled out in favor of the idea that inhomogeneous growth causes an initial buckling of elastic layers in the brain, followed by folding, wrinkling or creasing, after the bifurcation  \citep{Xu2010,Bayly2013,Tallinen2016}. This mechanism also explains the morphology of gut folding and intestinal villi \citep{Savin2011,Freddo2016}, as discussed in the Introduction. The kinematic theory that D'Arcy Thompson used to explain the forms of horns, antlers and shells also is, in essence, one of inhomogeneous growth. The growing literature on morphogenesis that was summarized in the Introduction is predicated on this idea, which seems to face no obstacle to explaining the general morphogenetic development of organs and features in three dimensions. The question of intrinsic length scale that arises for patterning phenomena also appears in morphogenesis. The equations of classical nonlinear elasticity do not possess such scales, which must arise therefore from some structural feature. It is commonly observed, and was demonstrated by \cite{BenAmar2010}, that the wavelengths of the wrinkles and folds that form in the post-bifurcated state is set by the thickness of the buckling layer. If there exists an interface energy, $\gamma$ between the buckling layer and substrate, a length scale $l_\mathrm{el} = \mu/\gamma$ arises, where $\mu$ is the elastic shear modulus \citep{Mora2011}. {However, this effect is more prominent at cellular length scales of deformation and has less of an influence at the organ scale.}

Inhomogeneous growth introduced via the elasto-growth decomposition of the deformation gradient tensor $\bF = \bF^\mathrm{e}\bF^\mathrm{g}$ \citep{Rodriguesetal1994,garikipatietal2004,garikipatietal2008,ambrosietal2011}, and a nonlinear, hyperelastic strain energy density function $\psi(\bF^\mathrm{e})$, such as the neo-Hookean strain energy density function, are the key ingredients that are combined in this treatment. 
\begin{subequations}
\begin{align}
\bF^\mathrm{e}(c) &= \bF\left(\bF^\mathrm{g}(c)\right)^{-1}\label{eq:growthFeFg}\\
\psi(\bF^\mathrm{e}) &=  \frac{1}{4}\lambda(\mathrm{det}\bF^{\mathrm{e}^\mathrm{T}}\bF^\mathrm{e} -1) - \frac{1}{2}(\frac{1}{2}\lambda +
  \mu)(\log\mathrm{det}\bF^{\mathrm{e}^\mathrm{T}}\bF^\mathrm{e}) + \frac{1}{2}\mu(\bF^\mathrm{e}\colon\bF^\mathrm{e} - 3)\label{eq:neohookean}
  \end{align}
  \end{subequations}
Here, $\lambda$ and $\mu$ are Lam\'{e} parameters, with the latter already introduced as the shear modulus. The first Piola-Kirchhoff stress tensor $\bP$ is obtained in the usual manner and is governed by the quasistatic balance of momentum equation.
  
  \begin{subequations}
  \begin{align}
  \bP &= \frac{\partial \psi}{\partial \bF^\mathrm{e}}\label{eq:firstPK}\\
  \mathrm{Div}\bP &= \bzero\label{eq:govelast}
\end{align}
\end{subequations}

\subsection{Diffusive cell migration and cortical folding}
\label{sec:diffcellmigbuckl}
\begin{figure}[hbt]
 \begin{minipage}[]{0.3\textwidth}
\centering
\includegraphics[width=0.3\textwidth]{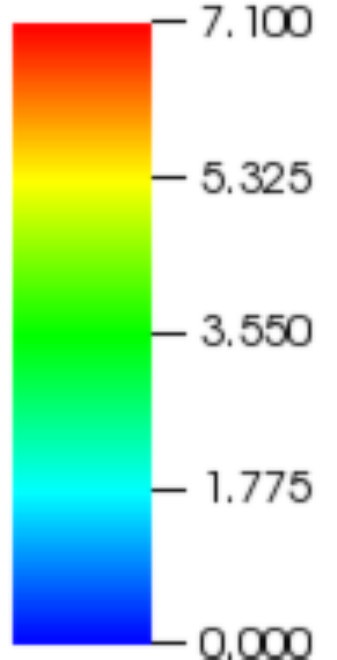}
\end{minipage}
\begin{minipage}[]{0.3\textwidth}
\centering
\subfloat[$t = 0$]{\includegraphics[width=1.1\textwidth]{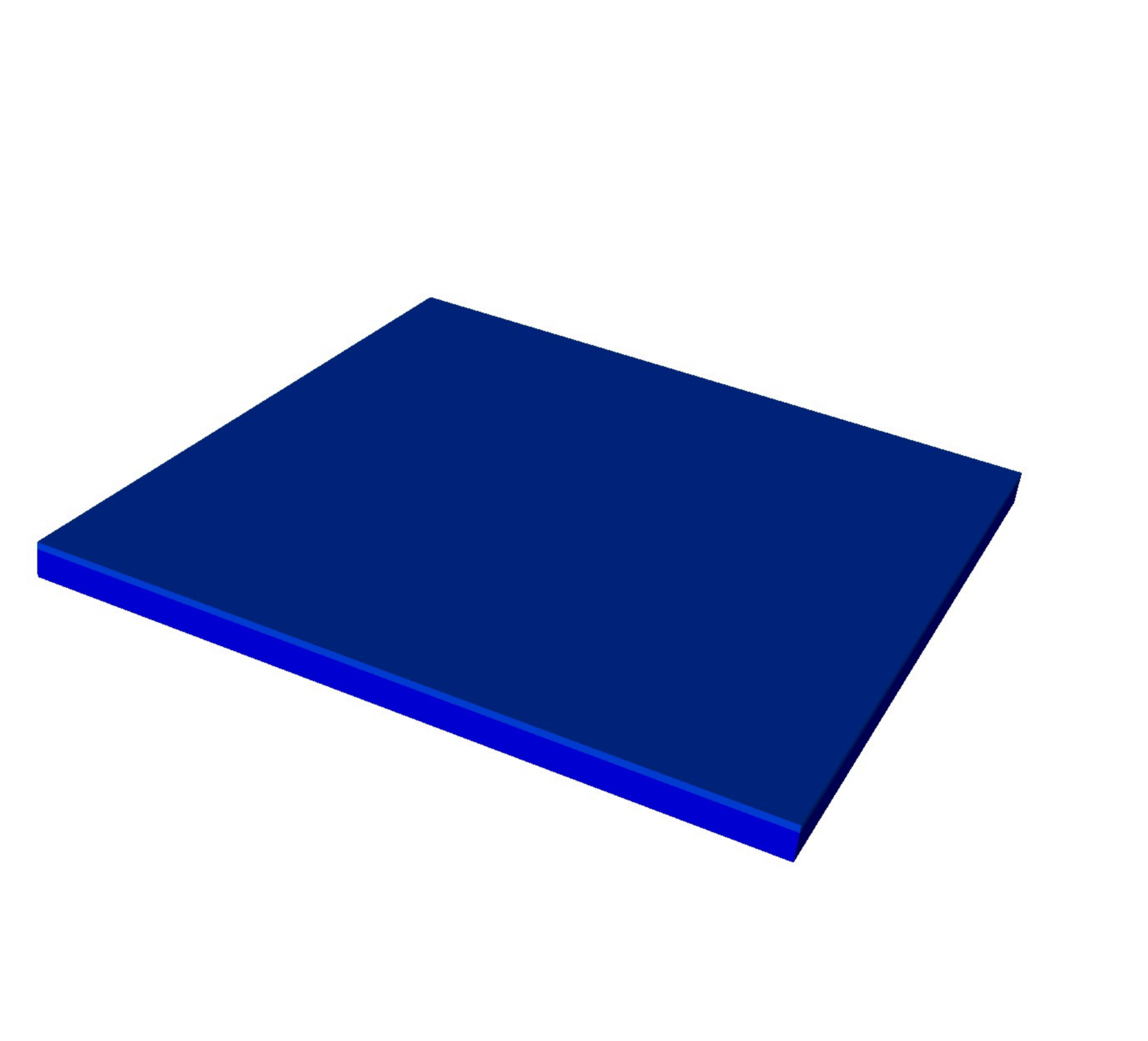}}
\end{minipage}
\begin{minipage}[]{0.3\textwidth}
\centering
\subfloat[$t = 31$]{\includegraphics[width=1.1\textwidth]{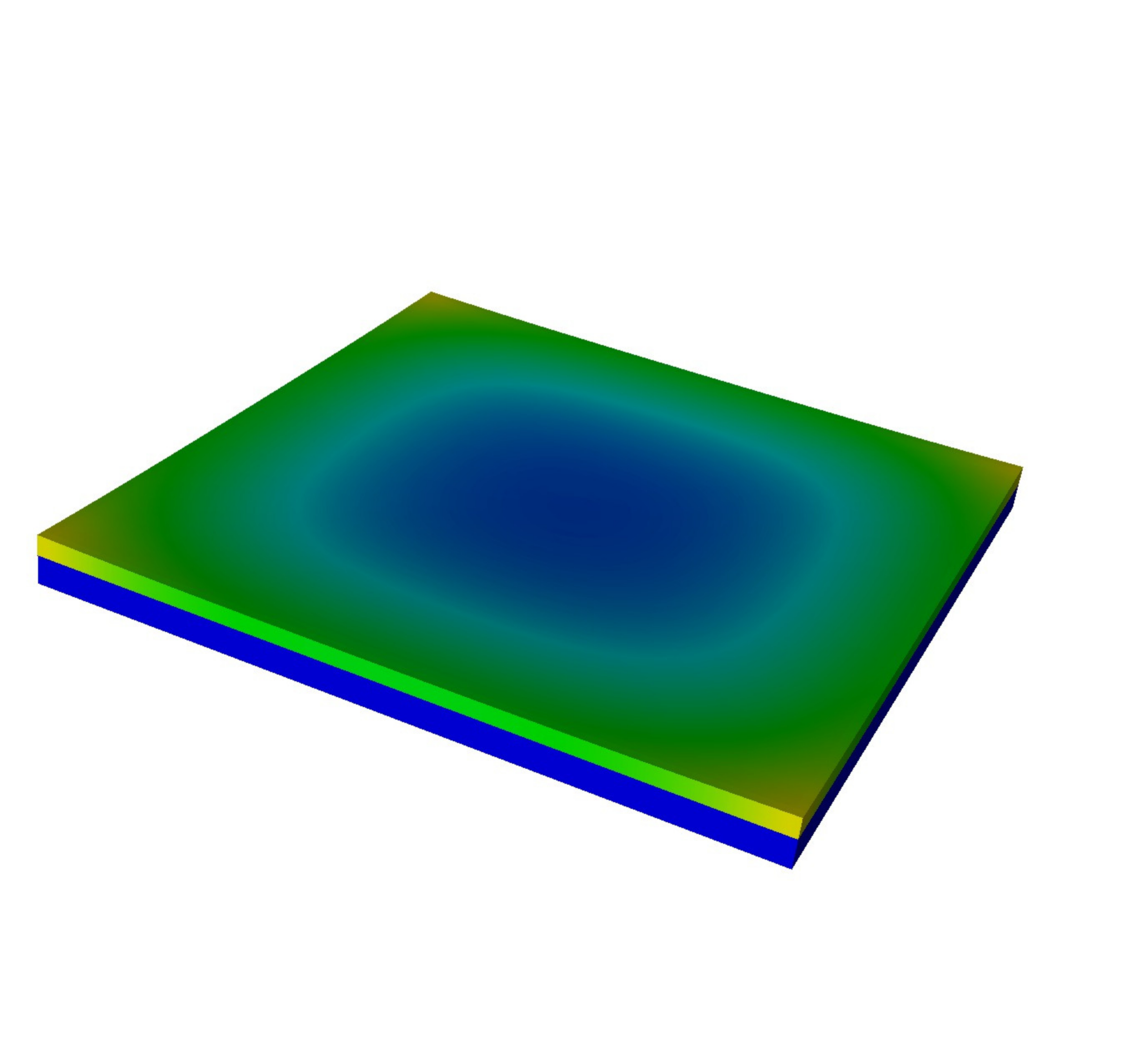}}
\end{minipage}\\
\begin{minipage}[]{0.3\textwidth}
\centering
\subfloat[$t = 55$]{\includegraphics[width=1.1\textwidth]{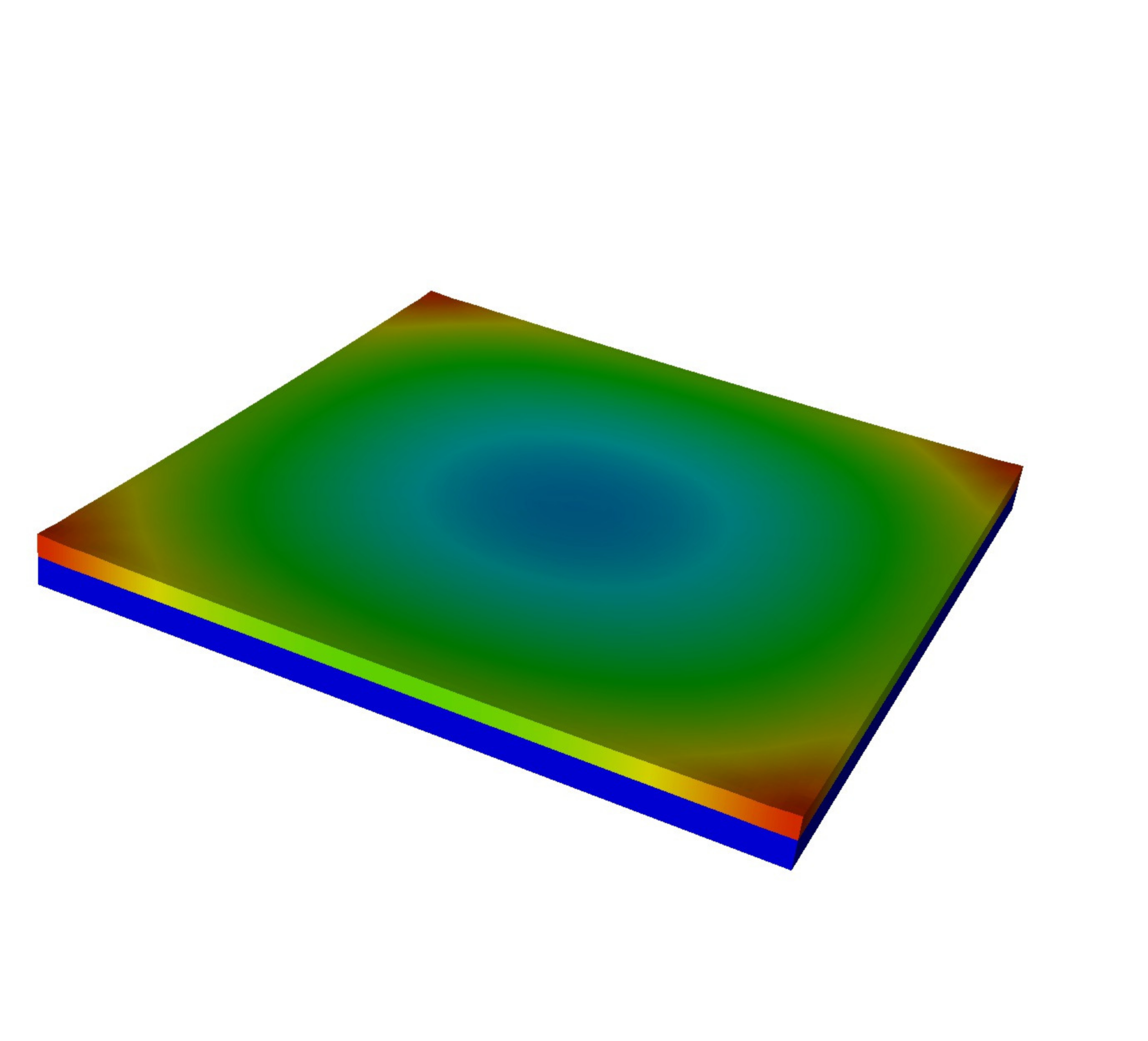}}
\end{minipage}
\begin{minipage}[]{0.3\textwidth}
\centering
\subfloat[$t = 56$]{\includegraphics[width=1.1\textwidth]{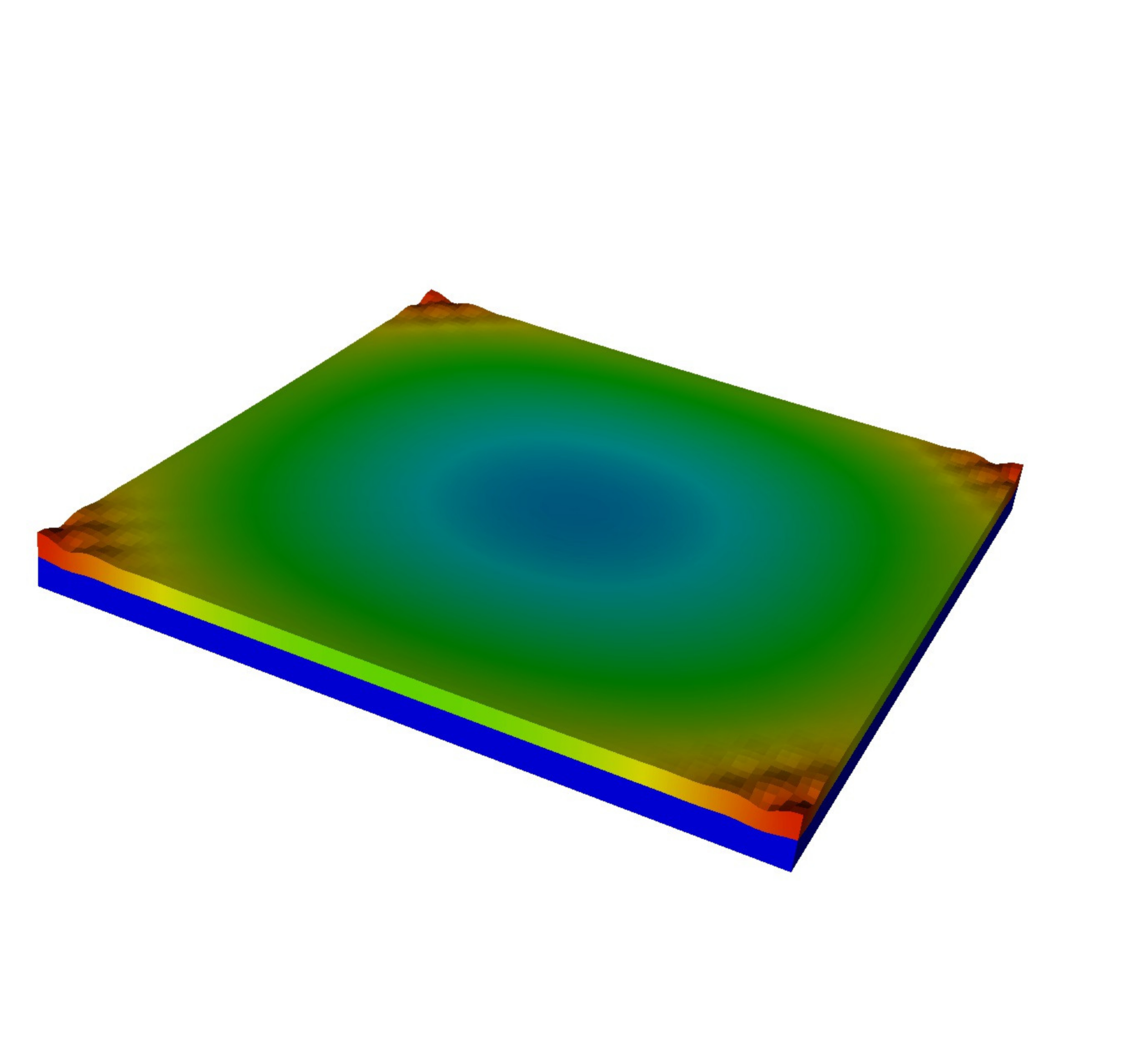}}
\end{minipage}
\begin{minipage}[]{0.3\textwidth}
\centering
\subfloat[$t = 62$]{\includegraphics[width=1.1\textwidth]{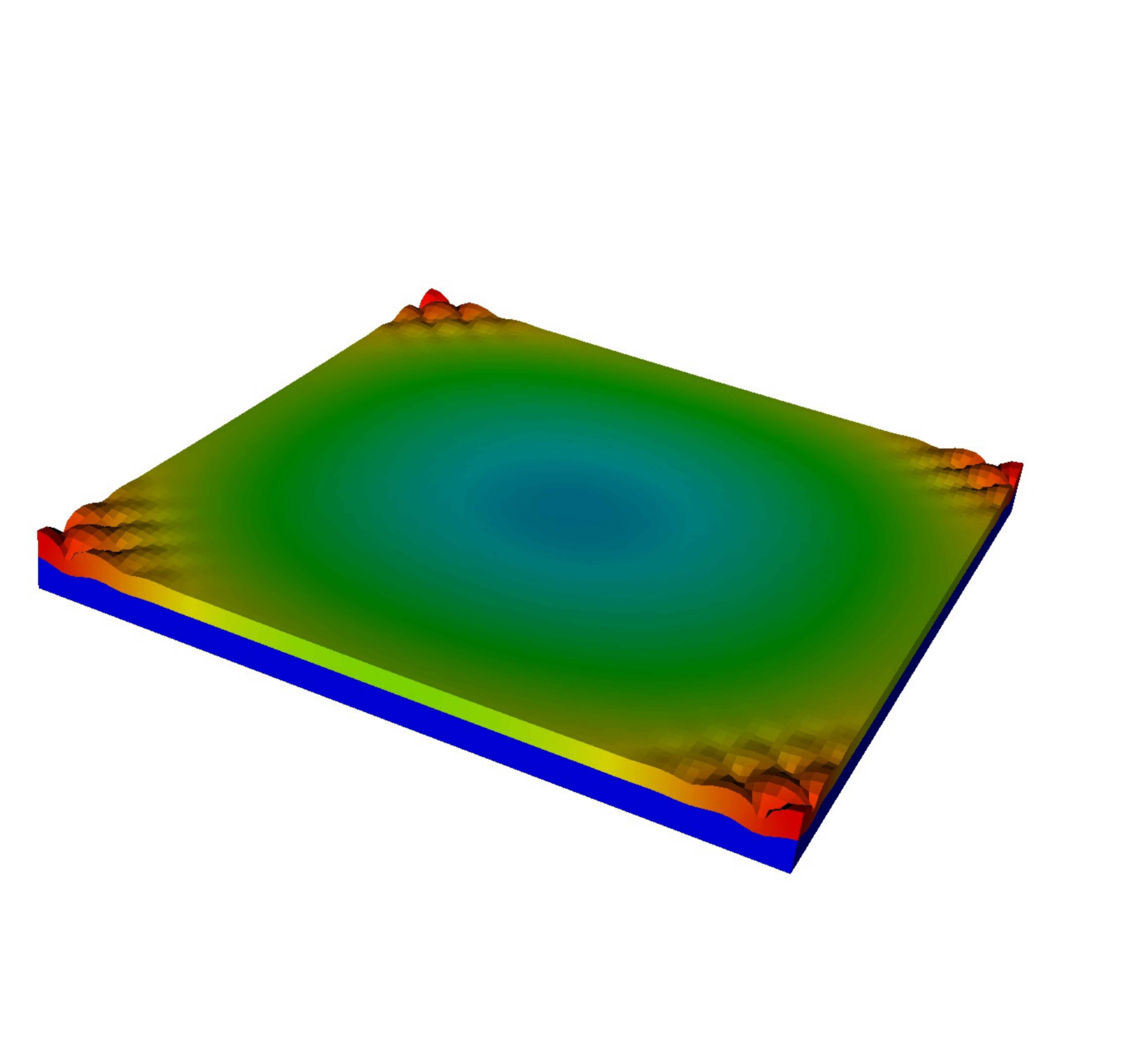}}
\end{minipage}
\caption{Buckling and post-bifurcation wrinkling/creasing of a flat layer on a substrate due to cell migration. Note the coincidence of the wrinkling/creasing and the corners of highest cell concentration. Also see Supplementary Movie S7.}
\label{fig:buckleflatlayer}
\end{figure}

A local growth tensor, $\bF^\mathrm{g}(c)$ has been modelled widely in the growth literature as an independently specified tensor function of cell concentration $c$, which is determined by local growth. However, in some developmental biological contexts at least, growth arises due to cell migration. Neuronal migration from the ventricles to the cortical layer appears to have an important connection to the densification and eventual buckling into gyri and sulci of the outer cortical layer of the brain \citep{Sun2014}. This introduces the coupling of cell distribution with morphogenesis by buckling and evolution of the post-bifurcated shape. Figure \ref{fig:buckleflatlayer} shows such a coupled evolution of cell transport and elastic buckling in a flat layer on a substrate. Cells migrate into the layer from the straight boundaries. Their diffusive motion is confined to this layer, which then buckles, wrinkles and creases into a post-bifurcated shape due to the local accumulation of cells. Also see Supplementary Movie S7. In this case, isotropic swelling was assumed.
\begin{equation}
\bF^\mathrm{g}(c) = \left(\frac{c}{c_\mathrm{crit}} \right)^{1/3}\bone,
\label{eqn:isotropicgrowth}
\end{equation}
where $c_\mathrm{crit}(\bx) = c(\bx,0)$ is set  to the initial cell concentration. Other parameters for this computation appear in Table \ref{tbl:buckleflatlayer}. Morphogenetic development is determined by the extent to which the diffusing influx has raised the local concentration beyond a threshold so that the growth tensor $\bF^\mathrm{g}(c)$ has induced buckling and a post-bifurcated shape of wrinkles and creases. While it serves as a fairly elementary illustration of morphogenesis following a diffusive concentration field, this example is not representative of the brain-like physiology because of its flat geometry and in-plane cell transport.

\begin{table}[h]
\centering
\caption{Parameters for the buckling and post-bifurcation wrinkling/creasing of a flat layer on a substrate due to diffusive cell migration.}
\begin{tabular}{ |c|c|c| c|}
\hline
 Parameter & $D$ & $\lambda$ & $\mu$ \\
 \hline
 Value & $1$ & $2887$  & $1923$ \\
 \hline
\end{tabular}
 \label{tbl:buckleflatlayer}
\end{table}

\subsection{Cell advection and cortical folding}
\label{sec:advcellmigbuckl}
\begin{figure}
\begin{minipage}[]{0.3\textwidth}
\centering
  \includegraphics[width=0.3\textwidth]{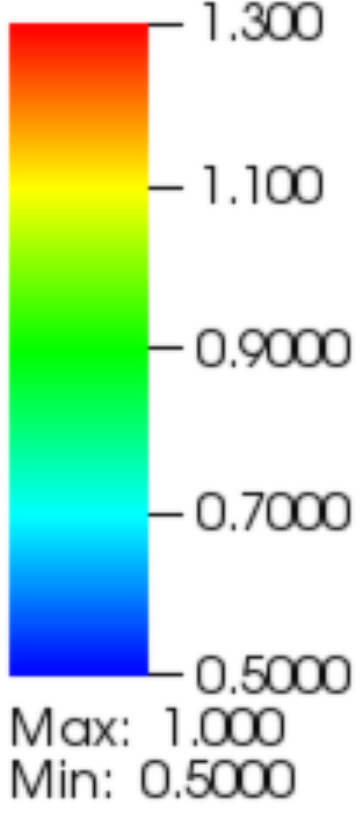}
\end{minipage}
\begin{minipage}[]{0.3\textwidth}
\centering
\subfloat[$t = 0$]{\includegraphics[width=1.5\textwidth]{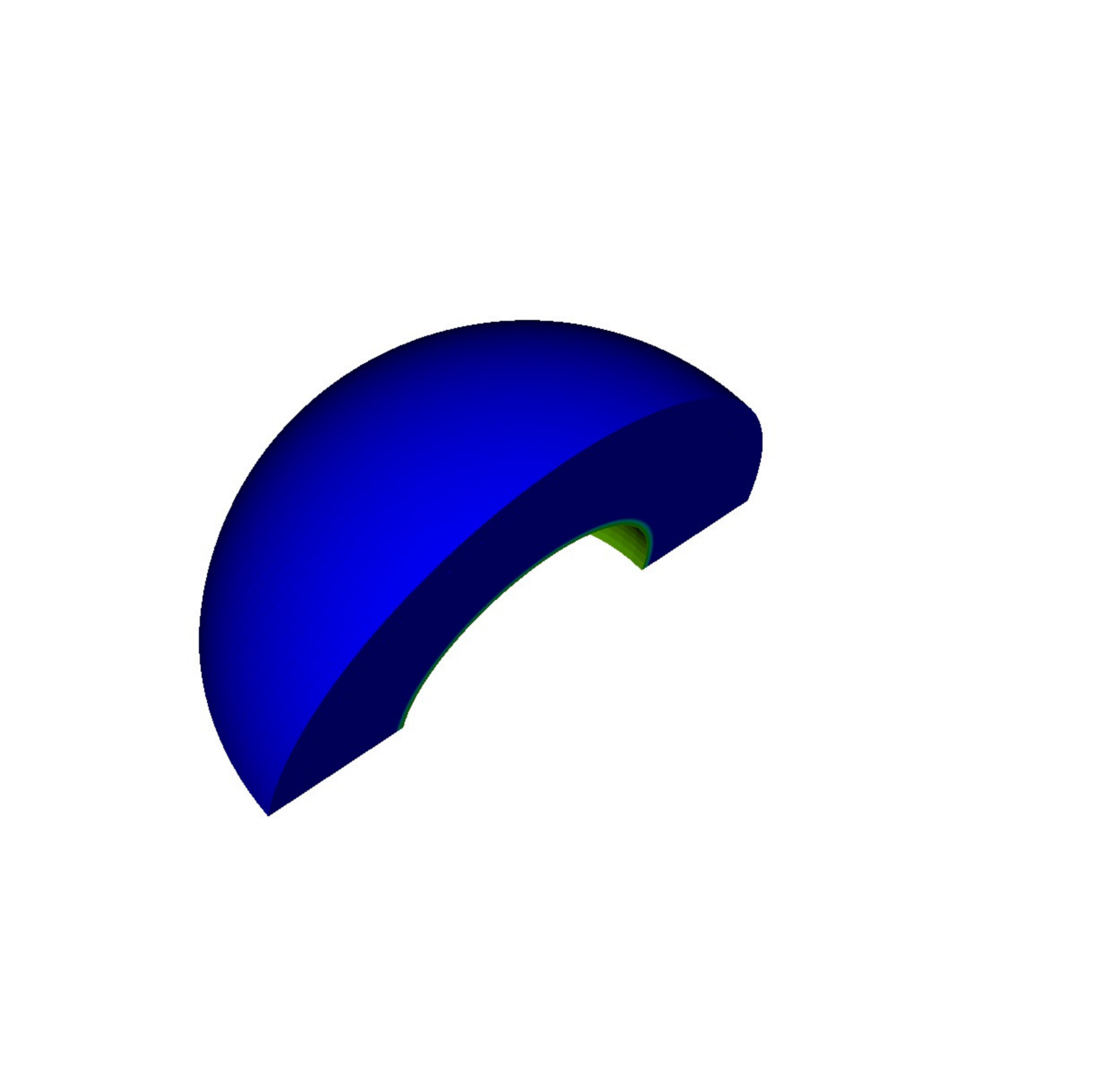}}
\end{minipage}
\begin{minipage}[]{0.3\textwidth}
\centering
\subfloat[$t = 80$]{\includegraphics[width=1.5\textwidth]{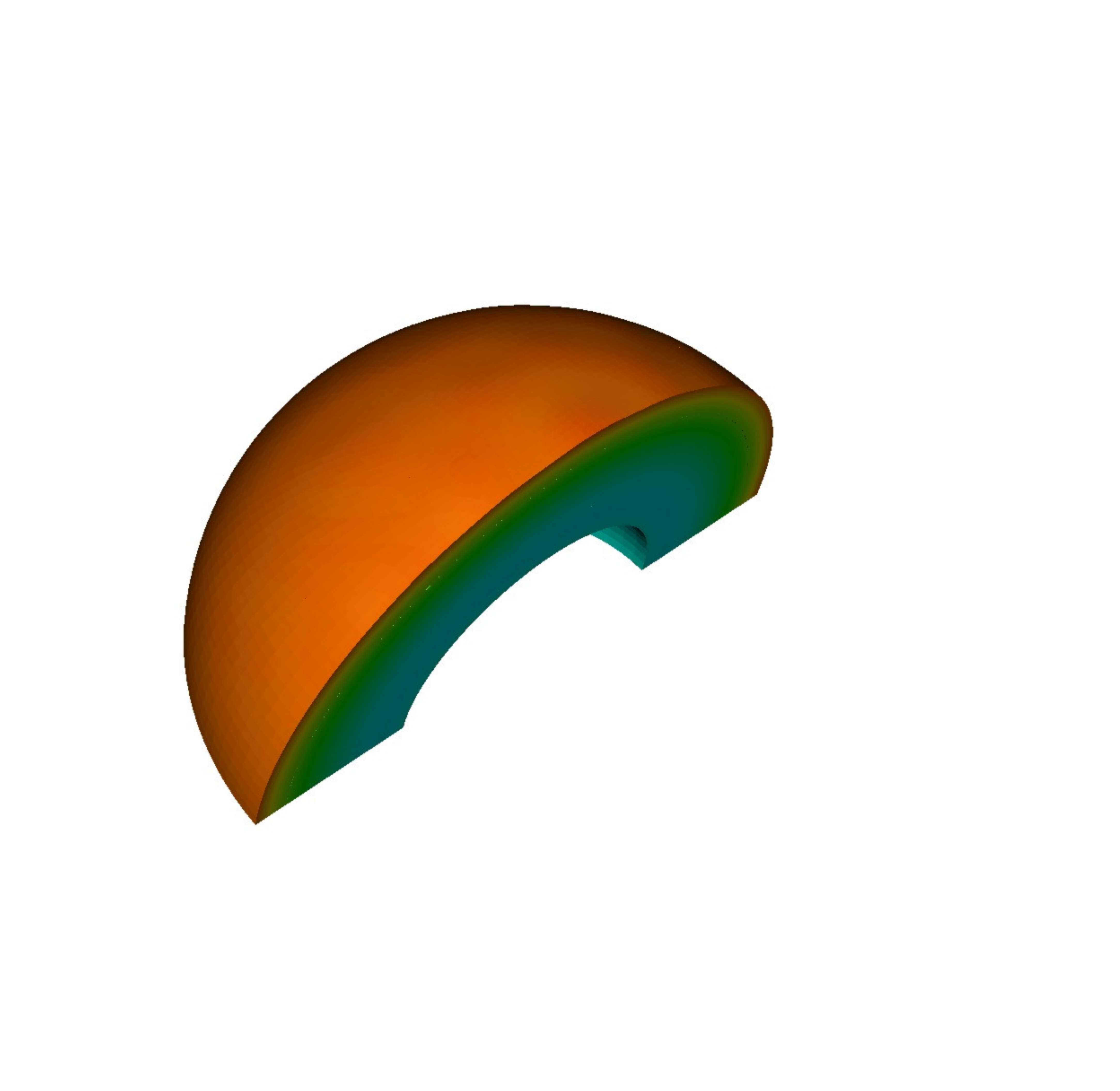}}
\end{minipage}\\
\begin{minipage}[]{0.3\textwidth}
\centering
\subfloat[$t = 100.5$]{\includegraphics[width=1.5\textwidth]{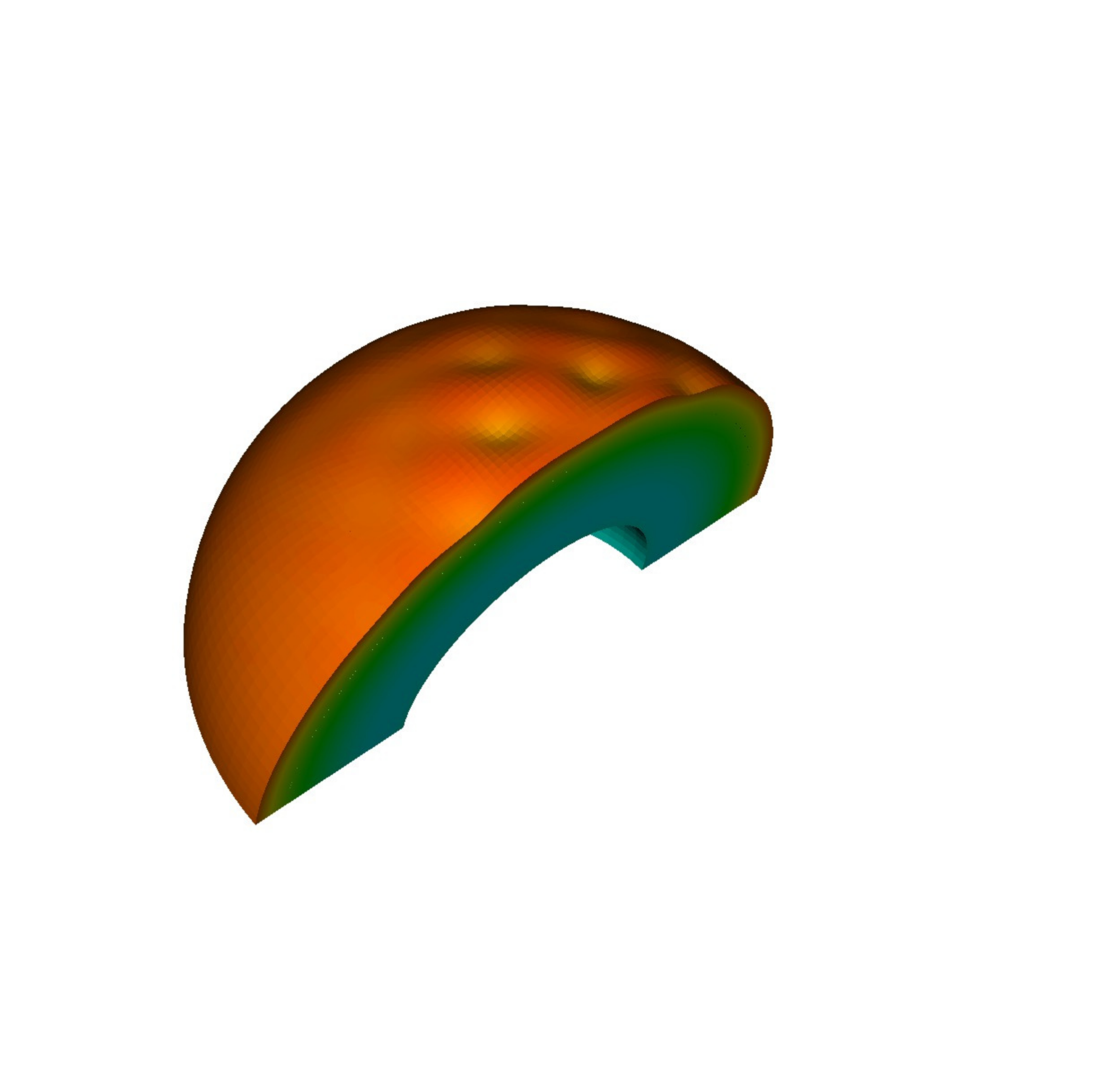}}
\end{minipage}
\begin{minipage}[]{0.3\textwidth}
\centering
\subfloat[$t = 103$]{\includegraphics[width=1.5\textwidth]{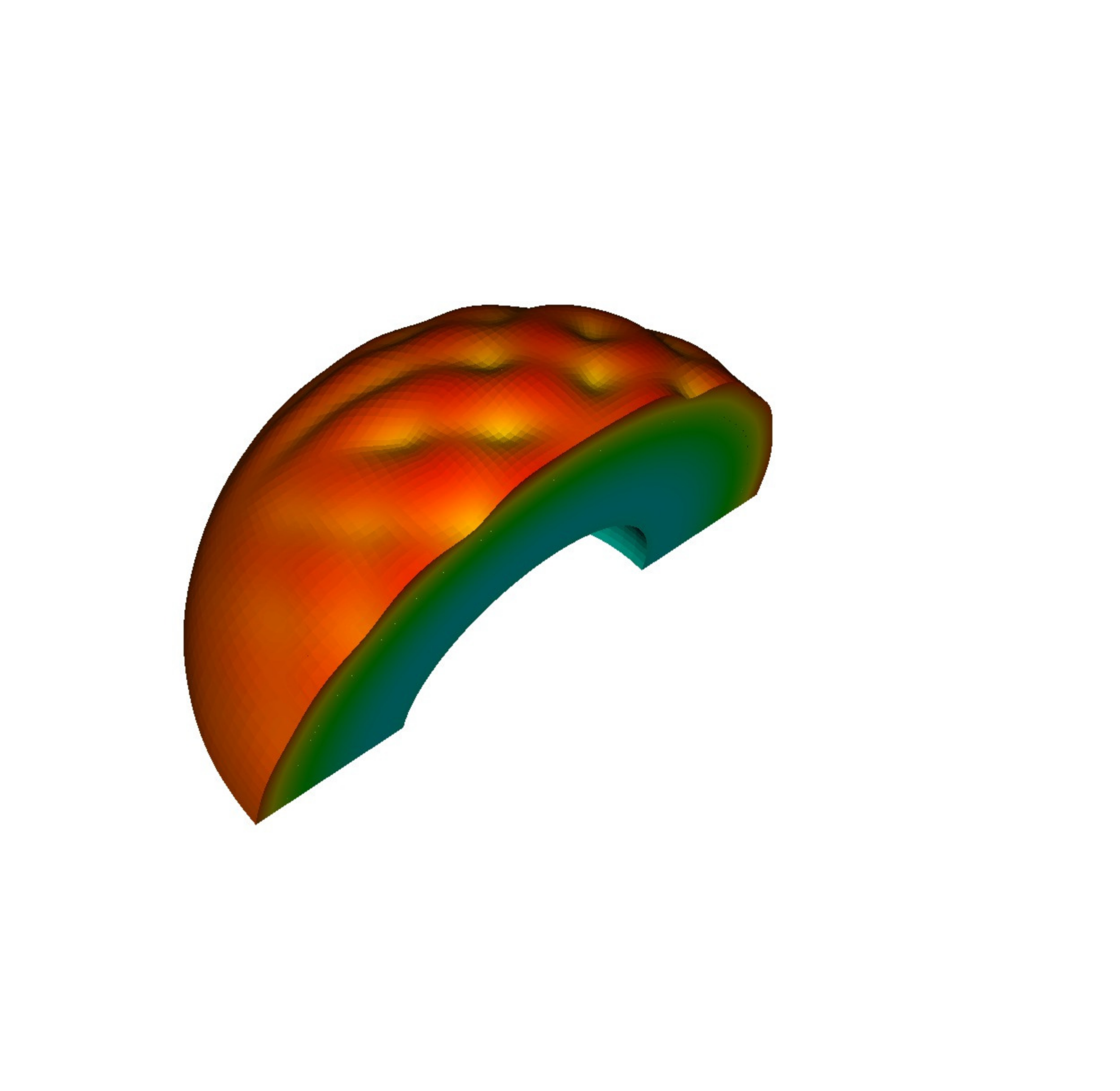}}
\end{minipage}
\begin{minipage}[]{0.3\textwidth}
\centering
\subfloat[$t = 104$]{\includegraphics[width=1.5\textwidth]{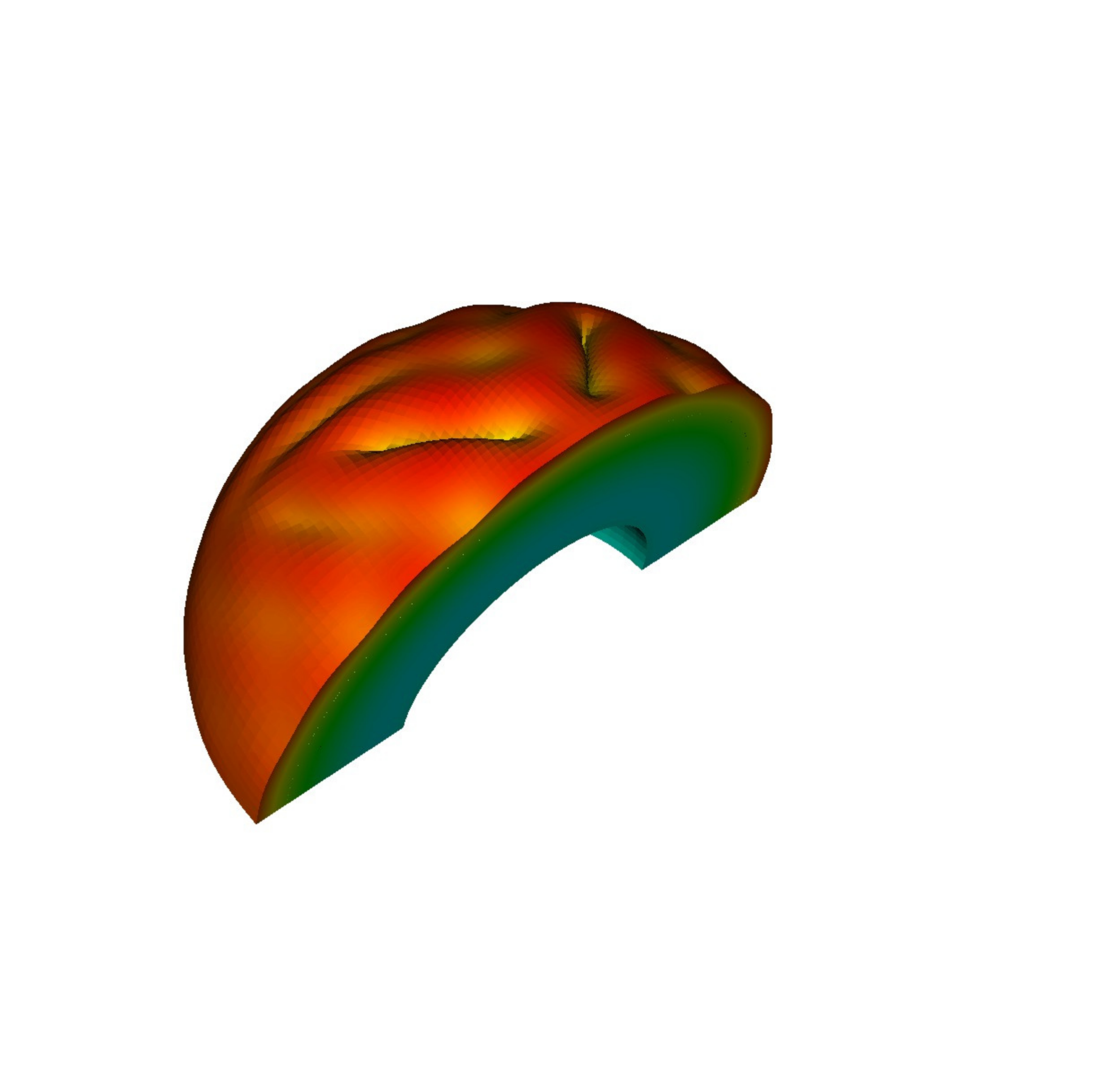}}
\end{minipage}
\caption{Sulcification and gyrification induced by advective cell migration. Also see Supplementary Movie S8.}
\label{fig:gyrification}
\end{figure}

Since migration of neurons from the ventricle to the cortex is directed by signalling of some nature, it is perhaps more appropriate to represent it as advective transport, with a small diffusive component:
\begin{subequations}
\begin{align}
\frac{\partial c}{\partial t} &= D\nabla^2 c - \bv\cdot\nabla c \quad\mathrm{in}\;\Omega\label{eq:advdiffA}\\
c &= \bar{c} \quad\mathrm{on}\;\partial\Omega_c\label{eq:advdiffB}\\
\nabla c\cdot\bn &= 0 \quad\mathrm{on}\;\partial\Omega_j\label{eq:advdiffC}\\
c(\bx,t) &= c_0(\bx) \quad\mathrm{at}\; t = 0\label{eq:advdiffD}
\end{align}
\end{subequations}
where $\bv$ is the advection velocity, and concentration boundary conditions have been reintroduced on $\partial\Omega_c$. When applied to the geometry of a thick hemispherical shell, it enables a first attempt at modelling the elaborate sequence of neuronal cell migration from ventricles to the cortex and the formation of sulci and gyri in the cortex by elastic instability-driven folding. The color contours in the sequence of images in Figure \ref{fig:gyrification} show the evolving cell concentrations as cells migrate into the cortex, here modelled as the outer tenth of the thick shell representing the brain. The hemisphere has been clipped here to show the cell concentration contours. The inner surface, $\partial\Omega_c$, represents the ventricle, at which the cell concentration boundary condition $c = \bar{c}$ has been applied to represent a constant level maintained by cell birth. Also see Supplementary Movie S8. Circumferential swelling has been assumed in this case: 
\begin{equation}
\bF^\mathrm{g} = \left(\frac{c}{c_\mathrm{crit}} \right)^{1/3}\left(\bone - \bn_\mathrm{r}\otimes\bN_\mathrm{R}\right)
\label{eqn:circumgrowth}
\end{equation}
where $\bN_\mathrm{R}$ and $\bn_\mathrm{r}$ are, respectively, the radial directions in reference and swollen configurations. The buckling instability and bifurcation occur once the cell concentration reaches a critical value in the cortex, which is $c_\mathrm{crit} \sim 1.3$ in this example. { The buckling initiates at points where the mesh density changes abruptly, and the post-bifurcated shape cannot be considered to represent the actual distribution of gyri and sulci. Instead, it points to the role of sharp perturbations and discontinuities in initiating folding and creasing, and thereby influencing the post-bifurcated shape.} The remaining parameters for this computation appear in Table \ref{tbl:gyrification}. Gyri and sulci develop in the post-bifurcation regime here. Note that the wavelength and depth of gyri and sulci is $\sim 0.1\times$ the shell thickness, which is the cortical thickness in this example. 

The ability to model brain folding as a consequence of neuronal migration brings back the question of how unsymmetrical morphologies of gyrification and sulcification may result. Recall that this was alluded to in the Introduction. In other computations, not shown here, non-uniform, but smoothly varying cell concentration fields at the ventricular (inner) radius do remain non-uniform upon advection into the cortical layer (outer radius), but do not induce unsymmetric post-bifurcated mode shapes. The bifurcation is a global phenomenon in the sense that a high compressive circumferential stress develops in the cortical layer to cause the initial buckled mode shape. This mode shape is controlled by the boundary conditions { and sharper perturbations, such as the above abrupt changes in mesh density, rather than any local, smooth fluctuations} that may arise from the cell concentration field. However, there remains the possibility of the post-bifurcated shape being controlled by a subsequently inhomogeneous cell distribution.

\begin{table}[h]
\centering
\caption{Parameters for sulcification and gyrification induced by cell migration.}
\begin{tabular}{ |c|c|c|c|c|}
\hline
 Parameter & $D$ & $\bv$ & $\lambda$ & $\mu$ \\
 \hline
 Value & $1$ & $0.1\times\bn_\mathrm{r}$ & $2887$  & $1923$ \\
 \hline
\end{tabular}
 \label{tbl:gyrification}
\end{table}

\subsection{Morphogenesis by cell segregation and elastic buckling}

\begin{figure}
 \begin{minipage}[]{0.3\textwidth}
\centering
\subfloat[$t = 0$]{\includegraphics[width=1.1\textwidth]{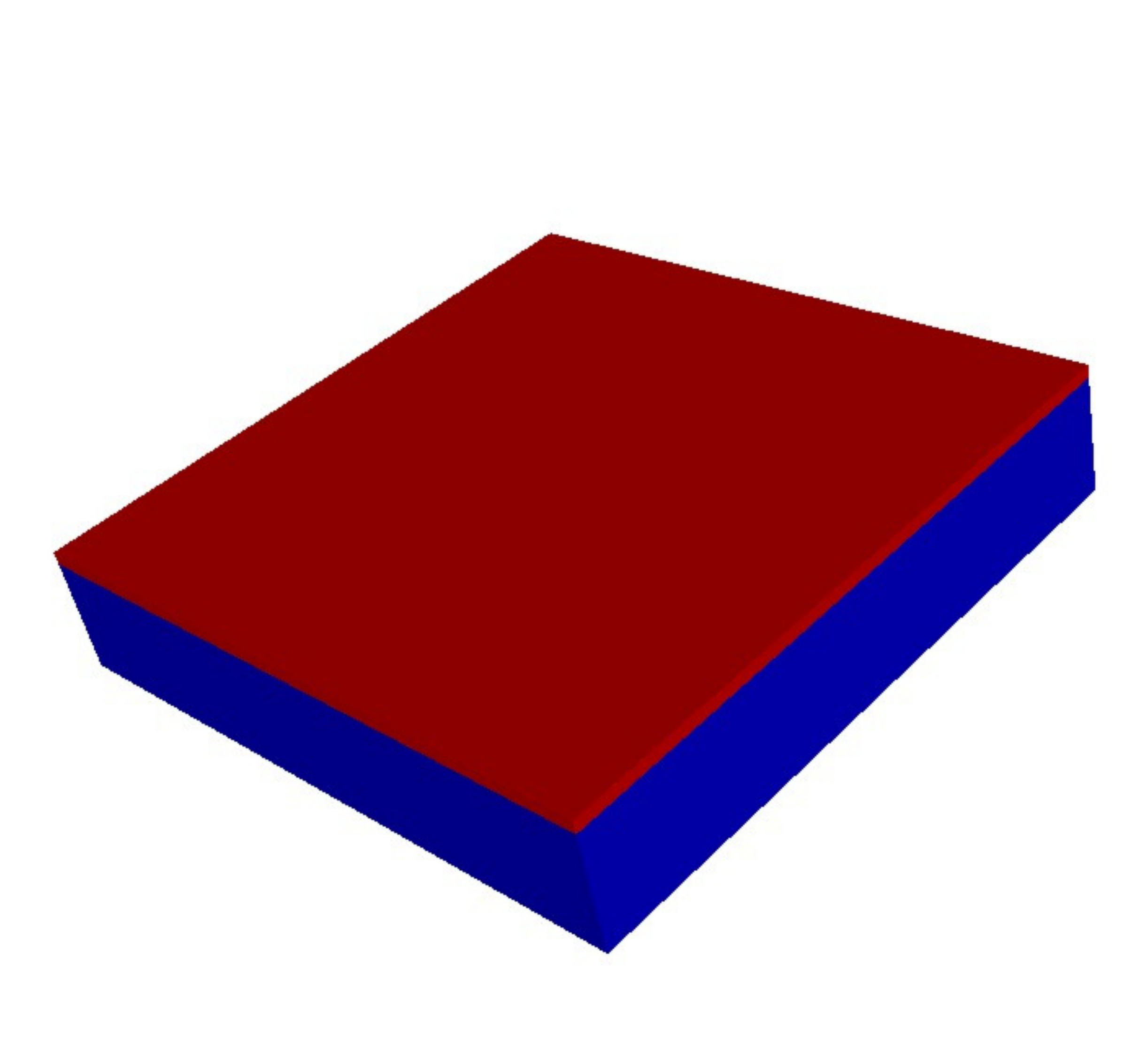}}
\end{minipage}
\begin{minipage}[]{0.3\textwidth}
\centering
\subfloat[$t = 20$]{\includegraphics[width=1.1\textwidth]{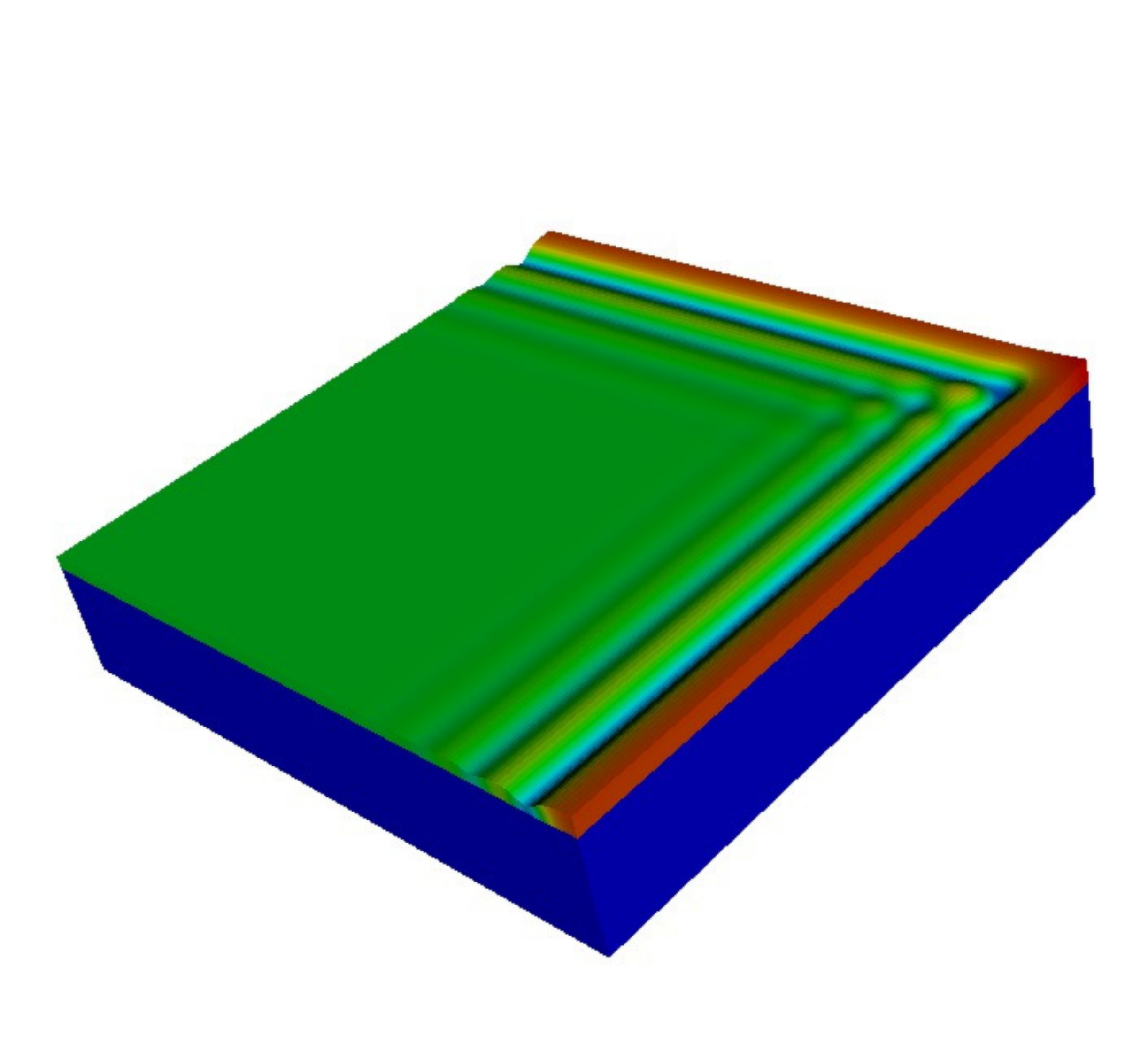}}
\end{minipage}
\begin{minipage}[]{0.3\textwidth}
\centering
\subfloat[$t = 48$]{\includegraphics[width=1.1\textwidth]{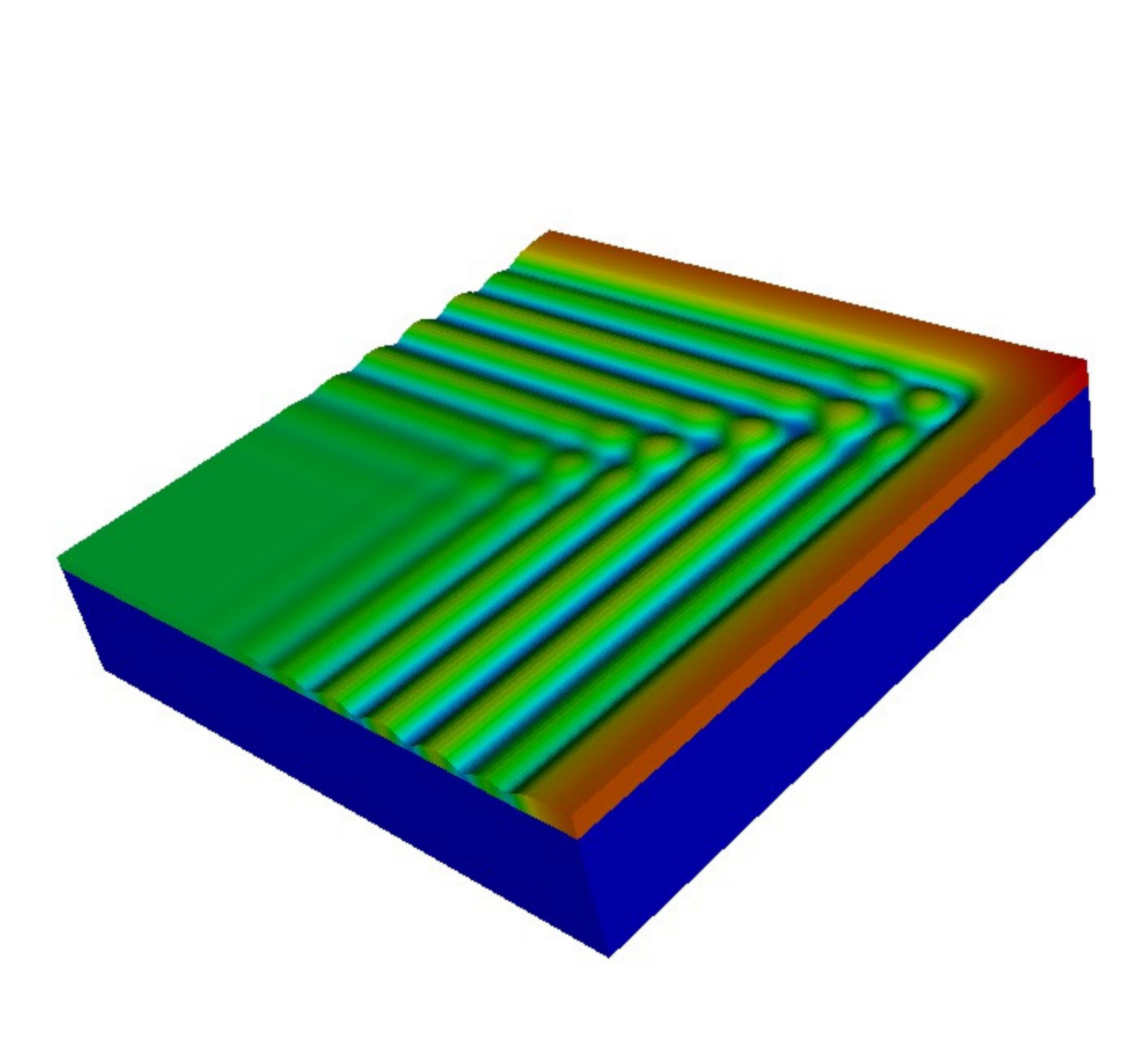}}
\end{minipage}\\
\begin{minipage}[]{0.3\textwidth}
\centering
\subfloat[$t = 72$]{\includegraphics[width=1.1\textwidth]{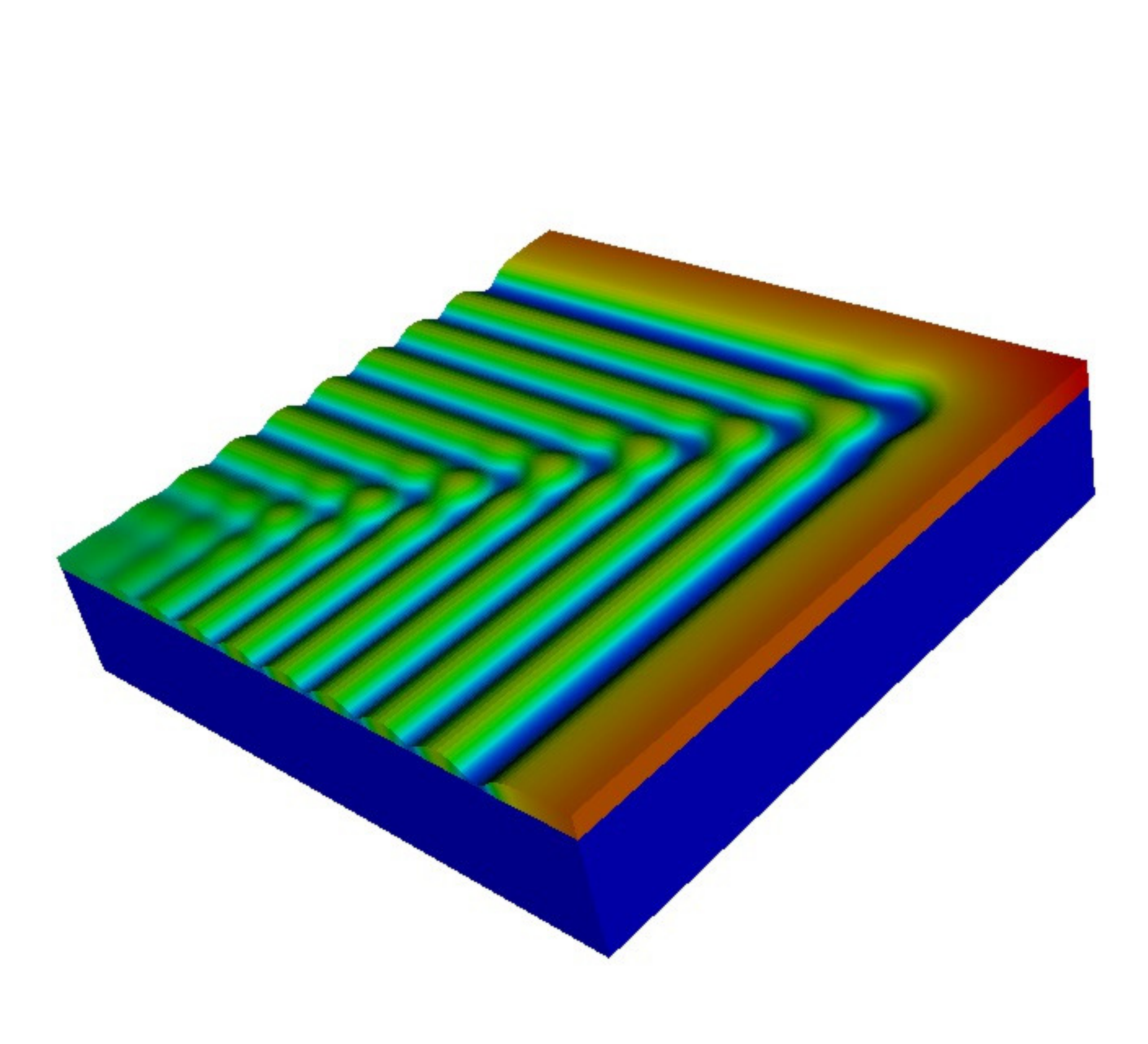}}
\end{minipage}
\begin{minipage}[]{0.3\textwidth}
\centering
\subfloat[$t = 96$]{\includegraphics[width=1.1\textwidth]{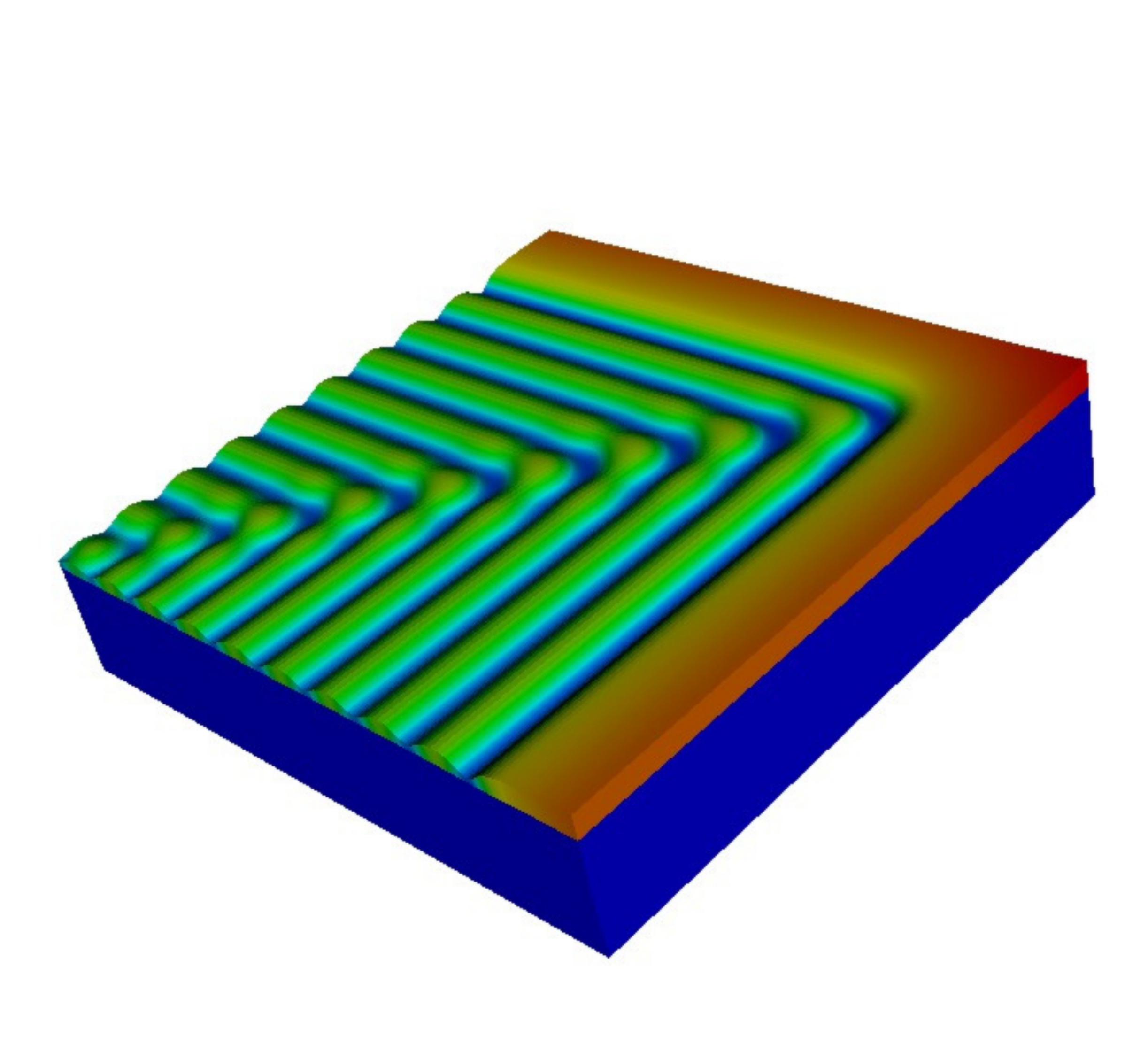}}
\end{minipage}
\begin{minipage}[]{0.3\textwidth}
\centering
\subfloat[$t = 120$]{\includegraphics[width=1.1\textwidth]{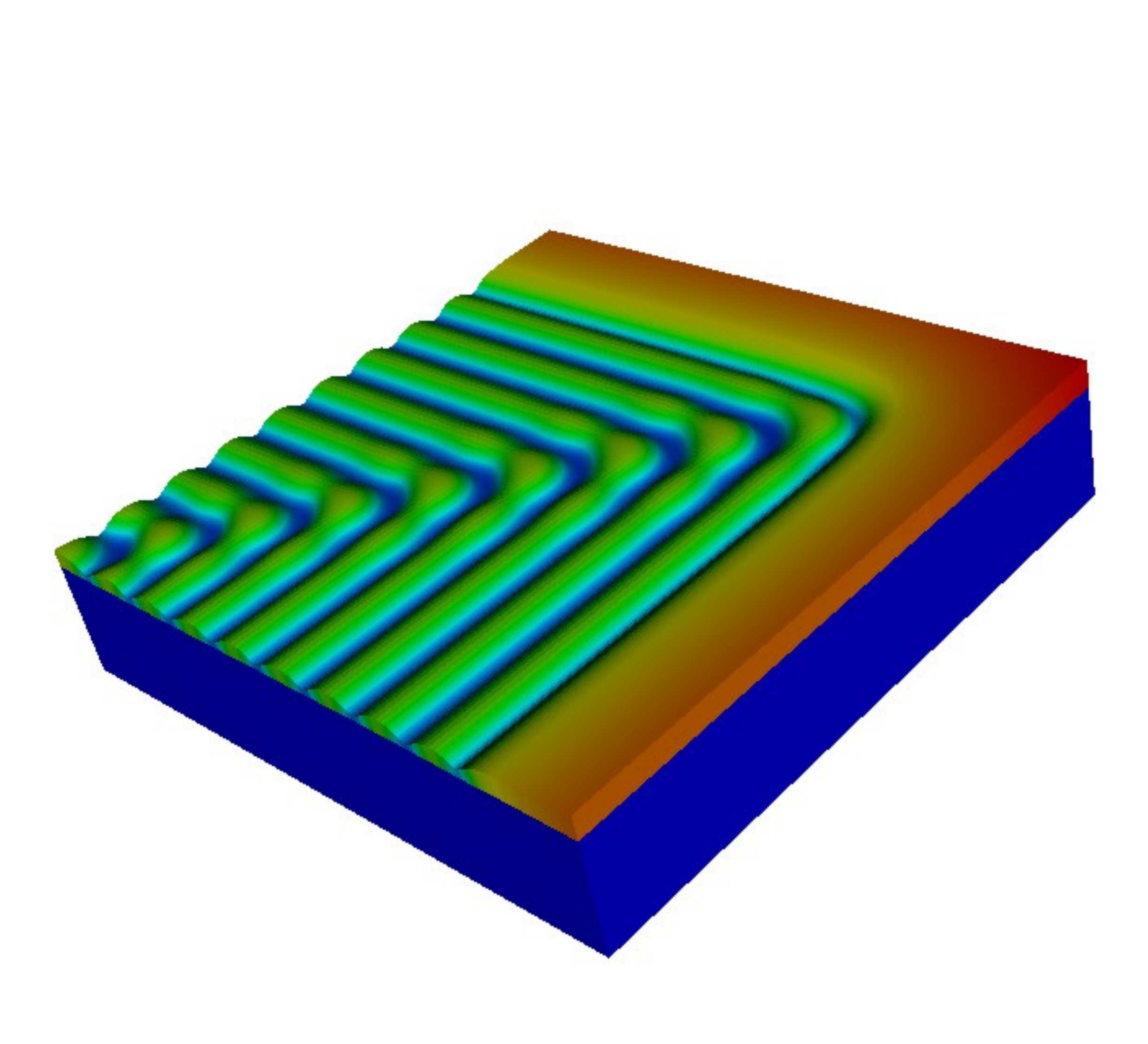}}
\end{minipage}
\caption{Patterning by cell segregation, coupled with morphogenesis by elastic bifurcation. Also see Supplementary Movie S9.}
\label{fig:chelast}
\end{figure}

Sections \ref{sec:diffcellmigbuckl} and \ref{sec:advcellmigbuckl} demonstrated phenomena in which cell transport laid down a scalar field by advection-diffusion, which then led to elastic buckling. The transport equations did not lay down a pattern of any sort, apart from delivering a high concentration of cells to cause the bifurcation. The final example in this section demonstrates inhomogeneous morphogenesis developing by elastic bifurcation, which is controlled by patterns forming in the underlying scalar field of cell concentration. The cells in this example are introduced in a thin layer lying on an elastic substrate, and undergo transport only within this layer (see Figure \ref{fig:chelast}). They form patterns governed by the cell segregation models of Section \ref{sec:phaseseg}, specifically the two-cell type tissue represented by Equations (\ref{eq:homogenergy}--\ref{eq:chD}). Due to the influx along two adjacent boundaries,  ridge-like patterns of cell segregated tissue develop. A high concentration front that develops near the flux boundaries reaches the spinodal regime and segregates into ridges of cells at the high concentration, $c_\beta$, and low concentration, $c_\alpha$, values. The cells of type $\beta$ being at a higher concentration than a threshold value, cause local elastic bifurcation and buckling into a bulge, while the $\alpha$ cells are at a lower concentration than this threshold and promote local contraction that aids the deformation into the corresponding valleys. A near steady state structure of orthogonal ridges develops, but as the boundary flux is maintained, the high concentration $\beta$ cells invade the valleys near the boundary, which also undergo buckling and from bulges. The ridges that develop bear similarities to fingerprint patterns (Figure \ref{fig:patternSizepartPosition}). Also see Supplementary Movie S9. Parameters for this computation appear in Table \ref{tbl:chelast}

\begin{table}[h]
\centering
\caption{Parameters for patterning by cell segregation, coupled with morphogenesis by elastic bifurcation..}
\begin{tabular}{ |c|c|c|c|c|c|c|c|}
\hline
 Parameter & $\omega$ & $c_\alpha$ & $c_\beta$ & $\kappa$ & $M$ & $\lambda$ & $\mu$\\
 \hline
 Value & $1$ & $0.05$ & $0.95$ & $1$ & $0.1$ & $2887$ & $1923$ \\
 \hline
\end{tabular}
 \label{tbl:chelast}
\end{table}

\subsection{Control of size and position by elastic buckling}
\label{sec:reacDiffSizePos}

In buckling phenomena of nonlinear elasticity, the initial buckling profile does not determine the final morphological form because of the dominance of post-bifurcation deformation. Very convoluted final forms are possible. Broadly, the relation between robustness of field position of the morphological form and boundary conditions that was noted before holds: far from boundaries, or on periodic 2-manifolds, such as the internal surfaces of intestines, field positions of morphological forms are not fixed. In contrast, the positions of prominent sulci and gyri that delineate neurological centers are controlled either by boundary effects from the end of the cortical layer at the spine or crown of the skull, or from the geometry of the underlying white matter \citep{Tallinen2016}. The same observation can be made about how the uniqueness of fingerprints may be controlled by boundary conditions defined at curves on the finger's surface where the print disappears. The same boundary effect is visible in defining the ridges of Figure \ref{fig:chelast}. However, it is not clear from these arguments how the positions of primordia of morphological forms that eventually give rise to facial features may be controlled. The locations of boundaries are not obvious. This invites the conjecture that the fine details of morphological form, such as on display in the external ear, seem unlikely to be determined by a single generation of elastic instabilities. Instead, primordia (buds) could be formed by initial elastic instabilities controlled by a reaction-diffusion or cell segregation pattern. These primordial morphological forms could be further sculpted by secondary elastic instabilities, themselves controlled by additional reaction-diffusion or cell segregation fields.

\section{Summary and outlook}

The primacy of partial differential equations for the representation of position and size, and therefore to control patterning and morphogenesis has been amply laid out in the literature, and summarized in the Introduction. While the dominant patterning models have been based on reaction-diffusion equations, it is worth noting that for tissue patterning at least, cell segregation and the resulting system of fourth-order transport equations present a compelling alternative rooted in a derivation that makes a direct connection with differential cell adhesion energies. The inherent stability of this system of equations and its evolution toward equilibrium, are also attractive features. Interesting questions arise around the roles of equilibrium states versus steady states in representing persistent patterns, and have been discussed in the text. While the multi-well tissue adhesion energy density function in a single scalar field does not provide an equilibrium solution to robustly represent tissues with more than two cell types, a two-field formulation can be employed to represent three cell types, as demonstrated here with three wells. This treatment is directly extendable to an arbitrary number of cell types, requiring only a two-parameter free energy function with as many wells in the plane as cell types. It to be pointed out is that the two fields in this description do not stand for cell types themselves, but any cell type can be described as a suitable linear combination of the two fields. The phase field literature is replete with formulations that model more complex structures \citep{Choksi2012}, including lipid bilayers \citep{Dai2013} and tubular structures \citep{Kraitzman2015}. Rather than review those works this perspective has chosen to dwell on a few broader observations. Phase segregation models also present a tempting representation of the epithelial to mesenchymal transition by a suitably parameterized transformation of the tissue adhesion energy density function  from double- to single-welled.

Three-dimensional morphogenesis of form is well-described by nonlinearly elastic bifurcation driven by inhomogeneous growth. This perspective has sought to connect it to patterning at the outset by focusing on cell migration as the mechanism for local growth in cell concentration. This has direct relevance to the process of neuronal migration that has been linked to eventual cortical gyrification and sulcification. However, tissue patterning is not a part of this process. When included via the cell segregation model, it leads to greater control over the positions of folds with local bulges and valleys. This last example appears to be a model that could be extended to morphogenesis of more complex features such as on a human face. Segregation of cell types in a tissue lays down a complex pattern, which by local peaks and valleys in concentration can induce at least the initial buds via elastic bifurcation that develop post-bifurcation forms of greater complexity such as the nose, lips and other features. Here it is also conjectured that further generations of reaction-diffusion and cell-segregation phenomena could themselves control subsequent elastic bifurcations and lead to fine details of morphological forms. 

{It is also important to acknowledge that the possibilities considered here for the coupling of patterning and morphogenesis are only one-way, with cell patterning defining how elastic bifurcations and post-bifurcation phenomena can shape tissues. J.D. Murray, in his seminal book \citep{Murray2003} has argued strongly for bidirectional, mechano-chemical coupling leading to closed-loop control of patterning and morphogenesis, without which robustness of form cannot be ensured. This represents a final form of the framework for mechano-chemical control of patterning and morphogenesis; the only reason it has not been included here, is that the author is not aware of well-described phenomena that suggest the influence of elastic effects on reaction-transport or cell segregation.}

\bibliographystyle{elsart-harv}
\bibliography{references}
\end{document}